\documentclass[fleqn,usenatbib]{mnras}
\usepackage[T1]{fontenc}
\usepackage{ae,aecompl}
\usepackage{graphicx}
\usepackage{amsmath}
\usepackage{amssymb}
\usepackage{txfonts}

\title[ES {\it vs} E]{Early-type galaxy speciation: Elliptical (E) and ellicular (ES)
  galaxies in the $M_{\rm bh}$-$M_{\rm \star,sph}$ diagram, and a
  merger-driven explanation for the origin of ES galaxies,
  anti-truncated stellar discs in lenticular (S0) galaxies, and the
  S\'ersicification of E galaxy light profiles}

\author[Graham]
{
Alister W.\ Graham$^{1}$\thanks{E-mail: AGraham@swin.edu.au}
\\
$^1$ Centre for Astrophysics and Supercomputing, Swinburne University of
Technology, Hawthorn, VIC 3122, Australia 
}

\date{Accepted XXX. Received YYY; in original form ZZZ}
\pubyear{2024}

\begin{document}
\label{firstpage}
\pagerange{\pageref{firstpage}--\pageref{lastpage}}
\maketitle

\begin{abstract}
In a recent series of papers, supermassive black holes were used to discern
pathways in galaxy evolution. By considering the black holes' coupling with
their host galaxy's bulge/spheroid, the progression of mass 
within each component has shed light on the chronological sequence of galaxy
speciation. Offsets between the galaxy-morphology-dependent $M_{\rm
  bh}$-$M_{\rm\star,sph}$ scaling relations trace a pattern of 'punctuated
equilibrium' arising from merger-driven transitions between galaxy types, such
as from spirals to dust-rich lenticulars and further to `ellicular' and
elliptical galaxies. This study delves deeper into the distinction between the
ellicular galaxies --- characterised by their intermediate-scale discs --- and
elliptical galaxies.  Along the way, it is shown how some anti-truncated
large-scale discs in lenticular galaxies can arise from the coexistence of a
steep intermediate-scale disc and a relatively shallow large-scale disc.  This
observation undermines application of the popular exponential-disc plus
S\'ersic-bulge model for lenticular galaxies and suggests some past bulge mass
measurements have been overestimated.  Furthermore, it is discussed how
merger-driven disc-heating and blending likely leads to the spheroidalisation
of discs and the conglomeration of multiple discs leads to the (high-$n$)
S\'ersicification of light profiles.  The ellicular and elliptical galaxy
distribution in the $M_{\rm bh}$-$M_{\rm\star,sph}$ diagram is explored
relative to major-merger-built lenticular galaxies and brightest cluster
galaxies.  The (super-)quadratic $M_{\rm bh}$-$M_{\rm\star}$ relations,
presented herein, for merger-built systems should aid studies of massive black
hole collisions and the gravitational wave background.  Finally, connections
to dwarf compact elliptical and ultra-compact dwarf galaxies, with their
100-1000 times higher $M_{\rm bh}/M_{\rm\star,sph}$ ratios, are presented.
\end{abstract}


\section{Introduction}

Over the past decade, the (black hole mass, $M_{\rm bh}$)-(spheroid stellar
mass, $M_{\rm \star,sph}$) diagram \citep{1995ARA&A..33..581K, 1998AJ....115.2285M,
  1999ApJ...519L..39W, 2001ApJ...553..677L} has steadily revealed clues about the coevolution of
galaxies and their central black hole.  For example, the 
dust-poor, low-mass lenticular (S0) galaxies appear to be primarily faded
primaeval S0 galaxies that need never have sported a spiral pattern. They
define a quasi-quadratic $M_{\rm bh}$-$M_{\rm \star,sph}$ relation with a slope of
2.39$\pm$0.81 and have $M_{\rm bh}/M_{\rm \star,sph}$ ratios that are, on average, five times higher
than in spiral (S) 
galaxies with the same $M_{\rm \star,sph}$ \citep{Graham-triangal}.  This 
disfavours these S0 galaxies being faded S galaxies that lost their gas and spiral pattern.
This population of S0 galaxies has also been highlighted for having an order of magnitude
higher $M_{\rm bh}$-to-(galaxy stellar mass, $M_{\rm \star,gal}$) ratio than the S galaxies and
merger-built early-type galaxies (ETGs) that collectively 
defined the Jeans-Reynolds-Hubble sequence \citep{1919pcsd.book.....J,
  1925MNRAS..85.1014R, 1926ApJ....64..321H}.\footnote{Sir James Jeans embraced
and advocated for the spheroid-to-spiral evolutionary sequence developed in the
`nebular hypothesis' \citep{Swedenborg1734, 1755anth.book.....K, LaPlace1796}. Reynolds
introduced the notion of a bridging population of S0 galaxies, which
\citet{1936rene.book.....H} adopted when he updated his schema, which had
initially introduced the `early' and `late' galaxy terminology to reflect the
centuries-old evolutionary scheme \citep[see][]{2019MNRAS.487.4995G}.}

to be more closely connected with their morphology
than their $M_{\rm bh}/M_{\rm \star,sph}$ 
ratio, alleged to dictate active galactic nuclei (AGN) feedback, itself widely
reported to regulate star formation. 
Whereas bipolar jets from a central black hole may not directly
impact the bulk of a galaxy disc's interstellar medium (ISM) --- although
radio-mode feedback can maintain an X-ray hot halo that curtails disc star
formation \citep{2003ApJ...599...38B} --- 
mergers can have a galaxy-wide influence on star formation and
build bulges through the redistribution of disc stars, thereby setting the galaxy
morphology.  Therefore, galaxy morphology, which reflects a galaxy's
accretion/merger history, should probably receive more attention
than it has regarding galaxy-black hole coevolution.\footnote{There is benefit in separating the S0 galaxies from
the E galaxies, and further separating dust-poor primordial S0 galaxies
from S0 galaxies built via major-wet-mergers. Indeed, exploration of galaxy morphology
relations with other properties may be befuddled unless the S0s are
separated according to their place in the galaxy speciation chain. For
example, this is why S galaxies initially appeared as a bridging population between E and
(all, combined) S0 galaxies in the ($M_{\rm bh}/M_{\rm \star,sph}$)-$M_{\rm
  \star,sph}$ diagram \citep{2019ApJ...876..155S}.}

To continue, 
major gas-rich, aka `wet', S galaxy mergers can transform S galaxies into
high-mass, dust-rich S0 galaxies (e.g. NGC~5128, NGC~4753) that have been
observed to define their
own $M_{\rm bh}$-$M_{\rm \star,sph}$ relation separate from the low-mass
dust-poor S0 galaxies and separate from the S galaxies.  These high-mass S0
galaxies can have more massive bulges than observed in S galaxies
\citep{2005ApJ...621..246B, Graham-S0}, again disfavouring a faded S galaxy
origin for much of this population. 
\citet{2012MNRAS.423...49K}, 
see also \citet{2020ApJ...905..154Y}, report that two-thirds of nearby ETGs with
prominent dust lanes are morphologically disturbed, suggestive of a merger
origin, and they are accompanied by bluer $NUV-r$ colours. This population has
low-to-intermediate levels of star formation midway between that of 
elliptical (E) and S galaxies \citep{Graham-SFR}. 
An additional round of merging of these massive 
galaxies will likely produce E galaxies that leave the `green
valley' or `green mountain' \citep{2018MNRAS.481.1183E, Graham-SFR}.
The offset nature of the E galaxies from the merger-built S0 galaxies in the
$M_{\rm bh}$-$M_{\rm \star,sph}$ \citep[and $M_{\rm
    bh}$-$\sigma$:][]{2000ApJ...539L...9F} diagram confirms a picture of
relatively gas-poor, aka `dry' 
merging \citep{Graham:Sahu:22a, 2023MNRAS.518.6293G}, as do their depleted
stellar 
cores \citep{1980Natur.287..307B, 1997AJ....114.1771F, 2004ApJ...613L..33G, 2012ApJ...744...74G}.
Once these E galaxies reside in a hot X-ray gas halo (either of their own making due
to, say, collision-induced shock-heating, stellar winds and supernova
explosions, or from their group or cluster environment), and with the shelter
of a disc now gone, their dust and cold gas tend not to survive
\citep{1979ApJ...231...77D, 2003ApJ...599...38B, 2021A&A...649A..18G}. 

The above cascade of mergers \citep[e.g.][]{2011MNRAS.413..101G}
erases the ordered motion of stars in discs and builds  
up a galaxy's spheroidal component \citep[e.g.][]{1993ApJ...403...74Q} and
black hole \citep[e.g.][]{2007ApJ...671...53M, 2012MNRAS.422.1306K}. 
The ensuing galaxy sequence described in  \citet{Graham-triangal} 
(i) reverses the original direction of assumed
evolution in the Jeans-Reynolds-Hubble sequence,
(ii) adds a new population of primaeval S0 galaxies, and
(iii) includes both an accretion- and  merger-driven pathway
for these S0 galaxies that is also evident in the colour-magnitude
diagram \citep{Graham-colour}.
The galaxy family tree can be thought of as an increasing 
entropy sequence \citep{1988ApJ...327...82R, 1999MNRAS.309..481L} in which
ordered rotation becomes disordered.
Rather than following a near-linear $M_{\rm bh}$-$M_{\rm \star,sph}$ relation.
the major-merger-built galaxies
roughly follow an $M_{\rm bh}$-$M_{\rm \star,sph}^2$ (or steeper) relation
with (galaxy type)-dependent zero-points
This has relevance beyond
understanding and quantifying the coevolution of galaxies and black holes,
extending to, for example, gravitational wave surveys
\citep[e.g.][]{2015MNRAS.451.2417R, 2023LRR....26....2A, 2024ApJ...966..105A, 2023arXiv230716628T}.
Furthermore, additional refined details are already getting teased out of the $M_{\rm
  bh}$-$M_{\rm \star,sph}$ diagram, such as the offset nature of brightest
cluster galaxies (BCGs) from regular (non-BCG) E galaxies, revealing the BCGs have been,
on average, built from a major dry merger of two non-BCG E galaxies (or some
equivalent combination of mass, e.g.\ 1 E $+$ 2 S0 galaxies).  Moreover, with dry
mergers establishing the E and BCG galaxy-black hole scaling
relations \citep{Graham:Sahu:22a}, AGN feedback has been relegated in these systems to
simply a maintenance role.

Prior to the formation of pressure-supported (aka dynamically-hot) discless E
galaxies, one encounters the ellicular (ES) galaxies
\citep{1966ApJ...146...28L} with intermediate-scale discs that do not dominate
the light at large radii.
ES galaxies are not quite the same as `disc ellipticals' \citep[][aka `discy
  ellipticals']{1988A&A...195L...1N}, which refers to alleged E galaxies with
pointed isophotes indicative of a somewhat or fully edge-on disc.  The latter
designation can include S0 galaxies with a large-scale disc and exclude ES
galaxies with a somewhat face-on and, thus, overlooked disc.
As presented in \citet{Graham-triangal}, the ES,e subtype is   
more akin to E galaxies, while the ES,b subtype encompasses compact massive
galaxies such as NGC~1277 \citep{2014ApJ...780L..20T, 2016ApJ...819...43G}
that appear more like the bulges of high-mass S0 galaxies, themselves resembling
`red nuggets' at $z\approx2.5\pm1$ \citep{2005ApJ...626..680D,
  2009ApJ...695..101D, 2015ApJ...804...32G, 2022MNRAS.514.3410H, 2023MNRAS.519.4651H}.  
While it is known that E4-E7 galaxies are (overwhelmingly) S0 galaxies with 
(often missed) large-scale discs 
\citep{1966ApJ...146...28L, 1970SvA....14..182G, 1984A&A...140L..39M,
  1990MNRAS.242P..24C, 1990ApJ...348...57V}, 
there are still hidden discs in galaxies that are routinely designated E0-E3
or S0.
\citet{2011MNRAS.414..888E} recognised some of these galaxies
(e.g. NGC~4476, NGC~4528, NGC~5631, PGC~28887 and UGC~3960) that are fast
rotators within $R_{\rm e}/2$ but slow rotators at 1 $R_{\rm e}$
and 
\citet{2014ApJ...791...80A} reported on galaxies that are fast rotators at
1 $R_{\rm e}$ but slow rotators at 2~$R_{\rm e}$ and beyond. These are ES
galaxies. 

ES galaxies are likely a bridging population between the (once and
perhaps still) dust-rich
high-mass S0 and the E galaxies.
They are not E nor S0 galaxies; similarly, they cannot be binned as
either a slow or a fast rotator. To better help distinguish them from E
galaxies, \citet{2016ApJ...831..132G} introduced the term ellicular to capture
their halfway nature between `elliptical' and `lenticular'
galaxies.\footnote{They appear closer to E than S0 galaxies and, in 
past literature, were more commonly designated E than S0, hence the preference
for the name ellicular over lentical.}
Here, the $M_{\rm bh}$--$M_{\rm \star}$ diagram is explored for signs of
displacement between ES and (non-BCG) E galaxies.  In the process, the
kinematics, morphology and light profiles of the ES and E galaxies were
revisited.  This has led to some interesting observations into the nature of
(evolving) galaxy light profiles and, thus, galaxy components.  In particular,
an explanation for, and the connection between, embedded and anti-truncated
discs is offered.  Furthermore, the dynamical heating of discs and the
effective erosion or camouflage(?) of their exponential density profile is
regarded as leading to the production of \citet{1963BAAA....6...41S} $R^{1/n}$
light profiles, whose mathematical properties are reviewed in
\citet{2005PASA...22..118G}.

The data for this investigation are mentioned in Section~\ref{Sec_data},
where a case study of NGC~3379 is presented.  This galaxy was not only
long-regraded as
an E galaxy \citep{1926ApJ....64..321H} but as an exemplification of the standard light profile
\citep{1979ApJS...40..699D} for \citet{1953MNRAS.113..134D} $R^{1/4}$
model.\footnote{\citet{1968adga.book.....S} developed the generalised $R^{1/n}$ model
to broadly describe galaxies that he thought had an $R^{1/4}$-bulge
and an exponential disc.
If no disc was present, it was assumed the  $R^{1/n}$ model
would default to the $R^{1/4}$ functional form. 
From the get-go, the $R^{1/n}$ model was
designed to accommodate the presence of an outer disc in S and S0 galaxies.
In the latter decades of the 1900s, this model was referred to as the $R^{1/4}$
`{\em law}' as it had become so ingrained as a mainstay of galaxy structure.}
Here, it is revealed that the fast rotating galaxy NGC~3379 likely has two
stellar discs, reflective of its merger origin.
Appendix~\ref{App_fits} 
provides new galaxy decompositions for additional ETGs and a collective insight from their 
light profiles.  In particular, it is revealed that the single $R^{1/n}$ model of
ETGs is 
likely accounting for a bulge plus a combination of (dynamically-heated) discs. 
It is suggested here that this mixing leads to the S\'ersicification of galaxy light
profiles. 
To illustrate this, an analysis of three additional S0 galaxy light profiles is 
presented in Appendix~\ref{App_fits},
along with a pair of ES galaxy profiles, a pair of E galaxy
profiles, and one E BCG profile.  Patterns emerge that support the successive merger
origin of these galaxies. All light profiles and magnitudes herein are calibrated to
the AB rather than the Vega magnitude system.  The results of regression
analyses between $M_{\rm bh}$ and $M_{\rm \star,sph}$ (and $M_{\rm
  \star,gal}$) are presented in Section~\ref{Sec_anal}, and a discussion is provided
in Section~\ref{Sec_Disc}, along with connections to other
elliptical/spheroid-like galaxies, namely compact elliptical (cE) and
ultra-compact dwarf (UCD) galaxies.

\section{Data}
\label{Sec_data}

\subsection{Sample}

\citet{Graham:Sahu:22a} tabulate spheroid (and
galaxy) stellar masses for a sample of $\sim$100 galaxies with directly
measured black hole masses. The galaxy sample has been uniformly imaged at 
3.6~$\mu$m with the Infrared Array
Camera - channel 1 \citep[IRAC-1:][]{2004ApJS..154...10F}
aboard the {\it Spitzer Space Telescope} \citep{2004ApJS..154....1W} 
and quantified with multi-component decompositions of the light. 
As described in \citet{2019ApJ...876..155S}, roughly half of the galaxy sample
were analysed using images provided by the Spitzer Survey of Stellar Structure
in Galaxies \citep[S$^4$G:][]{2010PASP..122.1397S, s4g-data} 
project\footnote{\url{https://irsa.ipac.caltech.edu/data/SPITZER/S4G/}}, with
the remainder analysed by either mosaicking Spitzer images, as described in
\citet{2016ApJS..222...10S}, or studying images taken from the Spitzer
Heritage Archive
\citep[SHA:][]{2010ASPC..434...14W}.\footnote{\url{https://irsa.ipac.caltech.edu/applications/Spitzer/SHA/}}
The {\it SHA} has a smaller effective pixel scale of 0.60 arcseconds, and thus a different
photometric zero-point of 21.581 mag.\footnote{The zero-point is
$21.097+2.5\log(0.75/0.60)^2=21.581$.}  Post Basic Calibrated Data (pbcd, aka
level 2) image mosaics (maic.fits files) were used from the SHA. 

In the current investigation, the subsamples of E and ES galaxies are studied
more closely.
In the past, and still today, ES galaxies tend to be grouped with E galaxies. 
Although \citet{Graham:Sahu:22b} recognised the ES galaxies as something of a halfway
station between E and S0 galaxies, the ES galaxies were intentionally
grouped with the E galaxies in \citet{Graham:Sahu:22a} and for the regression analysis in
\citet{Graham-triangal}.
That sample of ES and E galaxies excluded the ten BCGs (from the parent sample
of $\sim$100 galaxies), nine of which reside to the right of (not on) the
quadratic $M_{\rm bh}$--$M_{\rm \star}$ relation defined by the E and ES,e
galaxies\footnote{The notation ES,e is used for those ES galaxies more akin to
(extended) E galaxies.}, as expected if BCGs are, on average, built from the
merger of E and/or ES,e galaxies.
Brightest group galaxies (BGGs) and non-E BCG, of which there were just two,
are generally  not
expected to have experienced the same level of mergers as E BCG.
This previous grouping of ES,e and non-BCG E galaxies was done because it was considered too much to
introduce/discuss their separation in that study. The author also suspected 
that a couple of the E galaxies may have been misidentified ES (or S0) galaxies.
This has been investigated here with recourse to kinematic data
to inform the galaxy decompositions further. The light profiles of 
four S0 and 1 ES,e galaxies requiring the addition of a (previously missed) disc are presented here,
and new spheroid (stellar) masses are derived.  Four additional galaxies (2 E, 1
ES,e, and 1 E BCG) have their light profile presented and analysed here, and a slightly improved
model is obtained due in part to the inclusion of data at smaller ($\lesssim$5$\arcsec$)
and larger radii than used before. 

Table~\ref{TableES_E} lists the sample of non-BCG E and ES,e galaxies, along with a
reference to where each galaxy's decomposition can be found and notes 
regarding their kinematics. Spheroid and galaxy stellar masses \citep[based on a
diet Salpeter IMF][]{2024MNRAS.530.3429G} are also provided there, along with
the galaxies' central black hole masses taken from \citet{Graham:Sahu:22a}.
For the forthcoming investigation of the $M_{\rm bh}/M_{\rm\star}$ diagram,
these E and ES,e galaxies are now regarded as two separate populations.

\begin{table*}
\centering
\caption{ES and E galaxy sample}\label{TableES_E}
\begin{tabular}{lllll}
\hline
Galaxy    & $\log(M_{\rm bh}/M_\odot)$ & $\log(M_{\rm \star,sph}/M_{\odot})$ &
$\log(M_{\rm \star,gal}/M_{\odot})$ & (Source of decomposition). Kinematic notes \\
\hline
\multicolumn{4}{c}{(10 $-$ 1 S0 $+$ 1 former ES $=$) 10 ES,e galaxies} \\ 
NGC~0821  &  7.59$\pm$0.17 &  10.84$\pm$0.15  & 10.90$\pm$0.14  &  (GS23).   Regular (fast) rotator within $\approx$1~$R_{\rm e}$ (E+11, K+11, B+17).  Misaligned rotation\\
          &                &                  &                 &            at larger radii \citep{2009MNRAS.394.1249C, 2010ApJ...721..369T, 2014ApJ...783L..32S}. \\ 
NGC~1275  &  8.88$\pm$0.21 &  11.56$\pm$0.18  & 11.60$\pm$0.17  &  (SGD19).  BCG/cD in Perseus Cluster with KDC and undigested component.\\ 
          &                &                  &                 &            Merger \citep{2001AJ....122.2281C}.  $V/\sigma=48/246=0.20$ \citep{1985ApJ...299...41H}.\\ 
NGC~3377  &  7.89$\pm$0.03 &  10.30$\pm$0.14  & 10.36$\pm$0.13  &  (GS23).   Regular (fast) rotator (E+11, K+11, B+17). Twist in direction of rotation at \\
          &                &                  &                 &            larger radii \citep{2009MNRAS.394.1249C}. Possible merger \citep{2015AdSpR..55.2372N}.\\
NGC~3414  &  8.38$\pm$0.09 &  10.95$\pm$0.19  & 10.98$\pm$0.18  &  (SG16).   Nuclear disc, counter-rotating core (K+11) gives illusion of slow rotation.\\
          &                &                  &                 &            Polar ring \citep{1990AJ....100.1489W}. Merger \citep{2006MNRAS.370..891J}.\\ 
NGC~3585  &  8.49$\pm$0.13 &  11.38$\pm$0.15  & 11.39$\pm$0.14  &  (SG16).   High-rotation \citep{1995AandA...293...20S}. \\  
NGC~3607  &  8.16$\pm$0.18 &  11.29$\pm$0.18  & 11.37$\pm$0.17  &  (This work). Regular (fast) rotator (E+11, K+11). Merger \citep{2009AandA...505...73R}. \\ 
NGC~4291  &  8.51$\pm$0.37 &  10.68$\pm$0.19  & 10.72$\pm$0.18  &  (This work). $V/\sigma=68/278 = 0.24$ \citep{1994MNRAS.269..785B}. Previously E galaxy.\\ 
NGC~4473  &  8.07$\pm$0.36 &  10.75$\pm$0.13  & 10.83$\pm$0.13  &  (SG16).   Fast rotator (E+11, B+17). Counter-rotating inner discs (K+11).  \\ 
NGC~4552  &  8.67$\pm$0.05 &  11.01$\pm$0.16  & 11.07$\pm$0.16  &  (SGD19).  Non-regular (slow) rotator (K+11). Twisted isovelocity contours (E+11).\\
          &                &                  &                 &            Disc with inner ring and undigested component. Shells. \citep{1983ApJ...274..534M}.\\
          &                &                  &                 &            BGG in NGC~4552 Group. Well on its way to becoming an E galaxy.\\ 
NGC~4621  &  8.59$\pm$0.06 &  11.24$\pm$0.16  & 11.28$\pm$0.15  &  (SG16).   Regular (fast) rotator (E+11, K+11). Rare prolate kinematics.\\
\hline 
\multicolumn{4}{c}{(17 $-$ 3 S0 $-$ 1 ES $=$) 13 non-BCG E galaxies} \\
IC~1459  &   9.38$\pm$0.20 &  11.69$\pm$0.17  & 11.69$\pm$0.17  &  (SG16).   Counter-rotating core or discs \citep{1988ApJ...327L..55F}\\
         &                 &                  &                 &            \citep{2019MNRAS.488.1679P}. $V/\sigma=40/302=0.13$ \citep{1989ApJ...344..613F}. Ionised gas disc\\
         &                 &                  &                 &            with spiral pattern \citep{1990AandA...228L...9G}.  Shells \citep{1995AJ....109.1576F}.\\ 
NGC~1407 &   9.65$\pm$0.08 &  11.60$\pm$0.17  & 11.66$\pm$0.16  &  (SGD19).  KDC. Merger and slow rotator (B+17).\\
         &                 &                  &                 &            \citep{2014ApJ...783L..32S, 2018MNRAS.480.3215J}.\\
NGC~1600 &  10.25$\pm$0.04 &  12.06$\pm$0.13  & 12.06$\pm$0.13  &  (SGD19).  $V/\sigma=7/312 = 0.02$ \citep{1994MNRAS.269..785B}. Possible merger.\\
         &                 &                  &                 &            Shells \citep{1992MNRAS.254..723F, 1999MNRAS.310..879M}.\\
NGC~3608 &   8.30$\pm$0.17 &  10.98$\pm$0.14  & 10.98$\pm$0.14  &  (SG16).   Slow rotator (E+11, B+17). Counter-rotating core (K+11).  Marginally\\
         &                 &                  &                 &            misaligned phot./kin.\ axis. Shells. \citep{1992MNRAS.254..723F, 2009MNRAS.394.1249C}.\\
NGC~3923 &   9.47$\pm$0.13 &  11.55$\pm$0.17  & 11.55$\pm$0.17  &  (SGD19).  Slow rotator \citep{1997MNRAS.288....1P}. Merger. Shells \citep{1988ApJ...326..596P}, \\
         &                 &                  &                 &            \citep{1995ApJ...443..570Z, 2016AandA...588A..77B}.\\ 
NGC~4261 &   9.20$\pm$0.09 &  11.52$\pm$0.16  & 11.54$\pm$0.15  &  (SG16).   Non-regular slow rotator (E+11, K+11). Twisted isovelocity contours\\
         &                 &                  &                 &            (E+11). Past mergers \citep{2012MNRAS.421.2872B, 2013ApJ...773...87D}.\\
NGC~4374 &   8.95$\pm$0.05 &  11.61$\pm$0.15  & 11.61$\pm$0.15  &  (SG16).   Non-regular slow rotator (E+11, K+11, B+17). Misaligned phot./kin.\ axis\\ 
         &                 &                  &                 &            \citep{2009MNRAS.394.1249C}. Many(?) mergers \citep{2004AandA...415..499G, 2013ApJ...768..137Y}. \\   
NGC~5077 &   8.85$\pm$0.23 &  11.27$\pm$0.21  & 11.37$\pm$0.20  &  (This work). KDC, ring, and accreted gas/dust disc. \citep{2021AandA...650A..34R}.\\  
         &                 &                  &                 &             Possible ES,e galaxy.\\
NGC~5576 &   8.19$\pm$0.10 &  10.90$\pm$0.16  & 10.90$\pm$0.16  &  (SG16).   Non-regular slow rotator (E+11, K+11).\\
NGC~5846 &   9.04$\pm$0.06 &  11.55$\pm$0.15  & 11.55$\pm$0.15  &  (SG16).   Non-regular slow rotator (K+11, B+17). Marginally misaligned phot./kin.\\\
         &                 &                  &                 &            axis \citep{2009MNRAS.394.1249C}. Filamentary dust \citep{1997AJ....113..887F}.\\  
NGC~6251 &   8.77$\pm$0.16 &  11.80$\pm$0.16  & 11.84$\pm$0.15  &  (This work). $V/\sigma=40/293 = 0.14$ \citep{1985ApJ...299...41H}; \\
         &                 &                  &                 &              $V/\sigma=54/288 = 0.19$ \citep{1992AandA...262...52B}.  Dust lane \citep{1983MNRAS.203P..39N}. \\
NGC~7052 &   9.35$\pm$0.05 &  11.46$\pm$0.13  & 11.46$\pm$0.13  &  (SGD19).  $V/\sigma=40/270 = 0.15$ \citep{1988AandA...195L...5W}. \\
NGC~7619 &   9.36$\pm$0.09 &  11.69$\pm$0.14  & 11.71$\pm$0.13  &  (SG16).   Regular borderline-slow rotator$^{\dagger}$, $V/\sigma=0.16$ \citep{2022MNRAS.515.1104L}. \\  
\hline 
\multicolumn{4}{c}{4 (previously classified E and ES,e) S0 galaxies} \\
NGC~4697 &   8.26$\pm$0.04 &  10.20$\pm$0.16  & 10.86$\pm$0.14  &  (This Work).  Regular (fast) rotator (E+11, K+11, B+17). BGG in NGC~4697 Group. \\
         &                 &                  &                 &                Merger \citep{2006AJ....131..837S, 2015MNRAS.452...99S}. Possible ES,e galaxy\\ 
NGC~3091 &   9.62$\pm$0.08 &  11.25$\pm$0.22  & 11.70$\pm$0.20  &  (This work).  $V/\sigma=71/286$ \citep{1989ApJ...344..613F}; $V/\sigma=96/259$ \citep{1992AandA...262...52B}.\\
NGC~3379 &   8.62$\pm$0.13 &  10.27$\pm$0.20  & 10.89$\pm$0.18  &  (This work).  Regular (fast) rotator $V_{\rm rot}\gtrsim 83$ km s$^{-1}$ (K+11), \citet{2007ApJ...664..257D}.  \\
NGC~4649 &   9.66$\pm$0.10 &  10.81$\pm$0.16  & 11.48$\pm$0.14  &  (This work).  Regular (fast) rotator $V_{\rm rot}\gtrsim 114$ km s$^{-1}$ (K+11; B+17),\\   
         &                 &                  &                 &                 \citep{2001ApJ...546..903D}.\\
\hline
\multicolumn{4}{c}{Brightest Cluster Galaxy (E galaxy)} \\
NGC~4486 &   9.81$\pm$0.06 &  11.58$\pm$0.15  & 11.58$\pm$0.15  &  (This work).  Non-rotating galaxy (K+11).\\
\hline
\end{tabular}

Black hole masses and most stellar masses are sourced from \citet{Graham:Sahu:22a}.
References showing decompositions of the galaxy light are
SG16 = \citet{2016ApJS..222...10S}, 
SGD19 = \citet{2019ApJ...876..155S}, 
and GS23 = \citet{Graham:Sahu:22b}.
New spheroid and galaxy stellar masses are provided in `This work' for nine galaxies. 
Kinematic references: 
K+11 = \citet{2011MNRAS.414.2923K}; 
E+11 = \citet{2011MNRAS.414..888E}; and
B+17 = \citep{2017MNRAS.467.4540B}.  
K+11 and E+11  measurements pertain to roughly the inner 1~$R_{\rm e}$, where embedded discs reside.
KDC = Kinematically decoupled core.
$^{\dagger}$Something seems amiss with the plotting of NGC~7619 in the $\lambda_{\rm e}$-magnitude diagram in
  \citet[][their figure~5]{2022MNRAS.515.1104L}. 
  
\end{table*}

\begin{figure*}
\begin{center}
\includegraphics[trim=0.0cm 0cm 0.0cm 0cm, height=0.26\textwidth,
  angle=0]{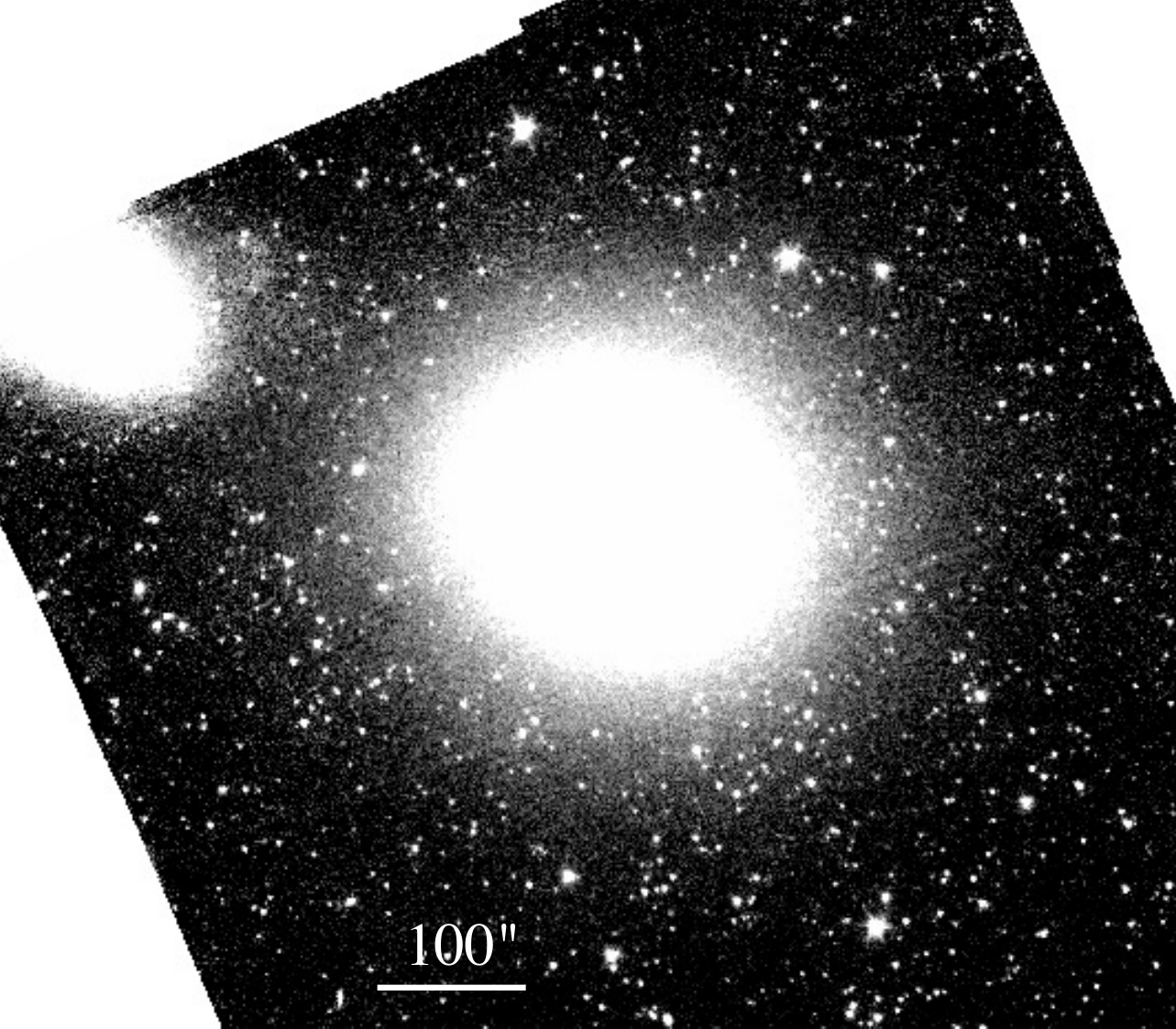} 
\includegraphics[trim=0.0cm 0cm 0.0cm 0cm, height=0.26\textwidth,
  angle=0]{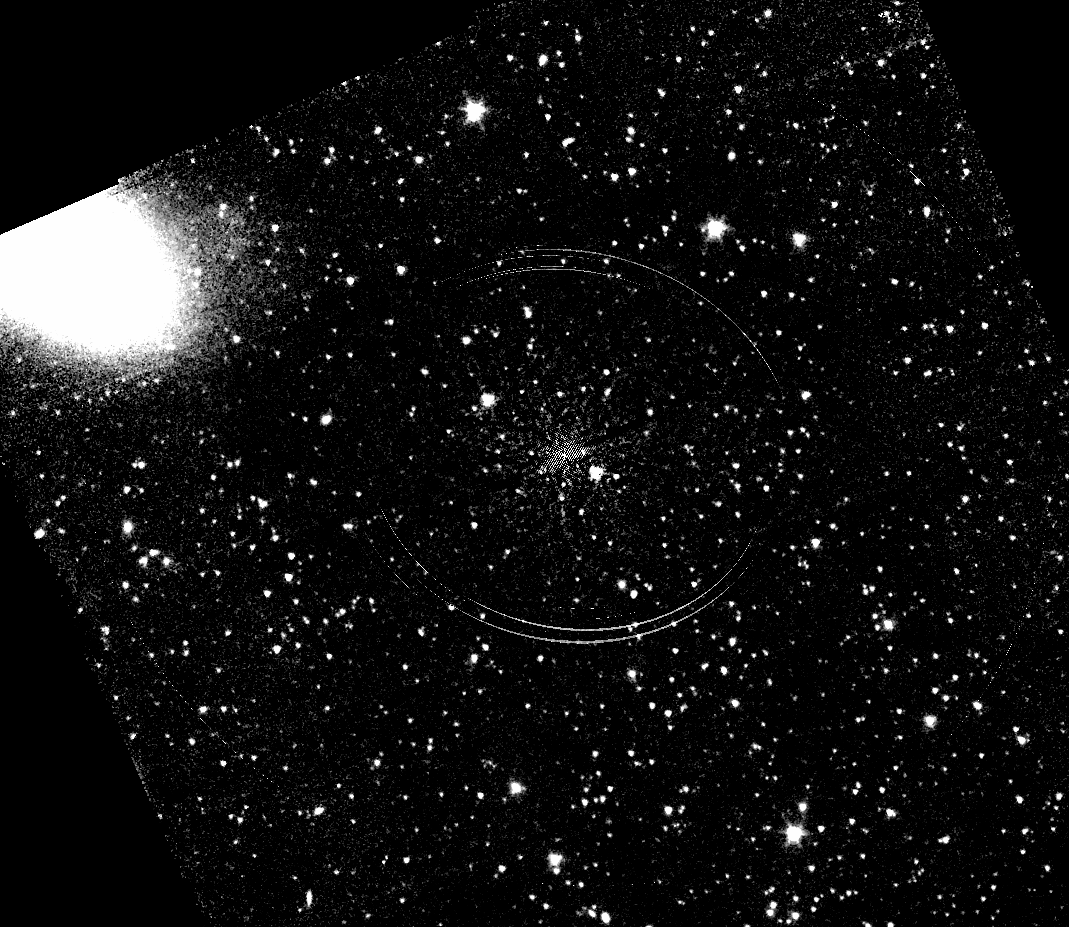}
\includegraphics[trim=0.0cm 0cm 0.0cm 0cm, height=0.26\textwidth,
  angle=0]{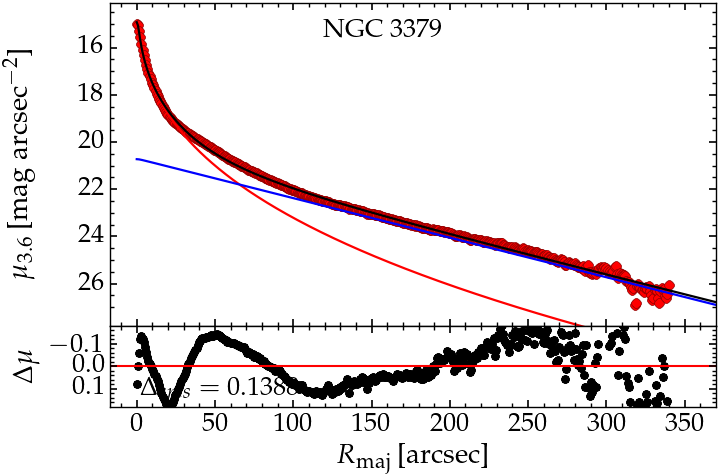} 
\end{center}
\caption{Left:  {\it Spitzer/IRAC1} 3.6~$\mu$m image of NGC~3379, courtesy of
 S$^4$G \citep[pipeline 1 image:][]{2015ApJS..219....3M}. 
The scale bar is 100$\arcsec$ $=$ 5.3~kpc long. 
 North is up, and east is left. 
 Middle panel:  The galaxy light of NGC~3379 and 
  the `sky-background' have been independently determined and subtracted.
Right panel: 
S\'ersic spheroid (red curve, $R_{\rm e,maj}=20\arcsec.9$,
$\mu_{\rm e}=18.90$ mag arcsec$^{-2}$, and
$n_{\rm maj}=2.74$) and exponential disc (blue line: $\mu_0=20.70$ mag
arcsec$^{-2}$, $h=65\arcsec = 3.45$~kpc). 
The disc is also evident in the kinematic map from \citet[][their figure~C1]{2011MNRAS.414.2923K}.
}
\label{Fig_N3379} 
\end{figure*}

\begin{figure}
\begin{center}
$
\begin{array}{c}
  \includegraphics[trim=0.0cm 0cm 0.0cm 0cm, width=0.85\columnwidth,
  angle=0]{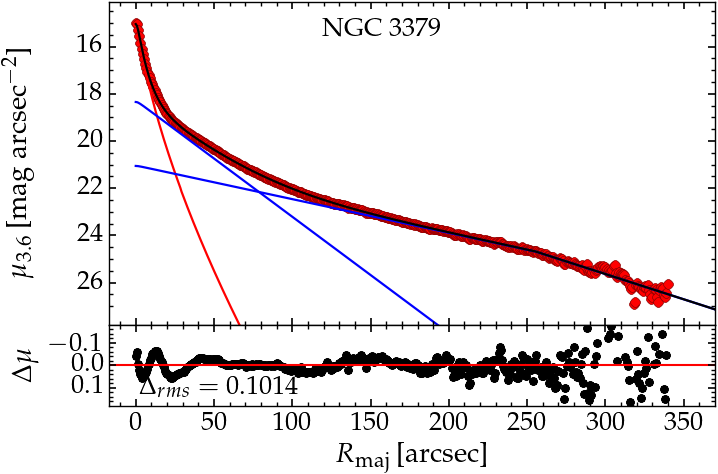} \\
\includegraphics[trim=0.0cm 0cm 0.0cm 0cm, width=0.85\columnwidth,
  angle=0]{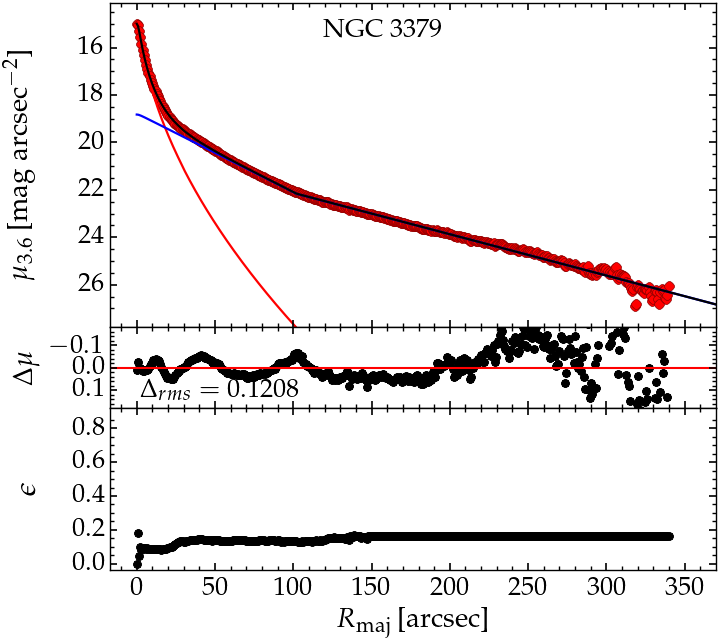} \\
\end{array}
$
\end{center}
\caption{Preferred decompositions for NGC~3379. 
  Top panel: S\'ersic spheroid (red curve: $R_{\rm e,maj}=6\arcsec.4$,
  $\mu_{\rm e}=17.10$ mag arcsec$^{-2}$, and
  $n_{\rm maj}=1.44$) plus intermediate-scale disc (straight
  blue line: $\mu_0=18.23$ mag arcsec$^{-2}$, $h=21\arcsec.5 = 1.14$~kpc) and
  weakly-truncated (at $r=261\arcsec$) large-scale disc (bent blue line:
  $\mu_0=20.97$ mag arcsec$^{-2}$, $h_{\rm inner}=74\arcsec = 3.9$~kpc,
  $h_{\rm outer}=50\arcsec = 2.65$~kpc).
  Bottom panel: S\'ersic spheroid (red curve: $R_{\rm e,maj}=8\arcsec.5$,
 $\mu_e=17.55$ mag arcsec$^{-2}$,  and  $n_{\rm maj}=1.87$)
  plus anti-truncated disc (bent blue line: $\mu_0=18.77$ mag arcsec$^{-2}$,
  $R_{\rm bend}=102\arcsec$, $h_{\rm inner}=32\arcsec.6 = 1.73$~kpc,
  $h_{\rm outer}=62\arcsec.3 = 3.3$~kpc). 
  Note: as seen in the upper panel, 
  an additional downward bend at $r\approx 260\arcsec$ would improve the
  fit in the lower panel. 
}
\label{Fig_dd}
\end{figure}

\subsection{ES,e galaxies} 

As noted above, ES galaxies, including many `disc ellipticals' \citep{1988A&A...195L...1N,
  1988oseg.proc..150C, 1988A&AS...74...25M}, have tended to be grouped with the E
galaxies, and this convention was used in \citet{Graham:Sahu:22a}.  However,
the ES,e (and ES,b) galaxies in \citet{Graham:Sahu:22a} had been modelled as bulge$+$disc systems, 
and half of these were previously identified as such
by \citet{1988A&A...195L...1N}.  The present sample's ES,e galaxies are 
NGC~821,   
NGC~1275 (BCG in the Perseus Cluster), 
NGC~3377, 
NGC~3414, 
NGC~3585, 
NGC~3607 (recognised by \citet{2011MNRAS.414.2923K} as having a fast embedded disc within a slower disc), 
NGC~4473 \citep[][counter-rotating stars,
 `2$\sigma$ galaxy']{2011MNRAS.414..888E},
NGC~4552 (aka M89; BGG in NGC~4552 Group of 12 within the Virgo~B Cluster), 
NGC~4621 \citep[a prototype E galaxy in ][]{1926ApJ....64..321H}, and 
NGC~4697 (BGG in NGC~4697 Group of 37).  Further details are provided in
Table~\ref{TableES_E}. 

\citet{2014ApJ...780...69L} had four of the above galaxies in their sample 
and modelled NGC~821, NGC~3377, and NGC~4697 with a thin embedded disc,
missing only the disc in the galaxy NGC~4473, which they classed as an E
galaxy.  The embedded discs in NGC~821, NGC~3377, and NGC~4473 are evident
in these galaxies' modified/local stellar spin profiles \citep[][their
  figure~2]{2017MNRAS.470.1321B}.  \citet{2014ApJ...780...69L} additionally
regarded NGC~3115 and NGC~4342 \citep{1998MNRAS.293..343V} --- from the larger
sample of $\sim$100 galaxies --- as having embedded 
discs contributing $\sim$10 per cent of the total flux.
NGC~4342 was modelled with an embedded disc in \citet{2019ApJ...876..155S} but
is problematic due to the galaxy's stripped nature. It is, therefore, not
included in the current ES,e galaxy data set.
As in \citet{Graham:Sahu:22b}, NGC~3115 is also considered an ES galaxy with
an embedded disc, but it is a compact ES,b galaxy (like Mrk~1216, NGC~1271 and
NGC~1277, NGC~1332, NGC~5845 and NGC~6861) and therefore it is included with
the (merger-built) dust-rich S0 galaxies because it is more like the {\it
  bulge} of a massive S0 galaxy.
In the literature, such ES,b galaxies have tended to be grouped with the S0
galaxies, which was done by \citet{Graham:Sahu:22a} for the four ES,b
galaxies identified there.  These are excluded from the ES,e sample.

\citet{Graham:Sahu:22a} and \citet{Graham:Sahu:22b} identified 17 (non-BCG,
non-cD) E galaxies, ten suspected ES,e galaxies, and an additional 8 E (plus 1
S0) cD/BCGs.
These E and ES,e classifications have been checked via a re-evaluation of the
light profiles, with recourse to published kinematic data.  Changes are
detailed in the following subsection and Appendix~\ref{App_fits}.

\subsection{Misclassified ES and S0 galaxies} 

Stellar discs are often missed in ETGs. 
This oversight effects studies concerned with galaxy morphology in the 
$M_{\rm bh}$--$M_{\rm \star}$ diagram.  For example, if S0 galaxies
are modelled as discless E galaxies, this can overestimate the spheroid 
mass by a typical factor of 4 \citep{2005MNRAS.362.1319L}.  This affects the
$M_{\rm bh}$--$M_{\rm \star,sph}$ scaling relations and impacts the
reported sizes and densities of the galaxies' spheroidal component.  

\citet{Graham:Sahu:22a} mistakingly regarded NGC~3379 \citep[from][]{2016ApJS..222...10S} and
NGC~3091 and NGC~4649 \citep[from][]{2019ApJ...876..155S} as E galaxies.  This
was because additional components, such as discs, were only included in the modelling of
the galaxies' light if convincing evidence was at hand. 
These three galaxies are remodelled here to reveal their S0 nature better. 
NGC~4697 \citep[from][]{2016ApJS..222...10S} is also remodelled here as an S0
rather than an ES galaxy.  However, arguably, the outer thick disc in this
galaxy is better regarded as a slowly rotating 
spheroid, as is the case with, for example,  NGC~821 and NGC~3377 \citep{2014ApJ...791...80A,
  2017MNRAS.470.1321B}. This is discussed later in regard to the
metamorphosis of galaxies.  Finally, NGC~4291 
\citep[from][]{2016ApJS..222...10S}  is remodelled as an ES,e galaxy rather
than an E galaxy, although the change to the spheroid mass is only 0.12~dex. 

The following subsection presents the reanalysis of NGC~3379, while the 
additional remodelled galaxies are presented in Appendix~\ref{App_fits} to aid
readability.  Collectively,
a double disc structure is revealed for the S0 galaxies that is well captured with an
anti-truncated disc model.
A double disc structure likely reflects the (e.g.\ S $+$ S) merger origin of
these S0 galaxies. 
Appendix~\ref{App_fits} also remodels a further four 
galaxies but does not alter their morphological classification.
As shall be seen, the profiles and their decompositions prove
to be highly illustrative as to the nature and evolution of ETGs.  The
evolving nature of the ETG light profiles shown in Appendix~\ref{App_fits}
reveals how the double disc structure of massive S0 galaxies gives way to the
embedded discs of ES galaxies and the apparent absence of discs in E
galaxies.  As seen in Section~\ref{Sec_anal}, this picture is
accompanied by transitions of galaxy morphology across the $M_{\rm
  bh}$--$M_{\rm \star,sph}$ diagram. 

The revised sample of (17 $-$ 3 S0 $-$ NGC~4291 $=$) 13 non-(cD/BCG) E galaxies are listed in
Table~\ref{TableES_E} 
along with (10 $+$ NGC~4291 $-$ NGC~4697 $=$) 10 ES,e galaxies.

\subsubsection{NGC~3379 (M105): double disc or anti-truncated disc}
\label{Sec3379} 

NGC~3379 is supposed to be the personification of the $R^{1/4}$ model 
  \citep{1979ApJS...40..699D}.  Here, it is demonstrated that the light
  profile of this
  fast-rotating galaxy \citep{1985AJ.....90..169D, 2011MNRAS.414.2923K}
  is best represented by a bulge plus a double-disc, or antitruncated disc, structure.
  
A 3.6~$\mu$m image for NGC~3379 was taken from S$^4$G and has a
pixel scale of 0.75 arcseconds and a photometric zero-point of 21.097
mag.  The Image Reduction and Analysis Facility (IRAF) task {\sc
  Isofit} \citep{2015ApJ...810..120C}, which replaces the flawed {\sc
  ELLIPSE} task \citep{1987MNRAS.226..747J}, was used to extract the
major-axis and geometric mean radius, $r=\sqrt{ab}$, light profile, known as the
equivalent-axis light profile\footnote{It is equivalent to a light profile
extracted from circularised isophotes, initially with major and minor axis
lengths $a$ and $b$.}, while also capturing position angle
twists, ellipticity gradients and radially-dependent, higher-order
Fourier harmonic terms describing isophotal departures from a pure
ellipse.  The new IRAF task {\sc Cmodel} \citep{2015ApJ...810..120C}
was used to generate a two-dimensional model of the galaxy that was
subtracted from the image.  The associated galaxy light profile that
accounts for all of this information was modelled with the Python code
{\sc Profiler} \citep{2016PASA...33...62C}, which can fit 
multiple components.  The result is shown in Fig.~\ref{Fig_N3379} when
using a traditional S\'ersic-bulge plus exponential-disc model.

NGC~3379 is the second brightest galaxy in the M96 (aka Leo I) Group, a
group of at least 36 galaxies \citep{2015AstBu..70....1K}, and it is the brightest
galaxy in the smaller NGC~3779 Group.\footnote{Observations have come a long
way since the Herschels observed what was GC~2203 \citep{1864RSPT..154....1H},
now NGC~3379 \citep{1888MmRAS..49....1D}, with the 30-inch metal 
reflecting telescope that they purchased in 1793 from the estate of John Michell, 
who had famously predicted black holes \citet{1784RSPT...74...35M}.}
It is not optimally 
modelled in \citet{2016ApJS..222...10S} due to an overlooked disc
\citep{1991ApJ...371..535C, 1999AJ....117..839S, 2011MNRAS.414.2923K}
with an ellipticity on the sky of just 0.14
\citep{2016ApJS..222...10S}.  Modelling NGC~3379 to nearly six arcminutes
(Fig.~\ref{Fig_N3379}) yields a total 
(model-extrapolated) apparent magnitude $\mathfrak{M}_{\rm 3.6,gal} = 8.74$ mag 
and a bulge magnitude $\mathfrak{M}_{\rm 3.6,sph} = 9.31$ mag when including a single, exponential, large-scale disc. 
This yields a bulge-to-total ($B/T$) ratio of 0.59.
Using a Galactic-extinction-corrected \citep{1998ApJ...500..525S, 2011ApJ...737..103S} $(B-V)_{\rm Vega}$ colour of 0.94 mag
\citep[RC3:][]{1991rc3..book.....D}, courtesy of the {\it NASA/IPAC
  Extragalactic Database}\footnote{\url{https://doi.org/10.26132/NED5}} 
({\it NED})\footnote{\url{https://ned.ipac.caltech.edu/}}, gives a
(stellar mass)-to-light ratio $M_\star/L_{3.6} = 0.80$ \citep[][their equation~4]{Graham:Sahu:22a}.  This is based on a
diet-Salpeter IMF \citep{2003ApJS..149..289B}, as are all the stellar masses
presented herein.  At a luminosity distance of 10.9$\pm$1.6~Mpc
\citep{2015AstBu..70....1K}, NGC~3379's total 3.6~$\mu$m absolute magnitude
$\mathfrak{M}_{3.6,gal}= -21.45$ mag, and using an absolute magnitude for the
Sun of $\mathfrak{M}_{\odot,3.6} =
6.02$ mag \citep[AB:][]{2018ApJS..236...47W}, the logarithm of the galaxy's 
stellar mass is 10.89$\pm$0.20 dex.
As in \citet{Graham:Sahu:22a}, the uncertainty on the mass comes from adding,
in quadrature, the uncertainty on the distance, $M/L$ ratio, and magnitude
\citep[see equation~9 in][]{2019ApJ...876..155S}. As such, the only change in
the formal uncertainty discussed in \citet{Graham:Sahu:22a} is the increased
uncertainty on the spheroid magnitudes from 0.15 to 0.25~mag.  For
E galaxies redesignated as ES galaxies, this change is from 0.15 to 0.20 mag. 
The above galaxy stellar mass is just 0.08~dex smaller than that reported in
\citet{Graham:Sahu:22a}.  However, the spheroid mass fraction of 0.59, which
is not yet applied, may still be too high.

In the top panel of Figure~\ref{Fig_dd}, a (weakly truncated, down-bending) large-scale disc
plus an intermediate-scale exponential disc is used to decompose 
NGC~3379's light distribution, 
along with the S\'ersic spheroid function, to improve the fit shown in
Fig.~\ref{Fig_N3379}. 
This improvement is evidenced by the wholesale removal of the snake-like pattern
in the residual light profile seen in Figure~\ref{Fig_N3379}, whose curvature
highlighted a mismatch in form between the model components and the actual galaxy.
With the new decomposition, the galaxy's apparent magnitude remains 8.74 mag 
while the spheroid mag dims from 9.31 to 10.28 mag.
The $B/T$ flux ratio is now 0.24.
A remarkably similar outcome is observed with NGC~4649 (Appendix~\ref{App_fits}).
Curiously, this $B/T$ ratio of $\sim$1/4 is typical of S0 galaxies when modelled with 
bulge$+$bar$+$disc components \citep{2005MNRAS.362.1319L}. 
The ellipticity profile of NGC~3379 does not suggest a bar is present. 

Simulations suggest that NGC~3379 may be a merger remnant, resembling
what the Antennae galaxies (NGC~4038/4039) will 
look like in 3 Gyr from now \citep{2018MNRAS.475.3934L}.
The presence of two discs (plus spheroid) support the merger
origin of this galaxy from the union of two disc galaxies.
 
While the (potential) detection of two
discs is at first a surprising new result for NGC~3379 (and NGC~4649 and
others in Appendix~\ref{App_fits}),
it should perhaps not be unexpected given that ETGs with double discs are known to exist.
Indeed, while some ETGs are known to have two discs, e.g. NGC~4494
\citep{2011MNRAS.414.2923K, 2017MNRAS.467.4540B}, sometimes one even 
counter-rotates with respect to the other, e.g. NGC~4550 \citep{1992ApJ...400L...5R} and
NGC~4528 \citep{2011MNRAS.414.2923K}.
It is plausible, if not likely, that in NGC~3379 we are witnessing the ongoing reassignment of disc stars into spheroid
stars, with an additional dry major merger required to complete the
creation of a pure E galaxy.  This is in accord with observed
galaxy-type-dependent $M_{\rm bh}/M_{\rm \star,sph}$
ratios \citep{Graham:Sahu:22a, Graham-triangal}. 

The double disc structure in the upper panel of Figure~\ref{Fig_dd} suggests
that a single anti-truncated disc model \citep{2002A&A...392..807P,
  2005ApJ...626L..81E}  may be applicable, and this is shown in the
lower panel of Fig.~\ref{Fig_dd}.
The transition from bulge to disc at 20-30$\arcsec$ meshes with the slight 
jump in the ellipticity profile from 0.09 to 0.13 over this range, and a
second slight jump can be seen at 110-140$\arcsec$ that is associated with the outer
disc. The anti-truncated nature of (some)
S0 galaxy discs may, therefore, be due to the merging of galaxy discs.
This scenario need not be the origin of all anti-truncated discs; indeed, in some S
galaxies, a more prosaic explanation is often the occurrence of an outer spiral
arm or ring.
Although the field-of-view of the frame containing NGC~3379 extends south-west
to twice the extent shown in Fig.~\ref{Fig_N3379}, the change in slope
of the light profile at $r\approx250\arcsec$ could conceivably be due to an
over-subtraction of the actual `sky background'.
The upward bend in the light profile of NGC~3379, starting at $r\approx110\arcsec$, is 
not due to an insufficient subtraction of the `sky background'. 
The `sky background' was measured from the histogram of the image's pixel values
\citep[][see their figure~1]{1993MNRAS.265..641A, 2019ApJ...873...85D, 2019ApJ...876..155S}.

Three additional S0 galaxies (NGC~3091, NGC~4649, NGC~4697) 
are presented in Appendix~\ref{App_fits}.  They
display the same behaviour in which two apparent discs can be modelled with a single
anti-truncated disc. 
If subsequent merging were to mess up the low-density outer disc --- moving
stars to higher vertical scale heights and larger orbital radii, expanding the
galaxy \citep[e.g.][]{1972AJ.....77..288G} --- but the
inner disc was to survive, then these S0 galaxies would be converted into ES
galaxies.  Erosion of both discs would produce an E galaxy.
This sequence and the associated erosion of
discs and build-up of the bulge (and bulge-to-disc ratio)\footnote{A caveat 
is that accretion and non-major mergers might build more disc than 
bulge.} is captured by
\citet[their figure~8]{1994A&A...283....1N}\footnote{In 
\citet{1994A&A...283....1N}, the `hidden discs' refers to face-on discs rather than
intermediate-scale discs.}  and \citet[figure~1]{Graham-colour}.
This is seen in Appendix~\ref{App_fits}, which also presents the light profiles for
a couple of ES galaxies (Section~\ref{Sec-App-E}) and a couple of E galaxies
(Section~\ref{Sec-App-E}) plus one BCG (Section~\ref{Sec-App-BCG}).

\subsection{Dust}

The division between dust-rich and dust-poor galaxies in \citet{Graham-S0} was
observed to mirror the galaxies' underlying origin.  Upon investigation of the
literature, the
bulk of the dust-rich S0 galaxies turned out to be known major wet
mergers, while the dust-poor S0 galaxies did not.
Consequently, (near and distant) newly-built ETGs can be
expected to have dust-dimmed supernovae and partly or wholly obscured AGN as
the gas/dust settles towards the galaxy centre. 

With time and the growth of galaxy mass comes the tendency to reside in an
X-ray hot gas cloud. 
Residence in such a cloud will destroy a galaxy's cold gas
and dust \citep{1998ApJ...501..643D}.  A galaxy's passage through a
galaxy-cluster-sized hot gas cloud will also result in ram-pressure stripping
of the cold gas and dust from the galaxy.  Therefore, not all
wet-major-merger-built S0 galaxies are expected to still retain a dust-rich
character, and subsequent merger/growth of these galaxies will further their 
widespread dust loss.  As such, the merger-driven evolution from dust-rich S0
galaxy to ES galaxy to E galaxy is expected to result in an apparent
brightening of their (Type Ia and other) supernova as the veil of dust thins.
Today, while many massive S0 galaxies contain strong dust features, less
strong but still widespread dust features can be seen in some E galaxies,
e.g., IC~1459, NGC~4374, NGC~5077, and NGC~5846. However, most fully-fledged E galaxies have
only a dusty nuclear disc or ring or no dust feature.

It is interesting to explore if the four reclassified E$\rightarrow$S0 galaxies contain dust.  
NGC~3379 (S0) is not dust-rich but has an $\sim$80~pc (radius) nuclear dust ring
\citep{1995AJ....110.2027V, 2000AJ....119.1157G}, and the bimodal colour distribution of 
its globular clusters suggests it was formed from hierarchical mergers that
produced the red, metal-rich globular clusters \citep{2004AJ....127..302R}. 
NGC~4649 (M60, S0) is also not dusty but is another suspected merger remnant \citep{2014ApJ...783...18D}.
NGC~3091 (S0) is the BGG in Hickson Compact Group No.\ 42.
Although not dusty, as with NGC~4649, it appears in the X-ray bright
catalogue of \citet{2010MNRAS.404..180D}, possibly explaining its absence of
dust.  
Finally, NGC~4697 is considered a relaxed merger remnant \citep{2003ApJ...599L..73Z,
  2006AJ....131..837S}.  It has a nuclear dust disc or ring.
Such dusty nuclear discs or rings are a common feature of massive ETGs
\citep[e.g.][]{1994AJ....108.1567J, 2001AJ....121.2431R, 2007ApJ...665.1084B}. 
The above four galaxies join the S0 galaxies NGC~1023, NGC~4762, and NGC~7332 in that
they, too, are mergers that do not display a dust-rich character today.

\section{Relations}
\label{Sec_anal}

\begin{figure*}
\begin{center}
\includegraphics[trim=0.0cm 0cm 0.0cm 0cm, width=0.6\textwidth, angle=0]{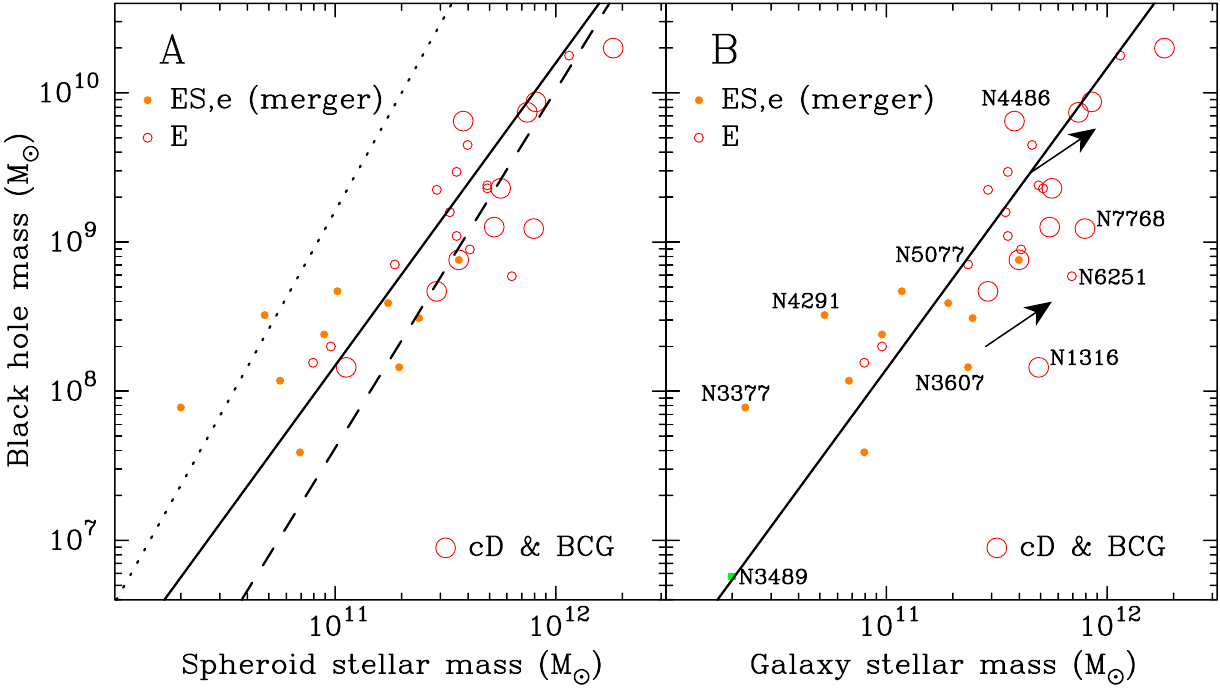} 
\caption{$M_{\rm bh}$-$M_{\rm \star,sph}$ and $M_{\rm bh}$-$M_{\rm \star,gal}$
  diagrams and relations for BCG (dashed line), non-BCG E galaxies (solid
  line), and non-BCG major-merger-built S0 galaxies (dotted line). The arrows
  show a mass doubling from an equal-mass dry-major-merger, taking galaxies
  rightward of the steeper than linear $M_{\rm bh}$-$M_{\rm \star,gal}$
  relation shown here. (The dust-rich S0 galaxy NGC~3489 happens to reside on
  the $M_{\rm bh}$-$M_{\rm \star,gal}$ relation defined by the non-BCG E
  galaxies.  A single $M_{\rm bh}$-$M_{\rm \star,gal}$ line for
  major-merger-built galaxies is provided in \citet{Graham-triangal}.) }
\label{Fig-M-M-M}
\end{center}
\end{figure*}

\begin{table}
\centering
\caption{$M_{\rm bh}$-$M_{\rm \star,sph}$ and $M_{\rm bh}$-$M_{\rm \star,gal}$ relations}
\label{Table-fit}
\begin{tabular}{lrccc}
\hline
Galaxy type      & $N$ &    slope (A)    & mid-pt (C) & intercept (B)  \\
\hline
\multicolumn{5}{c}{$\log(M_{\rm bh}/{\rm M}_\odot) = A[\log(M_{\rm \star,sph}/\upsilon\,{\rm M}_\odot) - C] +B$ (Fig.~\ref{Fig-M-M-M}A)} \\
S0/Es,b (dust$=$Y) & 18  &  2.63$\pm$0.78  & 10.70  &  8.42$\pm$0.18   \\
ES,e             & 10  &     ...         &  ...   &     ...           \\
E (non-BCG)      & 12  &  2.03$\pm$0.19  & 11.60  &  9.39$\pm$0.07    \\
BCG              & 10  &  2.40$\pm$0.40  & 11.78  &  9.49$\pm$0.13   \\
\multicolumn{5}{c}{$\log(M_{\rm bh}/{\rm M}_\odot) = A[\log(M_{\rm \star,gal}/\upsilon\,{\rm M}_\odot) - C] +B$ (Fig.~\ref{Fig-M-M-M}B)} \\
S0/Es,b (dust$=$Y) & 18  &  2.82$\pm$0.45  & 11.00  &  8.14$\pm$0.15  \\
ES,e             & 10  &     ...         &  ...   &     ...           \\
E (non-BCG)      & 12  &  2.02$\pm$0.21  & 11.60  &  9.36$\pm$0.06    \\
E (BCG)          & 08  &  2.78$\pm$0.86  & 11.78  &  9.43$\pm$0.16    \\
\hline
\end{tabular}

The sample size, $N$, of each galaxy type, is given in Column~2 and explained
in the text. 
The $\upsilon$ term equals 1.0 here and whenever using stellar mass estimates
consistent with those derived from eq.~4 in \citet{Graham:Sahu:22a}.
Data for the BCG and S0/Es,b (dust=Y) sample, explained in the text,
are provided in \citet{Graham:Sahu:22a}. 

\end{table}

From the initial sample of $\sim$100 galaxies with directly measured black
hole masses and Spitzer 3.6~$\mu$m images, there are 
13 ($=17-3$ S0 $-$1 ES) non-BCG E galaxies (Table~\ref{TableES_E}), 
10 ES,e galaxies (including the ES,e/BCG NGC~1275, Table~\ref{TableES_E})
plus an additional 8 E BCG and 1 S0 BCG \citep[][their
  table~2]{Graham:Sahu:22b}. 

The `cleaned' sample ---  
in the sense that 3 S0 galaxies and 1 ES
galaxy have been removed after the new galaxy decompositions performed here
--- of non-BCG E galaxies 
define a trend in the $M_{\rm bh}$-$M_{\rm \star,sph}$ diagram
(Fig.~\ref{Fig-M-M-M}). 
Excluding only the outlying supergiant E galaxy NGC~6251 --- to provide a more
robust result --- their distribution can be approximated as 
\begin{equation}
\log(M_{\rm bh}/M_\odot)= (2.03\pm0.19)[\log(M_{\rm \star,sph}/\upsilon M_\odot)-11.60]+(9.39\pm0.07).
\label{Eq1}
\end{equation}
This equation is somewhat comparable to the relation for (BCG, E, and ES,e) galaxies
without large-scale discs that was discovered by \citet{2019ApJ...876..155S}
and which has a slope of 1.90$\pm$0.20 and an intercept at $\log(M_{\rm
  \star,sph}/\upsilon M_\odot)=11.60$ dex equal to 8.96, albeit based on a
constant $M/L_{3.6}=0.6$ for all ETGs. 
The $M_{\rm bh}$-$M_{\rm \star,gal}$ diagram and relation for the cleaned
sample of non-BCG E galaxies can also be seen in Fig.~\ref{Fig-M-M-M} and Table~\ref{Table-fit}.
The result is similar to the $M_{\rm bh}$-$M_{\rm \star,sph}$ relation 
given that the spheroid contains the bulk, if not the entirety, of
the stellar mass in this sample, as can be deduced from Table~\ref{TableES_E}. 
 
The BCG sample is the ten BCGs 
identified in \citet{Graham:Sahu:22b}, while the E (BCG) sample excludes the
ES,e BCG NGC~1275 and the S0 BCG NGC~1316. 
The latter sample was used when deriving the $M_{\rm bh}$-$M_{\rm \star,gal}$
relation (Table~\ref{Table-fit}). 
Except for NGC~4486, the BCGs are offset to
the right of the $M_{\rm bh}$-$M_{\rm \star,sph}$ relation defined by the 
non-BCG E galaxies (solid line in Fig.~\ref{Fig-M-M-M}), 
as expected from a (mass-doubling) dry major merger of two non-BCG E
galaxies.  As such, the BCGs are not simply the high-mass end of the non-BCG E galaxy
mass-scaling relation.  Conversely, with the exception of
NGC~6251, the non-BCG E galaxies tend to be offset to the left of the $M_{\rm
  bh}$-$M_{\rm \star,sph}$ relation defined by the BCGs (dashed line in
Fig.~\ref{Fig-M-M-M}). 
The BCG with the lowest black hole mass in Fig.~\ref{Fig-M-M-M} is the S0 galaxy
NGC~1316 (Fornax Cluster), while the BCG with the third lowest black hole mass is the
ES,e galaxy NGC~1275 (Perseus Cluster).  The eight other BCGs are E galaxies.

The ES,e galaxies are clearly not drawn from the same population as the
non-BCG E galaxies; these discless E galaxies tend to have higher masses, likely
reflective of more (substantial) disc-destroying mergers than experienced by
the ES,e galaxies.  Following \citet{Graham-triangal}, the regression analysis
for the 
non-BCG major-merger-built S0 galaxies is also shown in Fig.~\ref{Fig-M-M-M},
by the dotted line. This sample and the regression line is updated here to
include the four reclassified S0 galaxies listed in Table~\ref{TableES_E}. 
The dust-rich (dust$=$Y) S0 and ES,b sample of non-BCGs is comprised of the 15 galaxies used in 
\citet{Graham-triangal}, minus the BCG NGC~1316 but with the addition of the four S0 galaxies listed in 
Table~\ref{TableES_E}. The initial sample of 15 galaxies are the 17 listed in table~2 of
\citet{Graham-S0} minus NGC~404 \citep[$M_{\rm
    bh}\approx6\times10^5$~M$\odot$:][]{2020MNRAS.496.4061D}
and NGC~3489 \citep{2010MNRAS.403..646N} 
due to their elevated weight at the low-mass end of the distribution. 
At this stage, removing one or two ES,e data points is enough to sway the
distribution and, therefore, no reliable relation is derived for this sample. 
Although a secure regression is not obtained for the ES,e galaxies, 
their bridging nature between the $M_{\rm bh}$-$M_{\rm \star,sph}$ relations
for the major-wet-merger-built S0 galaxies and 
major-dry-merger-built non-BCG E galaxies is apparent in the left-hand panel
of Fig.~\ref{Fig-M-M-M}. 
Among the ES,e galaxies, 
they tend to have 5--20 per cent of their stellar mass
in an intermediate-scale disc,
resulting in $\log(M_{\rm \star,gal}) - \log(M_{\rm \star,sph}) < 0.1$ dex.
As such, there is not a big shift for these galaxies between the
$M_{\rm bh}$-$M_{\rm \star,sph}$ and $M_{\rm bh}$-$M_{\rm \star,gal}$
diagrams.
The three $M_{\rm bh}$-$M_{\rm \star,sph}$ relations shown in the left-hand
panel of Fig.~\ref{Fig-M-M-M} are presented again in Section~\ref{Sec_Disc},
shown there in relation to black hole scaling relations for other
morphological types.

\section{Discussion and insights}
\label{Sec_Disc}

\subsection{Lessons learned from NGC~3379}

Several important observations 
are made below regarding what the remodelled light profile for NGC~3379 is revealing. 

{\it Bulge-to-total ratios:}
The remodelling of NGC~3379 raises the question asked in
\citet{Graham:Sahu:22b}: What is a bulge?  Traditionally, it was defined as
the inner excess above the inward extrapolation of the outer disc.  However,
multicomponent decompositions, such as the inclusion of a bar, can shave the bulge, sometimes
considerably so.
Furthermore, so can an anti-truncrated disc model.  Such a model was deemed
appropriate for describing NGC~4045 in \citet{2022MNRAS.514.3410H},
reducing the contribution from the bar in that galaxy.
While there is often clear evidence for multiple components, such as bars,
ansae and rings, counter-rotating discs, and so on, this is not the case for
NGC~3379.  Therefore, combining the two discs shown in Fig.~\ref{Fig_dd} into a single
anti-truncated disc may, for some readers, be more palatable than fitting
multiple components beyond a bulge and disc. Eitherway, the use of two discs
oran anti-truncated disc reduces the $B/T$ ratio to $\sim$1/4, down from
$\sim$1/2 when performing a traditional S\'ersic-bulge plus exponential-disc
fit. 

{\it Classical versus pseudobulges:}
When including a large-scale disc in the modelling of NGC~3379, the S\'ersic index of the
best-fitting spheroid is reduced from more than 2, specifically 2.74
(Fig.~\ref{Fig_N3379}), to less than 2 (specifically, 1.44 and 1.87 in
Fig.~\ref{Fig_dd}).  Given NGC~3379 is a wet-merger-built galaxy
\citep{2004AJ....127..302R}, it has a `classical' bulge, i.e., a merger-built
bulge, rather than a pseudobulge built from secular evolution of a single 
disc. This highlights the problematic nature of (the common practice of) using
a bulge's S\'ersic index to dictate the bulge's origin, a concern raised in
\citet{2014ASPC..480..185G} that bulges with $n<2$ can be built by mergers. 
 
{\it Rotation and dark matter fractions:}
The assumption that NGC~3379 is dominated by a pressure-supported
system would bias attempts
to measure the fraction of dark matter in this galaxy.  NGC~3379 contains a
significant, near-face-on, stellar disc.  If the globular cluster and
planetary nebula used to probe the kinematics at large radii are also
associated with a (somewhat heated) face-on, oblate disc structure \citep[a possibility
  noted at the end of Section~7.2 in][]{2007ApJ...664..257D}, it will lead to
a misleading estimate of the required amount of dark matter
\citep{2003Sci...301.1696R}.
Given the abundance of dominant large-scale discs in regular and low-mass ETGs,
studies using $\sigma^2 R_{\rm e,gal}/G$, where $\sigma$ is the stellar
velocity dispersion, as a proxy for the dynamical/total
mass in these galaxies will typically be in error.  This is because the rotation is
overlooked and $R_{\rm e,gal}$ does not apply to the virial theorem when
$R_{\rm e,gal}$ essentially tracks $R_{\rm e,disc}$. 
The subsequent combination of the above mass proxy with the stellar mass to estimate the
dark matter fraction is, therefore, expected to be erroneous and misleading
for (recognised and unrecognised) S0 galaxies. 

{\it Depleted stellar cores:}
A quantitative reanalysis of partially depleted cores in ETGs with
discs will be worth pursuing. The reduced S\'ersic index of the bulge model
--- when a disc component is included in the decomposition --- is known to
reduce the apparent core break radius, $R_{\rm b}$, and stellar mass deficit,
$M_{\rm def}$, of the depleted core relative to the inward extrapolation of
the S\'ersic model over the core region.  This is simply because, with a
reduced S\'ersic index, the inward extrapolation of the
bulge's revised S\'ersic model is not as centrally concentrated and steep.
This was the case with the S0 galaxy NGC~4382 \citep{2012ApJ...755..163D}, and
\citet{2014MNRAS.444.2700D} additionally already note that several other
previously misclassified (as E) S0 galaxies may need reinvestigating due to
overlooked discs.  The core-S\'ersic model \citep{2003AJ....125.2951G,
  2004AJ....127.1917T} was not employed here due to the 
2$\arcsec$ seeing, but it was used in \citet{2012ApJ...755..163D} to report
$R_{\rm b}=1\arcsec.03$ for NGC~3379, with \citet{2013AJ....146..160R} later
reporting $R_{\rm b}=1.09\pm0.04$ arcsec.  The Spitzer 3.6~$\mu$m light
profile for NGC~3379 was additionally modelled with a core-S\'ersic bulge, but
this had no significant impact and thus the simpler model was used.  If,
however, using a routine like {\sc GALFIT} \citep{2002AJ....124..266P}
that places maximum weight/trust in the highest signal-to-noise data points,
i.e.\ a galaxy's centre, then it would be necessary to use the more
sophisticated core-S\'ersic model \citep{2014PASP..126..935B} otherwise the
results from the S\'ersic model would be unduly biased/skewed. 
Reductions in S\'ersic indices  may resolve the unexpectedly high
$M_{\rm def}/M_{\rm bh}$ ratios between 5 and 50 reported for some galaxies
while most have ratios less than 4 \citep{2004ApJ...613L..33G,
  2013AJ....146..160R, 2014MNRAS.444.2700D}.

\subsection{Galaxy speciation} 

While widespread dust or a bar is a signpost of a disc in ETGs, for a century, 
large- and intermediate-scale discs have routinely been missed by visual
classification. Careful quantitative studies of images and light profiles, coupled with
confirmation from kinematic maps and profiles, has, however, broken this impediment and unlocked
great insight into the speciation of galaxies. 

Lessons learned from the other remodelled galaxies, when viewed in conjunction
with each other, suggest further insights into the structure and speciation of
galaxies. 
Here, the origin of galaxy species is briefly discussed, with insights stemming from
careful investigations into their morphology, backed by observations (reported
in the literature) 
of the galaxies' kinematics and merger history. The typically
overlooked population of ES galaxies has been one of the keys to disentangling
the wholesale origin of galaxy types. 
As with discs in S0
galaxies, discs in ES galaxies can be challenging to spot. 
However,
some studies did, and, as noted earlier, these galaxies were referred to as `disc ellipticals' as a
point of differentiation \citep{1988A&A...195L...1N} from the (pure) E galaxies. 
Nine examples were provided by \citet{1995AandA...293...20S}, who referred to them
as `discy ellipticals', 
and a dozen examples are displayed in  \citet[their figure~23]{2015ApJS..217...32B}.
\citet{2011ApJ...736L..26A} used kinematical information to reveal the
(fully embedded) intermediate-scale disc in the ES galaxy NGC~3115 previously presented in
\citet[their figure~10b]{1995A&A...293...20S} and later by
\citet{2015ApJS..217...32B} and \citet{2016ApJS..222...10S}. 
\citet{2014ApJ...791...80A} noted that
centrally-fast-rotating ETGs can have rapidly declining (or increasing) specific
angular momentum profiles \citep[see also][]{2017ApJ...840...68G,
  2017MNRAS.470.1321B}. That is, their inner discs may sometimes be embedded
in a larger disc.

\subsubsection{Mergers versus monolithic collapse}

According to a wealth of literature, the sample of dust-rich S0 galaxies with
directly measured black hole masses \citep[][table~2]{Graham-S0} have been
built by mergers. 
Mergers can build up a spheroid by delivering new stars (from
the infalling secondary galaxy) and disrupting pre-existing disc stars
(in the primary galaxy).  Dry mergers also readily explain the shift to larger
spheroids and black hole masses while simultaneously explaining the shift to
lower $M_{\rm bh}/M_{\rm \star,sph}$ ratios when transitioning from disc-dominated
to spheroid-dominated galaxies \citep{Graham:Sahu:22a}.  Furthermore,
\citet{2023MNRAS.518.6293G} reveals that 
they explain the observed bend in the $M_{\rm bh}$-$\sigma$ relation
\citep{2013ApJ...764..151G, 2018ApJ...852..131B, 2019ApJ...887...10S}. 
This reduces the role that AGN feedback plays in establishing the black hole
scaling relations. Furthermore, \citet{2024arXiv240711127B} report a reduced
role for AGN feedback in all galaxy types.

Since massive dust-rich S0 galaxies and ES/E galaxies are known merger
remnants, it discounts a monolithic collapse scenario
\citep{1962ApJ...136..748E} for the bulk of their 
stars. Indeed, as just noted, the $M_{\rm bh}/M_{\rm \star,sph}$ ratios of
E and ES galaxies are consistent with their creation from the dry merger of
massive S0 galaxies.  A potential exception worthy of note is that old, metal-poor globular
clusters may have formed from a mini/localised collapse, and once they did
condense out of a 
proto-galactic gas cloud, they became effectively collisionless, as was and is the dark matter.
While still embedded in a dense gas cloud, dynamical friction and drag will
inevitably lead to some of these globular clusters becoming the nuclei of
galaxies \citep{1975ApJ...196..407T} around which
large gas discs will form and turn into stellar discs, giving rise to the primaeval S0
galaxy population. These nuclei will be returned to in
subsection~\ref{subsec-UCD}.

\subsubsection{Accretion and merger driven metamorphosis}

\begin{figure}
\begin{center}
\includegraphics[trim=0.0cm 0cm 0.0cm 0cm, width=0.9\columnwidth, angle=0]{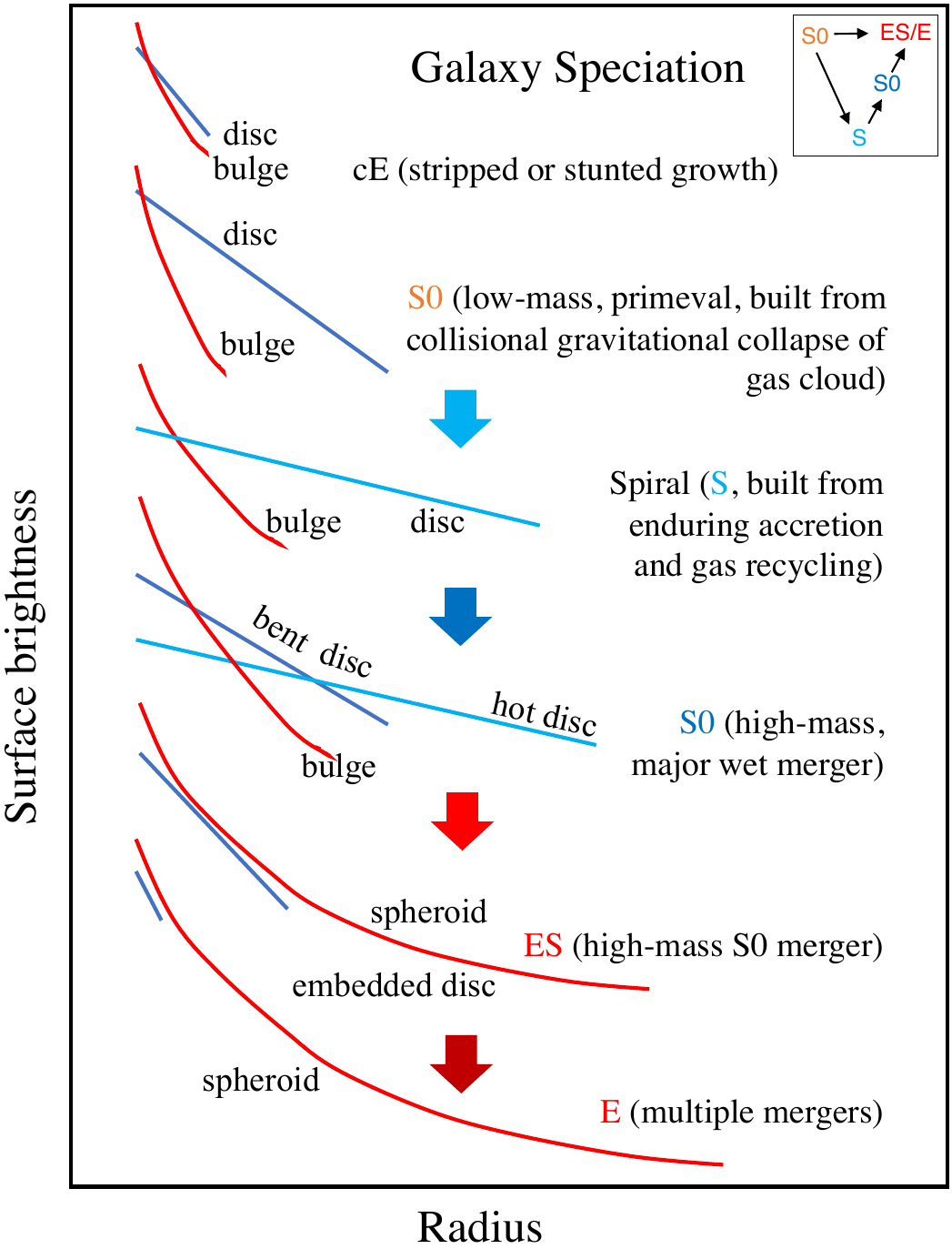}
\end{center}
\caption{Progressive speciation of common galaxy types, revealing the
  S\'ersicification of galaxy light profiles as S\'ersic $n\approx1$ discs give way to 
  higher-$n$ profiles that decline slowly at large radii. The schematic also helps to
  visualise/explain the construction of the
  $R_{\rm e,gal}$-$n$ relation \citep{1993MNRAS.265.1013C, 1994MNRAS.271..523D}, the 
  $M_\star$-$n$ relation \citep{1994ApJS...93..397C, 1994MNRAS.268L..11Y},
  and the $M_{\rm bh}$-$n$ relation \citep{2001ApJ...563L..11G, 2007ApJ...655...77G}.
Depleted galaxy cores that can form in dry merger events are not shown. 
The `Triangal' from \citet{Graham-triangal} is shown in the upper-right corner.
}
\label{Fig_prog}
\end{figure}

Fig.~\ref{Fig_prog} crudely\footnote{This representation is crude in the sense
that it does not (yet) incorporate additional features that might be
important, such as bar strength, spiral arm strength or 
winding angle.} displays the speciation of galaxies, focussing on
the spheroid and disc components.  The tendency for some discs with longer
scalelengths to have fainter central surface brightnesses, i.e.\ lower
densities \citep{1998MNRAS.300..469S, 2001MNRAS.326..543G}, has been
incorporated here but is not considered a strict rule.  While the
major-merger-built galaxies tend to have thick/heated discs (or no discs)
which do not endear themselves to bars or spiral patterns, these features can
inhabit the precursor population.  The demographics of those anatomical
features in galaxies across the $M_{\rm bh}$-$M_{\rm \star,sph}$ diagram will
be presented in future work.
For now, the upper right corner of Fig.~\ref{Fig_prog} includes the `Triangal' from
\citet{Graham-triangal}, developed as an alternative to the `Tuning Fork'
\citep{1928asco.book.....J, 1936rene.book.....H}, and the `Trident' or
`ATLAS$^{3D}$ Comb' 
\citep{1976ApJ...206..883V, 2011MNRAS.416.1680C}, which do not have
evolutionary arrows, 
do not recognise the two types of S0 galaxy
\citep[primaeval and wet-major-merger-built:][]{Graham-S0}, 
and have no direct path linking S galaxy mergers with S0 galaxies. 
The' Triangal' includes major mergers of both primaeval and
wet-major-merger-built (in general, low- and high-mass) S0 galaxies to create
ES and E galaxies. This key ingredient was missing from the previous 
morphology schemata, which did not account for `punctuated equilibrium' events
\citep{Graham:Sahu:22a}, such as major mergers  of S galaxies. Building on 
those schemata, the `Triangal' can bypass the gradual changes linking disc
galaxies possessing strong spirals (or bars or bulges) to those with weak
spirals and then those with no spiral pattern, jumping straight from S to S0,
thereby circumventing the limitations of past schema and capturing the major
evolutionary pathways.  Embellishments of spiral and bar strength will be added
in a forthcoming paper. 

Among the four remodelled S0 galaxies (Section~\ref{Sec3379} and
\ref{Sec_trio}) --- all of which are considered in the literature to have been built
from a major merger event --- they appear to have an inner disc with an
exponential scalelength of 2$\pm$1~kpc and an outer disc with an 
exponential scalelength of 5$\pm$2~kpc.  The exception is NGC~3091 in
Hickson Compact Group No.\ 42, with an inner (sole?) disc scalelength of 5~kpc
and an outer exponential (halo?) with a scalelength of 15~kpc.  Curiously,
\citet{2007MNRAS.378.1575S} discovered a tendency for halos to have
exponential light profiles. 

In Fig.~\ref{Fig_prog}, one can see how the `bulge' component of S0 galaxies,
when defined as the inner excess above the inward extrapolation of the outer
exponential profile, will yield a mismatch with the dynamically hot pressure-supported
`bulge' (shown by the red curves).  The dynamical heating and partial erosion
of the outer disc(s) can produce 
 ES galaxies whose disc is fully embedded within what is regarded as a slowly
 or non-rotating oblate ellipsoid rather than a disc.  This situation is seen in
 kinematic maps that extend beyond 1$R_{\rm e}$ \citep{2014ApJ...791...80A}.

While \citet{1964rcbg.book.....D} tabulated bright
galaxies, noting which were peculiar (`pec')\footnote{These are often
unrelaxed mergers.}, \citet{1959VV....C......0V} and
\citet{1966ApJS...14....1A} focussed on these peculiar and interacting
galaxies. They are transitional types, covering mergers-in-progress, and they 
encapsulate many recent wet and dry major merger remnants, such as the dust-rich S0
disc galaxies built from gas-rich mergers and E galaxies with stellar shells
\citep[e.g.][]{1985A&A...149..442H, 1988ApJ...331..682H, 2010RAA....10..220X,
  2021AandA...650A..34R, 2024MNRAS.tmp..477R}. 
Mergers are `part and parcel' of galaxy life and tend to disguise galaxy 
discs, in part by destroying the spirals (and bars) that would otherwise
reveal the presence of these discs.
The RC3 \citep{1991rc3..book.....D} is full of S0 galaxies that were
mislabelled as E galaxies.
This is because, from a casual glance at an image, spiral-less galaxy discs
tend to resemble spheroids unless they are viewed close to 
edge-on.

It took kinematic observations
\citep[e.g.][]{1998A&AS..133..325G, 2011MNRAS.414..888E} to confirm that the
actual number of ``dynamically hot'' galaxies is much lower than was
initially thought. 
\citet[][their figure~3]{1999ApJ...521...50M} showed that (true, slow
rotating) E galaxies only dominate the ETG population at magnitudes brighter
than $M_{B,Vega} \approx -20.6$ mag (H$_0=75$ km s$^{-1}$ Mpc$^{-1}$), while S0
galaxies dominate at fainter magnitudes.  This was supported by
\citet{2011MNRAS.414..888E} using rotational measures and by
\citet{2013MNRAS.432.1768K} using photometric decompositions of the galaxy
light.  A division at this magnitude had previously been seen between galaxies
with and without depleted cores \citep{1997AJ....114.1771F,
  2003AJ....125.2936G}, supporting the dominance of dry (gas-poor) galaxy
mergers at high masses \citep[e.g.][]{2006ApJ...648..268B}.  The scouring
action of a binary black hole in a dry merger erodes the central phase-space
of stars and leaves its signature as a depleted core
\citep{1980Natur.287..307B, 2001ApJ...563...34M, 2004ApJ...613L..33G, 2013ApJ...773..100K}, thereby
enabling the identification of galaxies built from gas-poor mergers.  The
scouring also explains the removal of the dense nuclear star clusters
\citep{2010ApJ...714L.313B} prevalent in the lower mass galaxies.
The prevalence of a central black hole in massive galaxies built from major
mergers, such as the black holes found in the BCGs in the upper right of
Fig.~\ref{Fig-grand}, suggests that the gravitational recoil from colliding
black holes \citep[e.g.][]{2007ApJ...659L...5C, 2008ApJ...682L..29B,
  2008ApJ...678..780G} is, in general, insufficient to remove supermassive
black holes from galaxies.  Arguably, some black hole binaries that are yet to
coalesce --- perhaps in a gas-poor spherical galaxy
\citep{2006ApJ...642L..21B, 2013ApJ...773..100K, 2015ApJ...810...49V} --- may
remain undetected in these galaxies due to the limiting spatial resolution of
the kinematic data used to infer the presence of the central mass.

\subsubsection{The decline of disc sizes}

As noted, S0 galaxies with anti-truncated discs may conceivably arise from the
merger of two disc galaxies in which neither disc entirely morphs into a
dynamically-hot spheroidal structure.  Suppose the outer component becomes
sufficiently hot, transitioning from a disc to an ellipsoid, aka spheroid,
but the inner, more robust, higher-density disc remains largely intact. In
that case, one will have an ES galaxy.  
The smaller disc outlives the more fragile, and more dispersed, larger disc.

An ES galaxy may also be bolstered by a gas-rich galaxy collision in
which the stars experience a non-collisional merger to produce a spheroid.  At
the same time, the gas and dust undergo a collisional merger.  This happens
because although the cross-section of a star is much greater than that of a
gas particle, the number of gas particles is so vastly higher that gas
interactions are much more frequent.  Therefore, wet galaxy collisions can
`heat' stellar discs to build bulges, while gas and dust form new discs due to gravitational
collapse akin to what would have occurred in proto-galaxies building primaeval
S0 galaxies.  At that time, it was only the dark matter that experienced a
non-collisional contraction. Presumably, having experienced a net loss of
angular momentum during the disc galaxy collision, the 
gas collapses to form a smaller disc, adding to the ES nature of the system.
Inner gas discs, therefore, need not arise from accretion alone but may also
arise out of major mergers. 

This process could explain some blue cores in ETGs and some nuclear discs in E
galaxies.  If nuclear discs in E galaxies arise from the major merger of
merger-built S0 galaxies, then it makes sense that, with less gas available in
these subsequent collisions and gas shocks robbing it of its angular momentum, the
ensuing generation of discs is smaller. These (10s to 100s of pc) nuclear discs may be
distinct from, and larger than, the spheroidal-shaped nuclear star clusters
seen in low-mass ETGs \citep[e.g.][]{2003AJ....125.2936G}.  One may then
expect to find some E galaxies having nuclear discs with stellar populations
that are younger than the galaxy at large, e.g.\ NGC~4261 and NGC~7052.
Thus, again, minor accretion events need not be the origin of these discs but rather past
(major) mergers. A distinguishing test could be that 
the heavily recycled gas from mergers is expected to be more metal-rich
than the relatively pristine gas brought in from accreted gas clouds.

\subsection{Counter rotation}

Four galaxies in the sample are known to have counter-rotating components:
NGC~3414 (ES,e), 
NGC~4473 \citep[ES,e, `2$\sigma$ galaxy'][]{2011MNRAS.414..888E, 2014ApJ...791...80A},
IC~1459 (E), and
NGC~3608 (E).
Although counter rotation could signal the presence of two large-scale
discs, and thus it would be interesting to check on a single 
anti-truncated disc, none of these four galaxies are designated S0 galaxies.
They were modelled in \citet{2016ApJS..222...10S}, where NGC~3414 was noted 
to additionally have a polar ring, as does NGC~5077. While the high
ellipticity in NGC~4473, at around 0.4 at $R_{\rm maj}=100\arcsec=7.3$~kpc, is
typical of a disc, it has been suggested that this outer ellipticity is the result of
multiple minor mergers \citep{2015MNRAS.452.2208A}.  However, it perhaps forms 
a dynamically hot, slowly rotating disc-like structure. The counter-rotation in
this galaxy stems from 
two counter-rotating stellar discs that occupy the inner region 
\citep{2011MNRAS.414.2923K}. 
The fast counter-rotating nuclear disc in IC~1459 contains less than 1 per
cent of the galaxy's total stellar mass \citep{1988ApJ...327L..55F}, and, as such,
IC~1459 is not regarded as an ES,e galaxy but instead a slowly rotating E galaxy. 
Interestingly, IC~1459 also contains a faint large-scale spiral pattern of 
ionised gas and stars \citep{1985LNP...232...27M, 1990AandA...228L...9G}, and
3 per cent of the stellar population is young \citep[$\sim$400
  Myr][]{2017ApJ...834...20A}.
Finally, NGC~3608, which resides in a diffuse X-ray halo
\citep{2003ApJS..145...39M}, has a slow ($\lesssim$40 km s$^{-}$) counter-rotating core
within the inner $\sim$5$\arcsec = 0.5$~kpc \citep{2011MNRAS.414.2923K}. 
The rise in ellipticity 
beyond $R_{\rm maj}\approx18\arcsec$ \citep{2016ApJS..222...10S} is
associated with the slowly ($\lesssim$40 km s$^{-}$) rotating main galaxy
\citep{2014ApJ...791...80A}. Rather than two discs, this galaxy would likely
be well-described by two S\'ersic functions, but the author has refrained from
performing a double S\'ersic fit in the interest of constraining the
complexity of the model used here.

\subsection{Making waves}

Although the sample size is low, the BCG appear to define a near-cubic $M_{\rm
  bh}$-$M_{\rm \star,gal}$ relation.  Their relation's steep slope
(2.78$\pm$0.86) is comparable with that seen among the sample of dust-rich S0 galaxies
(2.82$\pm$0.45), for which there are twice as many, and comparable
with the slope of the spiral galaxies
\citep[$2.68\pm0.46$:][table~1]{2018ApJ...869..113D, Graham-triangal} for
which the sample size is three times larger. The dust-poor, low-mass S0
galaxies also appear to follow a distribution in the $M_{\rm bh}$-$M_{\rm
  \star,gal}$ diagram with a slope steeper than 2
\citep[][figure~A4]{Graham-triangal}, although an expanded data set would be
valuable to confirm this.  These observations beg the question as
to whether the near-quadratic $M_{\rm bh}$-$M_{\rm \star,gal}$ relation for
the E galaxies underestimates the true slope.
Taking the E galaxy scaling $M_{\rm bh} \propto M_{\rm
  \star,gal}^{2.02\pm0.21}$ (Table~\ref{Table-fit}), which is very similar to
the $M_{\rm bh} \propto M_{\rm \star,gal}^{2.07\pm0.19}$ scaling for the ensemble of 
major-merger-built galaxies \citep{Graham-triangal}, and coupling it with the 
$M_{\rm bh} \propto \sigma^{8.64\pm1.10}$ scaling for massive galaxies with
core-S\'ersic light profiles likely built from dry major mergers
\citep{2019ApJ...887...10S}, one obtains $M_{\rm \star,gal} \propto
\sigma^{4.28}$, which is comparable to the canonical value of 4 for non-dwarf
ETGs \citep{1976ApJ...204..668F}.

The Illustris simulation \citep{2014MNRAS.445..175G, 2014MNRAS.444.1518V} 
has also yielded a quadratic-like distribution 
in the $M_{\rm bh}$-$M_{\rm \star,gal}$ diagram 
\citep[][their figure~1]{2020ApJ...895..102L}, 
see also the MassiveBlack simulated tracks \citep[][their
  figure~7]{2012MNRAS.423.2397K},
and an array of additional simulations indicating a quadratic or steeper
$M_{\rm bh}$-$M_{\rm \star,gal}$ scaling 
\citep{2012MNRAS.423.2397K, 2013ApJ...765L..33L, 2014MNRAS.437.1576B}. 

Interestingly, as noted above, the low-mass dust-poor S0
galaxies also appear to define their own steep $M_{\rm bh}$-$M_{\rm \star,gal}$
relation \citep[and $M_{\rm bh}$-$M_{\rm \star,sph}$
  relation:][]{2012ApJ...746..113G, 2013ApJ...764..151G, 2013ApJ...768...76S},
with a notably higher $M_{\rm bh}/M_{\rm \star,gal}$ ratio for a
given black hole mass than the major-merger-built ETGs.  It is postulated that
these low-mass galaxies are faded S0 galaxies. 
In passing, one may speculate if these galaxies are aged counterparts of 
the tentative `blue monsters' observed at $z>10$ \citep{2023MNRAS.520.2445Z}.
These local S0 galaxies are often overlooked. For instance, their low stellar masses
and low-to-zero star formation rates has often resulted in sample selection effects missing
many of them in plots of star formation rate versus galaxy stellar mass. Consequently,
some tales of galaxy evolution \citep[e.g.][and references therein]{2014MNRAS.440..889S}
are incomplete when focussed on the old ideas of making ETGs by either fading S galaxies
\citep{1951ApJ...113..413S, 1972ApJ...176....1G}
or colliding S galaxies 
\citep[e.g.,][]{1977egsp.conf..401T, 1979A&A....76...75R}. 
Addressing the often overlooked low-mass S0 galaxies that did not form from the above
two scenarios has led to the expanded picture of galaxy speciation
presented in \citet{Graham-SFR}. 

For practical purposes pertaining to studies of gravitational waves from 
colliding supermassive black holes, which are expected in BCGs and the other
major-merger-built galaxies, 
an 
$M_{\rm bh}$-$M_{\rm \star,gal}$ 
relation for merger-built ETGs (dust-rich, high-mass S0 galaxies and ES and E
galaxies) is provided in \citet{Graham-triangal}.
The use of $M_{\rm \star,gal}$ rather than $M_{\rm \star,sph}$ should be helpful for studies and simulations that do not separate
galaxies into bulge/disc components.  
Excluding the S and above mentioned faded/preserved dust-poor, low-mass (primaeval)
S0 galaxies that have not experienced much/any major merger action
yielded a slope of $\sim$2 rather than $\sim$1
for this $M_{\rm bh}$-$M_{\rm \star,gal}$ relation. 
This relation is not updated here because the reclassification of five
galaxies does not change the major-merger-built sample.  There are revised
$M_{\rm \star,gal}$ values for nine galaxies, but the changes are not large
enough to warrant a rederivation of the $M_{\rm bh}$-$M_{\rm \star,gal}$
relation.

\subsection{Compact elliptical (cE) and ultracompact dwarf (UCD) galaxies}
\label{subsec-UCD}

\begin{figure*}
\begin{center}
\includegraphics[trim=0.0cm 0cm 0.0cm 0cm, width=0.9\textwidth, angle=0]{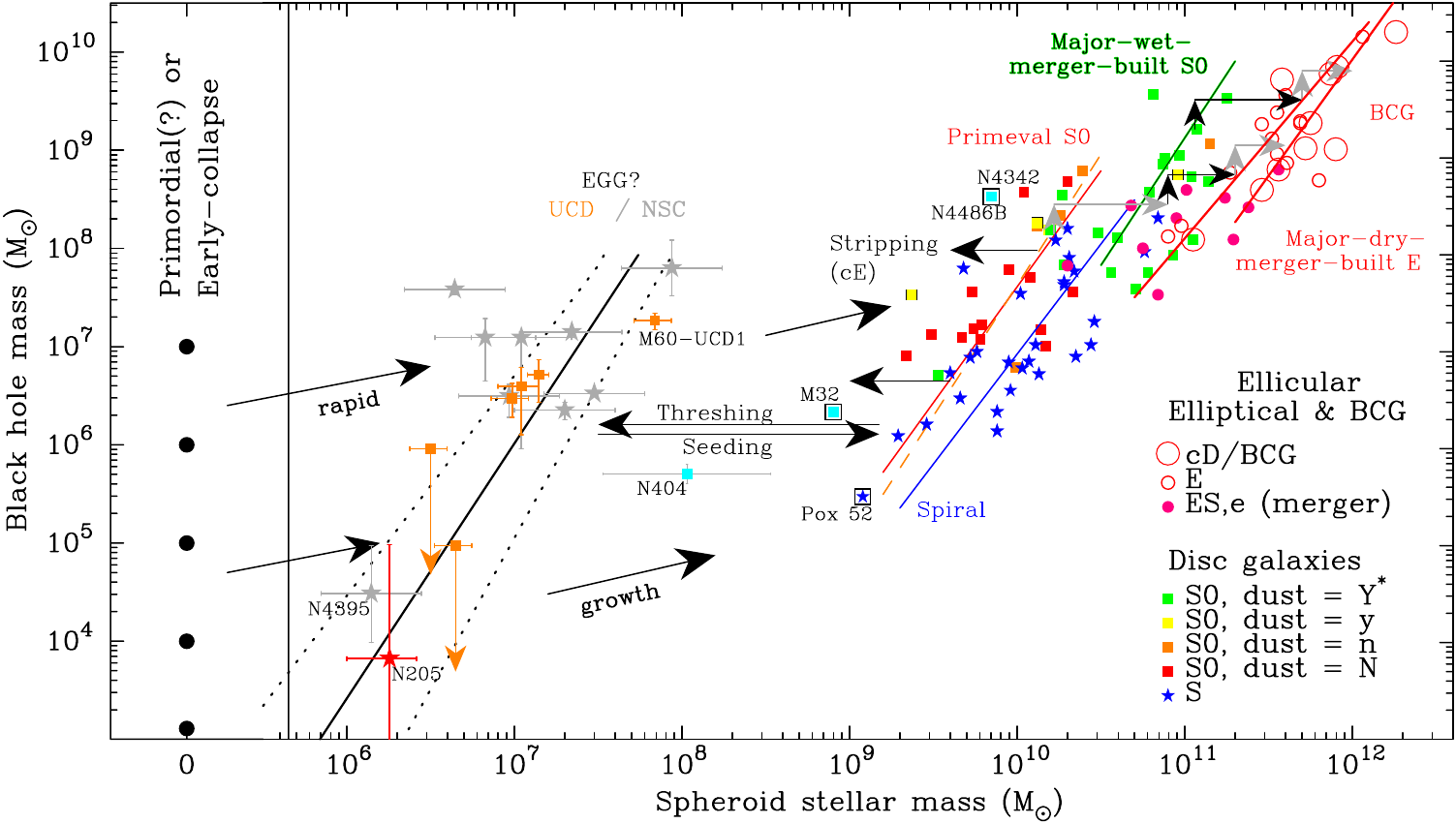}
\end{center}
\caption{
 Adaption of figure~6 from \citet{Graham:Sahu:22b}, incorporating
 figure~2 from \citet{2020MNRAS.492.3263G} for the nuclear star clusters (NSCs) 
and ultracompact dwarf (UCD) galaxies  plus an updated
  version of figure~1 from \citet{Graham-triangal} based on the remodelled
  galaxies presented herein. 
$^*$ Four merger-built but non-dust-rich galaxies remodelled
  and reclassified as S0 galaxies (see Table~\ref{TableES_E}) 
are coded green here as they were grouped with the dust-rich S0 galaxies for the regression. 
  Some low-mass, dust-poor S0 galaxies may be regarded as ancients or elders,
  representing the disc-dominated galaxies before a spiral pattern was established. 
  The UCD galaxies may be the NSCs of threshed (heavily stripped) disc galaxies 
  or old embryonic giant globulars (EGGs) that partially collapsed to form a
  central massive black hole. Some of these EGGs may never grow a
  fully-fledged galaxy around them, or 
  be accreted (at early or late times)
  into a galaxy and effectively seed it with a massive black hole.
}
\label{Fig-grand}
\end{figure*}

This paper has focussed on ETGs that are intrinsically
large and luminous, having been built by mergers. There are, however, a couple
of types of lower-mass elliptical-like galaxies.  These are the rare compact
elliptical (cE) galaxies that can form from tidal stripping
\citep{2001ApJ...557L..39B, 2002ApJ...568L..13G} and UCD galaxies that can
form from even more extreme tidal threshing \citep{2003MNRAS.346L..11B,
  2004ApJ...616L.107I, 2008MNRAS.385L..83C, 2012ApJ...755L..13K,
  2013MNRAS.433.1997P, 2023MNRAS.526L.136P}.
The cE galaxies tend to be overly metal-rich for their luminosity
\citep[e.g.][]{2009Sci...326.1379C, 2009MNRAS.397.1816P} and underluminous for
their (red) colour \citep[e.g.][their Figure~11, and references
  therein]{2019MNRAS.484..794G}, which is as expected if they had previously
evolved to a larger mass that was then pared back.

Instead of experiencing tidal stripping, one may ask if some dwarf cE galaxies
were instead denied growth due to a hungry, more dominant neighbour or simply
a lack of fuel in a void-like environment. 
Such a mechanism for dwarfism, i.e.\ born that way,
as suggested in \citet[][footnote~20]{2013pss6.book...91G} for some cE galaxies, 
implies that some evolutionary arrows flow to the left
{\em and} right in Fig.~\ref{Fig-grand}.
This leads to the notion that perhaps some
UCDs also need not be the compact nuclei of threshed disc galaxies.  Indeed,
the similarity of UCDs to large old globular clusters (GCs) has long been noted.
Strongly stunted growth in the early Universe might `freeze-in' GC and some UCD galaxies around bigger
galaxies.  Speculatively, some  may have condensed out of proto-galactic gas
clouds while the remaining gas continued to experience
collisional contraction, forming the main galaxy disc and thus no longer
providing gas drag in either the galaxy halo, to drive the GCs/UCDs inwards, or
within the GCs/UCDs, to aid runaway collisions and massive black hole growth.
Irrespective of this possibility, if galaxies start life by growing a cluster
of stars around a massive black hole --- with their larger-scale disc subsequently
forming  --- then today's NSCs and UCDs may be a
better comparison point for the high-$z$ AGN referred to as `little red dots'
\citep{2023arXiv230607320L, 2024Natur.627...59M} than today's more
fully-formed galaxies.\footnote{The author publicised this point at August 2024 conferences: 
\url{http://cosmicorigins.space/smbh-sexten}
and
\url{https://indico.ict.inaf.it/event/2784/}
}

Stepping things back even further, in terms of
increased $M_{\rm bh}/M_{\rm \star,sph}$ ratios, leads to 
massive black holes with no stellar association in Fig.~\ref{Fig-grand}.
Massive naked black holes might form from the direct
collapse of giant gas clouds \citep{1967SvA....11..233D, 1993ApJ...419..459U}
or through other mechanisms involving baryons that may have
created heavy black hole seeds, such as dark stars and Pop~III.1
stars \citep{2009ApJ...705.1031S, 2019MNRAS.483.3592B, 2023MNRAS.525..969S}.
If some black holes instead formed in the first few seconds of the
Universe, they would contribute to the `dark matter' of the Universe.  Although black hole masses of
$10^{5\pm2}$ M$_\odot$ are greater than the theorised primordial black hole
population \citep{1998PhLB..441...96A, 2002PhRvD..66f3505B,
  2021FrASS...8...87V}, lower-mass primordial black holes could still help to
later create heavy `seed masses'
for establishing the high-$z$ quasars with virial masses of $10^{8\pm1}$
\citep{2022ApJ...937L..30L, 2023arXiv230309391Y}. 
The cosmic X-ray background and the cosmic radio background does, however, place limiting constraints on
the numbers of 1-1000 M$_\odot$ primordial black holes
\citep{2022MNRAS.517.1086Z}.  Nonetheless, at least
some dark matter might be `black matter' as these black holes need not be made
of baryonic matter.\footnote{It would take one million $10^6$ M$_\odot$ black holes
to account for the missing mass in a (Milky Way)-sized galaxy. Although
this is $10^4$ times higher than the number of globular clusters, these black holes would
still have a sufficiently low volume (and surface) number density to make it
unlikely that current gravitational lensing studies in our Galaxy or the Magellanic
Clouds would have detected them \citep{2000ApJ...542..281A,
  2001ApJ...550L.169A, 2007A&A...469..387T}.}

Many globular clusters are old.  If giant globulars form early enough while
still immersed in a soup of gas, then the gas drag
force may lead to runaway collisions and the formation of an IMBH within the
system \citep{1965AZh....42..963Z, 1985ApJ...292L..41S}.  The
existence of massive Pop III stars \citep{1953Obs....73...77S}
may aid this process \citep{2003Natur.422..871U, 2017MNRAS.468..418F},
although the lack of robust IMBH detections in globular clusters currently
disfavours this mechanism. 

The total collapse
of a globular would lead to a naked black hole. On the other hand, if some
stars remain in the most massive globulars, 
one may observe a black hole as the nucleus of a UCD galaxy.
If the black hole becomes active and manages to sufficiently blow away the gas causing the 
drag on the stars, this may act to stabilise the system.  
Such black hole birth in a dense cluster of stars, a kind of embryonic giant
globular (EGG), would not resolve the need for dark matter to explain early
structure formation, as these clusters would be baryonic and not start to
collapse until after the epoch of recombination, some 380,000 years after the
Big Bang \citep{2020A&A...641A...6P}. 
If an EGG or three were to fall into a low-mass, disc-dominated
galaxy, the system might well join the black hole galaxy scaling
relation.\footnote{The high scatter for S galaxies in the $M_{\rm
  bh}$-$\sigma$ diagram \citep{2019ApJ...887...10S} might be related to this.}
Interestingly, this may be occurring in NGC~4424, with the capture of
`Nikhuli' \citep{2021ApJ...923..146G}.  This seeding phenomenon is labelled in
Fig.~\ref{Fig-grand}.  On the other hand, if an EGG continues to grow a disc
at high-$z$, the system will likely evolve into a nucleated dwarf ETG,
specifically a dwarf S0 galaxy, which
are common in galaxy clusters today \citep[e.g.][]{1985AJ.....90.1759S, 2003AJ....125.2936G}.

\subsection{A word on entropy} 

In the Introduction section, entropy was briefly noted in passing.
As \citet{1973PhRvD...7.2333B} proposed, 
massive black holes have a huge amount of entropy.  
As black holes become more massive, their surface area grows and their entropy
increases.  Entropy requires black holes to have a temperature and thus that they
radiate, which
\citet{1975CMaPh..43..199H} later proved 
following discussions by Vladimir Gribov and Yakov
Zel'dovich in 1972-1973 \citep{Anselm98, 2017eqft.proc....9A}.
The stellar component of galaxies also has a kind of entropy. 
When the large primordial (collisional) gas clouds contracted, they spun up to
conserve angular momentum and formed discs that were supported from further
gravitational collapse by their rotation.  These are expected to be the first
galaxies, some of which may be today's low-mass, dust-poor S0 galaxies.
The gas cooled within these discs and formed stars (collisionless particles). 
Repeated major collisions built subsequent generations of galaxies by dynamically 
heating the discs and building the bulges, thereby more fully filling
the phase space of a galaxy's halo and thus increasing the entropy: chaos
over order; `hot' pressure supported spheroids (random orbits)  over
dynamically-cold rotation-supported discs (ordered orbits).

\section*{Acknowledgements}

This paper is dedicated to the memory of Tom Jarrett (1963-2024).  I am
blessed to have recently enjoyed Tom's friendship and passion for
astronomy. Tom alerted me to the disc in NGC~4649, which prompted me to
re-check the E and ES galaxies in my sample. This reassessment resulted in
this paper's appendix and contributed to the conclusions.

I thank Ewan Cameron for his ongoing support of his R statistical software.
I am grateful to David Liptai, Conrad Chan and Robin Humble of the
Astronomy Data and Computing Services (ADACS) in Australia for keeping the
Image Reduction and Analysis Facility (IRAF) alive on the OzSTAR and `Ngarrgu
Tindebeek' supercomputer at Swinburne University of Technology.

Publication costs were funded through the 82783405 Australia and New Zealand
Institutions (Council of Australian University Librarians affiliated) Open
Access Agreement.
Some of this work was performed on the OzSTAR national facility at Swinburne
University of Technology. The OzSTAR program receives funding in part from the
Astronomy National Collaborative Research Infrastructure Strategy (NCRIS)
allocation provided by the Australian Government and the Victorian Higher
Education State Investment Fund (VHESIF) provided by the Victorian Government.
This work has used the NASA/IPAC Infrared Science Archive (IRSA)
and the NASA/IPAC Extragalactic Database (NED)\footnote{\url{https://doi.org/10.26132/NED1}},
funded by NASA and operated by the California Institute of Technology. 
This research has also used the SAO Astrophysics Data System
Bibliographic Services and the {\sc Rstan} package available at
\url{https://mc-stan.org/}.

\section{Data Availability}

The imaging data underlying this article are available in the NASA/IPAC
Infrared Science
Archive.\footnote{\url{https://irsa.ipac.caltech.edu/applications/Spitzer/SHA}}
The S$^4$G dataset Digital Object Identifier (DOI) is 10.26131/IRSA425. 
The spheroid and galaxy stellar masses are tabulated in
\citet{Graham:Sahu:22a}, with revisions (This Work) in Table~\ref{TableES_E}.

https://doi.org/10.26131/IRSA425

\bibliographystyle{mnras}
\bibliography{Paper-BH-mass5}{}

\appendix

\section{Uncovering hidden discs}
\label{App_fits}

\begin{figure*}
\begin{center}
\includegraphics[trim=0.0cm 0cm 0.0cm 0cm, height=0.23\textwidth, angle=0]{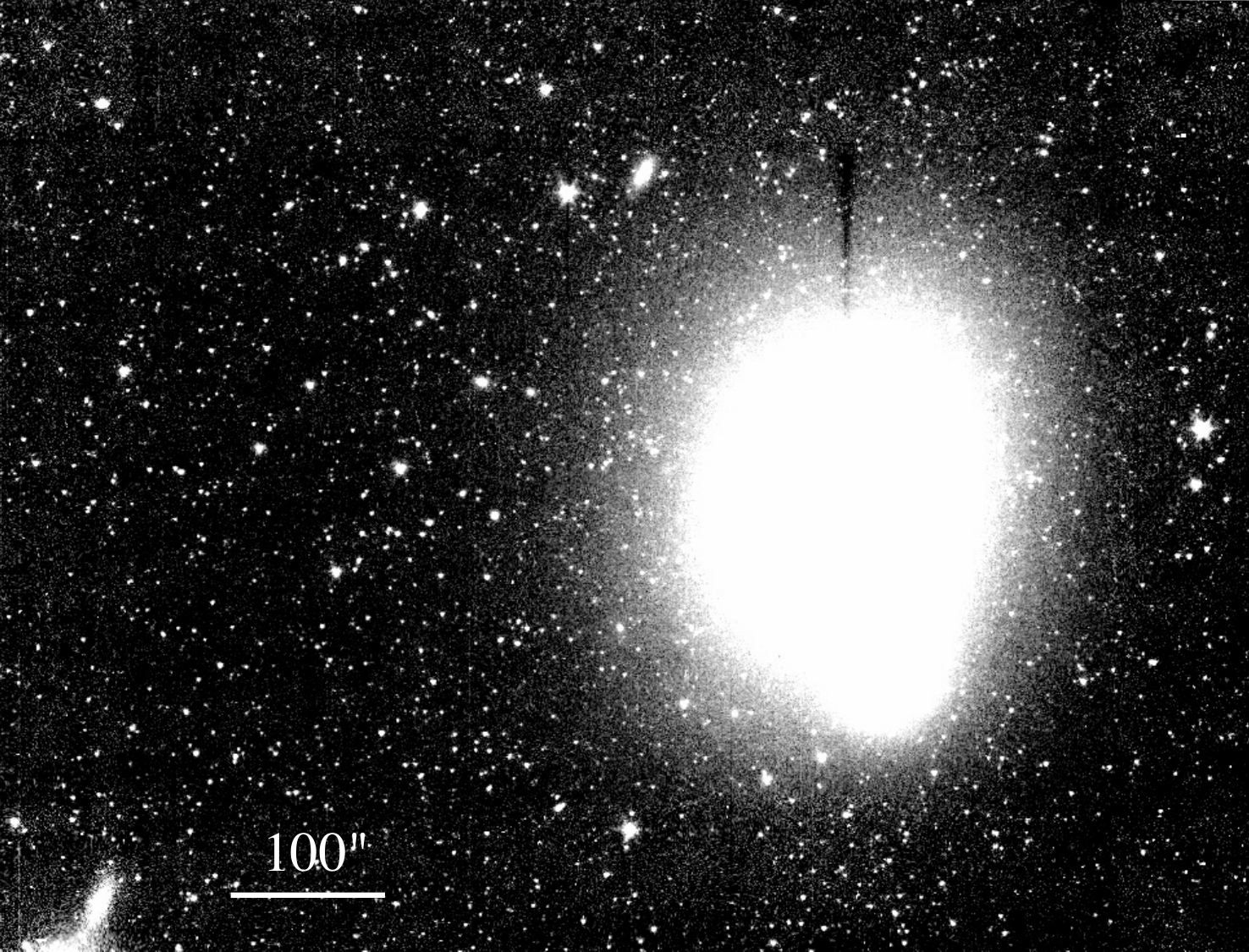} 
\includegraphics[trim=0.0cm 0cm 0.0cm 0cm, height=0.23\textwidth, angle=0]{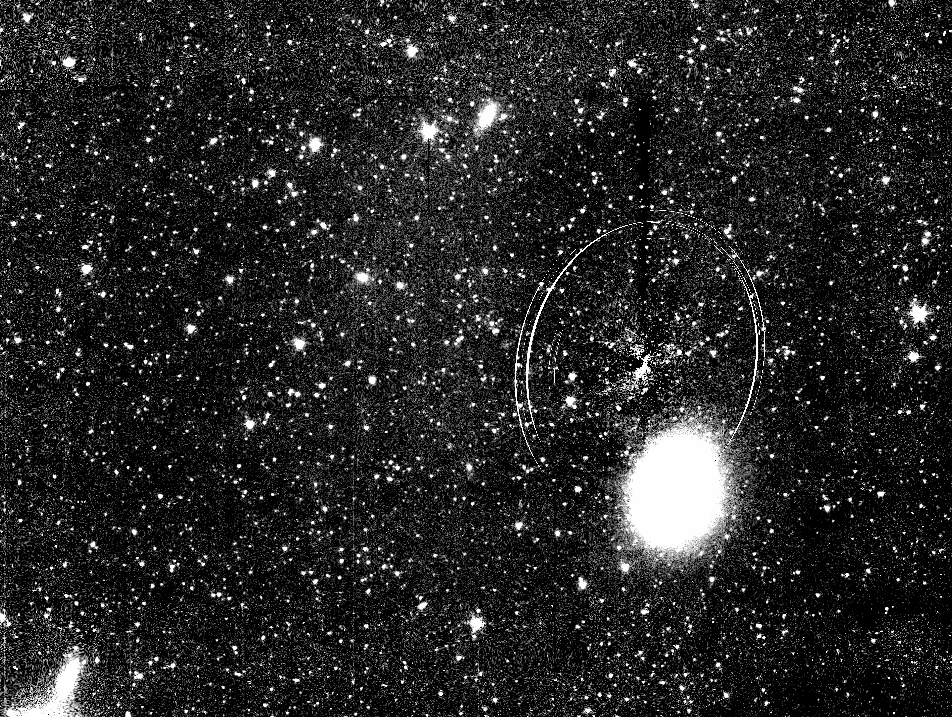}
\includegraphics[trim=0.0cm 0cm 0.0cm 0cm, height=0.23\textwidth, angle=0]{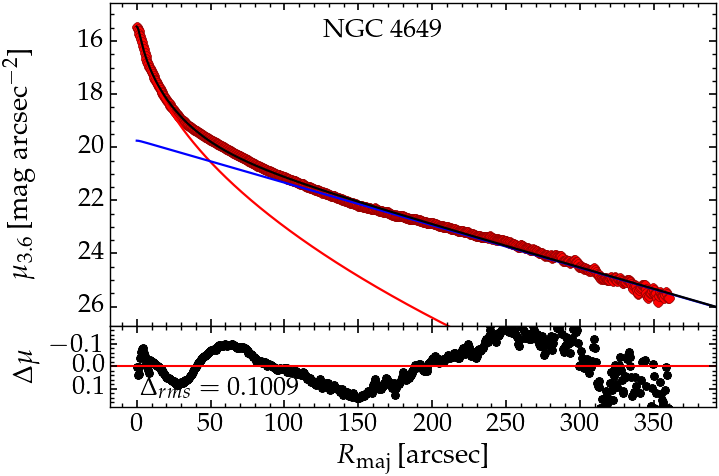} 
\end{center}
\caption{Left: Similar to Fig.~\ref{Fig_N3379} but for NGC~4649 (M60), which has a
  remarkably similar light distribution. 
    Image courtesy of {\it SHA}.  
The scale bar is 100$\arcsec$ $=$ 7.8~kpc long. 
    Middle: Residual image after removing the symmetrical light distribution 
in NGC~4649, thereby better revealing NGC~4647. 
Right: Major axis light profile 
decomposed into a S\'ersic spheroid (red curve: $R_{\rm e,maj}=22\arcsec.5$,
$\mu_{\rm e}=18.61$ mag arcsec$^{-2}$, and 
$n_{\rm maj}=2.04$) and exponential disc (straight blue line: $\mu_0=19.73$
mag arcsec$^{-2}$, $h_{\rm disc,maj}=68\arcsec = 5.3$~kpc). 
The disc is also evident in the kinematic map from \citet[][their figure~C1]{2011MNRAS.414.2923K}.
}
\label{Fig_N4649} 
\end{figure*}

\begin{figure}
\begin{center}
$
\begin{array}{c}
\includegraphics[trim=0.0cm 0cm 0.0cm 0cm, width=0.85\columnwidth,
  angle=0]{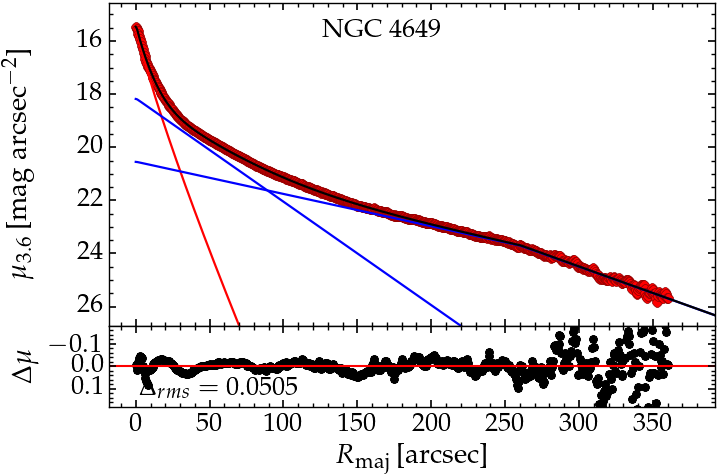} \\
\includegraphics[trim=0.0cm 0cm 0.0cm 0cm, width=0.85\columnwidth,
  angle=0]{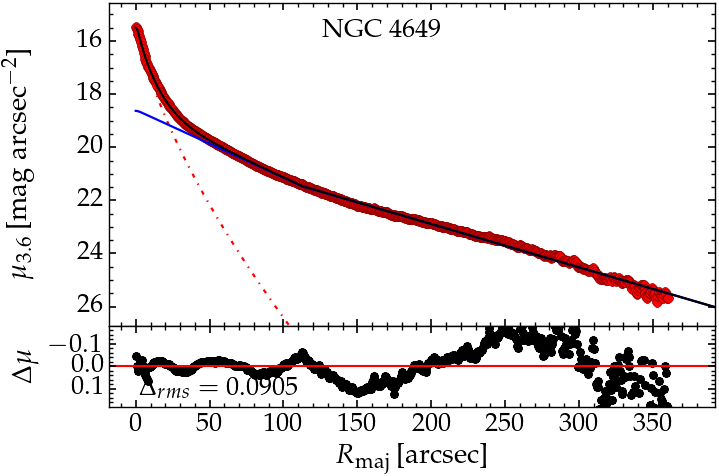} \\
\end{array}
$
\end{center}
\caption{Similar to Fig.~\ref{Fig_dd} but for NGC~4649.
  Top panel: S\'ersic spheroid (red curve: $R_{\rm e,maj}=9\arcsec.84$ and
  $n_{\rm maj}=1.33$) plus intermediate-scale disc (straight
  blue line: $\mu_0=18.20$ mag arcsec$^{-2}$, $h=28\arcsec.4 = 2.2$~kpc) and 
  weakly-truncated (at $r=260\arcsec$) large-scale disc (bent blue line:
  $\mu_0=20.58$ mag arcsec$^{-2}$, $h_{\rm inner}=90\arcsec = 7$~kpc,
  $h_{\rm outer}=55\arcsec = 4.3$~kpc).
  Bottom panel: Spheroid (red curve: $R_{\rm e,maj}=12\arcsec.7$, 
  $n_{\rm maj}=1.46$, and $\mu_e=17.84$ mag arcsec$^{-2}$)
  plus anti-truncated disc (bent blue line: $\mu_0=18.58$ mag arcsec$^{-2}$,
  $R_{\rm bend}=113\arcsec$, $h_{\rm inner}=42\arcsec.4 = 3.3$~kpc,
  $h_{\rm outer}=66\arcsec.6 = 5.2$~kpc).
  Note: as seen in the upper panel, 
  an additional downward bend at $r\approx 260\arcsec$ would improve the
  fit in the lower panel. 
} 
\label{Fig_dd2}
\end{figure}

\subsection{A trio of lenticular galaxies (S0)}
\label{Sec_trio}

\subsubsection{NGC~4649 (M60)}
\label{Sec_N4649}

NGC~4649 was not modelled to sufficiently large radii in
\citet{2019ApJ...876..155S} to clearly detect the large-scale disc-like
component whose kinematics were observed by \citet{2014ApJ...791...80A} to
extend to at least four arcminutes.  Comparing with the simulations of
\citet{2011MNRAS.416.1654B}, \citet{2017MNRAS.467.4540B} note that NGC~4649 is
consistent with being an E galaxy that is a fast-rotating merger remnant. The
system's net angular momentum has not been cancelled in the merger, and a (hot
disc-like) structure with significant rotation remains.  The passage of (old
or young) satellites through discs can also heat them
\citep{2016ApJ...823....4D}.
Remodelling the light profile of NGC~4649 to six arcminutes
(Fig.~\ref{Fig_N4649}) yields a galaxy (model-extrapolated) 3.6~$\mu$m
magnitude of 8.16 mag and a S\'ersic spheroid magnitude of 8.85 mag when
including a single large-scale exponential disc.  This yields a $B/T$ ratio of
0.53.
From an initial luminosity distance of 16.2$\pm$1.1~Mpc \citep{2001ApJ...546..681T},
and after a 0.083 mag correction/reduction to the distance modulus
\citep{2002MNRAS.330..443B, 2019Natur.567..200P}, 
and with $(B-V)_{\rm Vega}=0.95$ and thus 
$M_\star/L_{3.6} = 0.82$ \citep[][their equation~4]{Graham:Sahu:22a}, 
the logarithm of the galaxy stellar mass is 11.48 dex. 
This is just 0.08 dex less than previously obtained 
\citep{Graham:Sahu:22a}.  However, the spheroid mass may still be too high
for the same reasons\footnote{A dimmed galaxy magnitude is
a common occurrence when adding a large-scale disc because the extrapolated
model's light at large radii
coming from
an extrapolated exponential, i.e., a S\'ersic $n=1$, 
disc model is generally less than that from an extrapolated single-S\'ersic
galaxy model with $n>1$. This is one of the reasons why vast numbers
of automated 
single-S\'ersic fits to massive galaxies overestimate their luminosities.
Of course, while the two-component fit gives improved/stable galaxy
magnitudes, the component parameters can still be quite erroneous, as repeatedly
illustrated here in Section~\ref{Sec_N4649}.} 
as seen with NGC~3379 in Section~\ref{Sec3379}.

\begin{figure*}
\begin{center}
\includegraphics[trim=0.0cm 0cm 0.0cm 0cm, height=0.21\textwidth, angle=0]{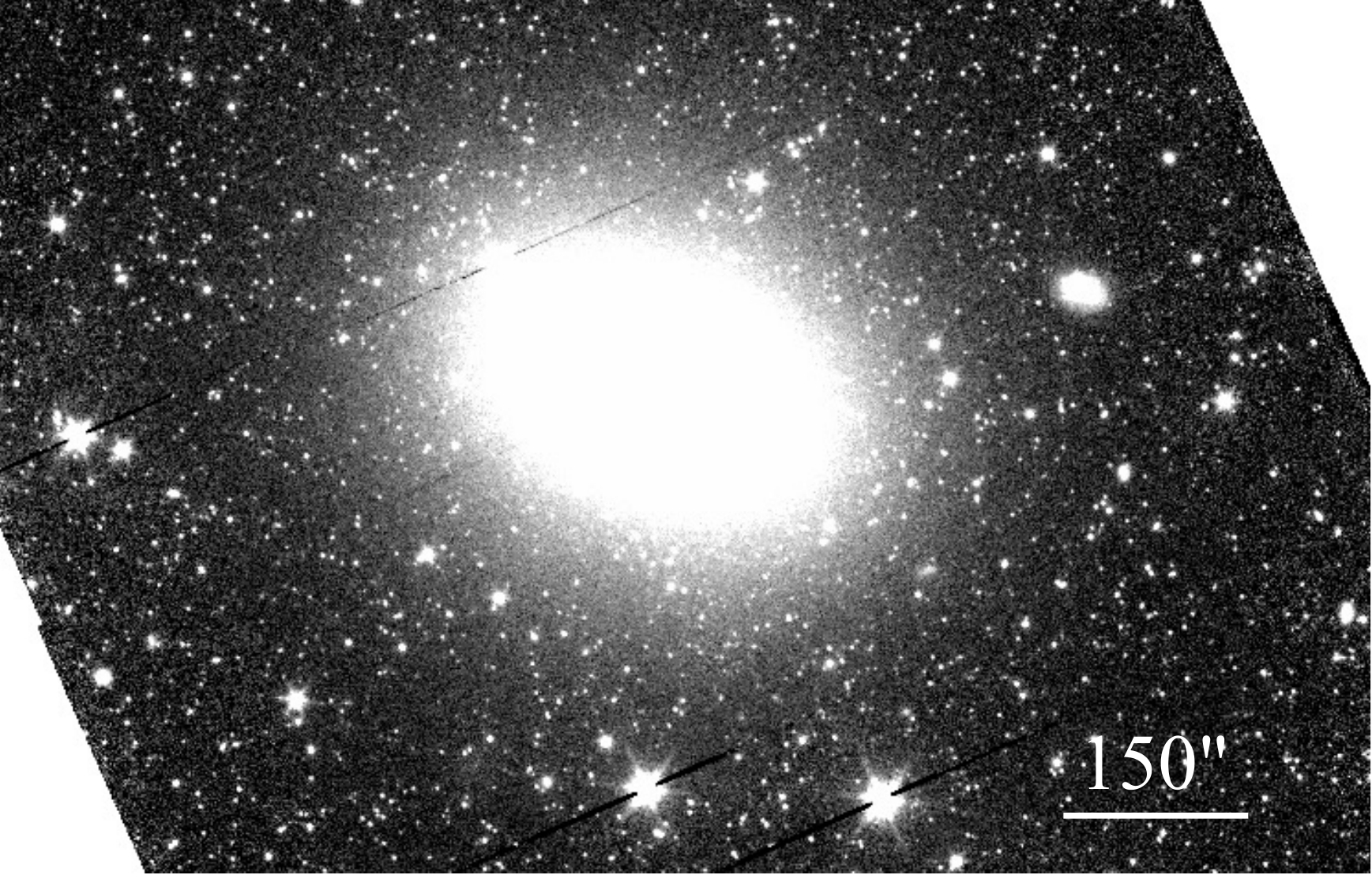} 
\includegraphics[trim=0.0cm 0cm 0.0cm 0cm, height=0.21\textwidth, angle=0]{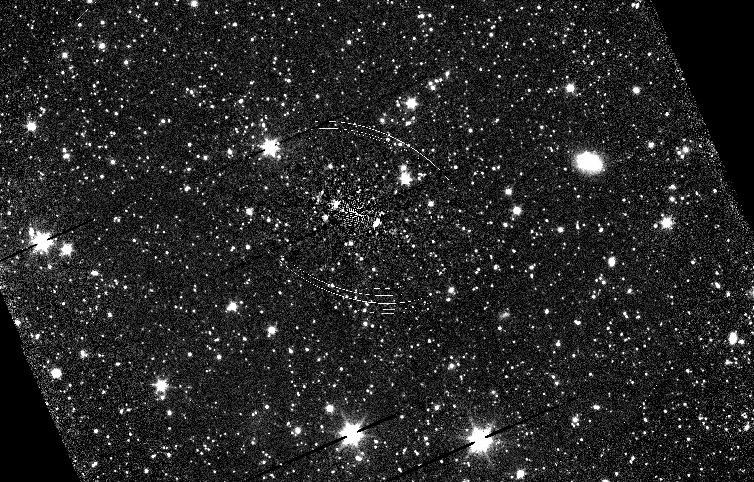}
\includegraphics[trim=0.0cm 0cm 0.0cm 0cm, height=0.21\textwidth, angle=0]{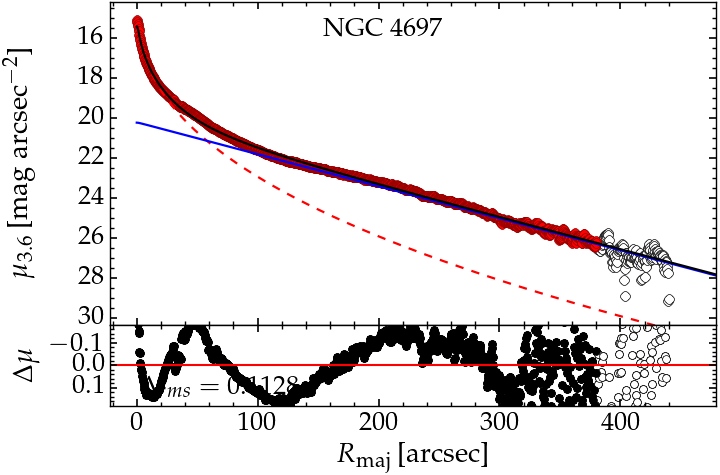} 
\end{center}
\caption{Similar to Fig.~\ref{Fig_N3379} but for NGC~4697. 
{\it Spitzer/IRAC1} 3.6~$\mu$m image courtesy of
 S$^4$G \citep[pipeline 1 image:][]{2015ApJS..219....3M}. 
The scale bar is 150$\arcsec$ $=$ 8.1~kpc long. 
Right panel: 
The open circles were excluded from the fit. 
For reference: $n_{\rm sph,maj}=2.4$, $R_{\rm e,sph,maj}=24\arcsec$,
$\mu_{\rm e,sph}= 18.98$ mag arcsec$^{-2}$, $\mu_{\rm 0,disc}=20.22$ mag
arcsec$^{-2}$, and $h_{\rm disc,maj}=68\arcsec$. 
}
\label{Fig_N4697} 
\end{figure*}

As with NGC~3379 (Section~\ref{Sec3379}), 
adding a second (intermediate-scale) disc to NGC~4649  does not affect the
total galaxy magnitude derived from the spheroid plus single exponential disc
fit above but improves the fit.  The residual light profile in the top panel of 
Fig.~\ref{Fig_dd2} is notably improved compared to that in 
Fig.~\ref{Fig_N4649}.  The spheroid 
magnitude changes from 8.85 to 9.82 mag, which is $\sim$1 
mag fainter than acquired when fitting a S\'ersic plus (single) large-scale
exponential
component.  The new $B/T$ flux ratio in NGC~4649 is 0.22. 
The lower panel of Fig.~\ref{Fig_dd2} reveals how an anti-truncated disc 
effectively captures the inner, intermediate-scale disc and the outer
large-scale disc.  The hump in the
residual light profile from $\sim$150 to $\sim$350 arcseconds, previously seen in
Fig.~\ref{Fig_N4649}, can be removed with the introduction of a second bend 
in the new disc model, matching the single bend at $\sim$260$\arcsec$ in the
truncated exponential disc model used in the upper panel of
Fig.~\ref{Fig_dd2}.  However, as with the light profile for NGC~3379,
this second bend at $r\sim$260$\arcsec$ commences at $\mu_{3.6}\approx 24$ mag
arcsec$^{-2}$ and could result from an over-subtraction of the `sky background'
for this galaxy, the third brightest in the Virgo Cluster, or perhaps due to 
contamination from NGC~4647. 
Finally, as with the discussion around NGC~3379, the need for a large
partially depleted core in NGC~4649 \citep{2006ApJS..164..334F,
  2014MNRAS.444.2700D} is diminished after adding disc component(s) that
reduce the bulge S\'ersic index from $>2$ to $<$2. This is evident from the
inner-most region of the residual light profiles in Fig.~\ref{Fig_N4649} and Fig.~\ref{Fig_dd2}

\begin{figure}
\begin{center}
$
\begin{array}{c}
  \includegraphics[trim=0.0cm 0cm 0.0cm 0cm, width=0.85\columnwidth,
  angle=0]{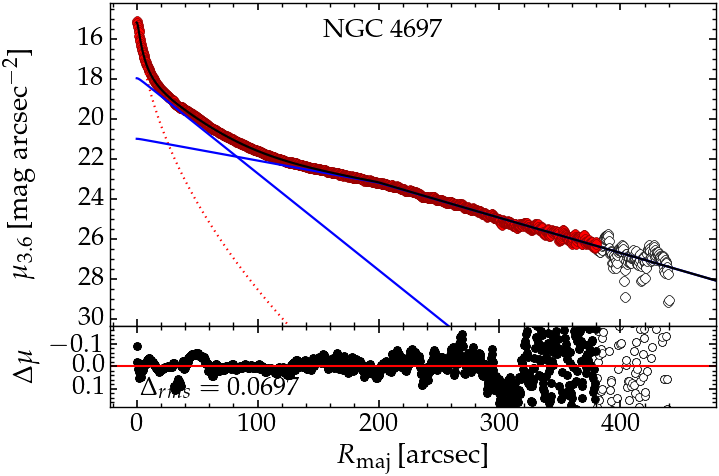} \\
\includegraphics[trim=0.0cm 0cm 0.0cm 0cm, width=0.85\columnwidth,
  angle=0]{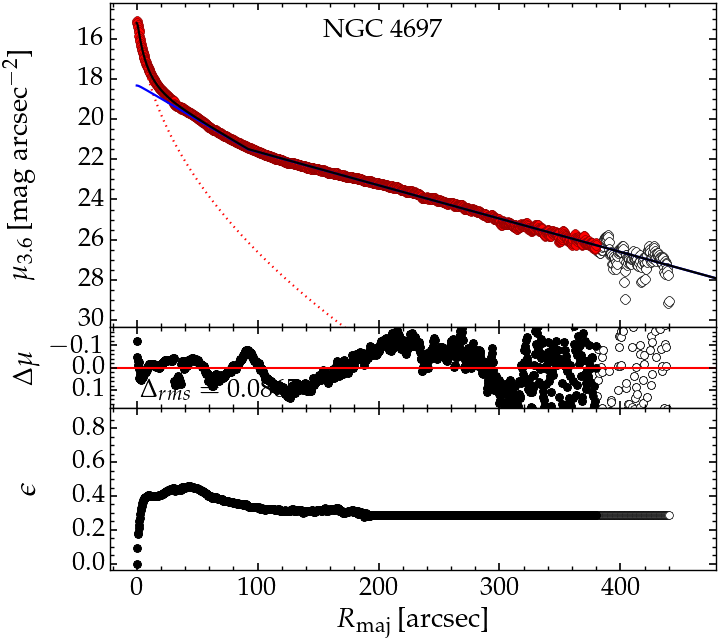} \\
\end{array}
$
\end{center}
\caption{Similar to Fig.~\ref{Fig_dd} but for NGC~4697. 
Lower panel: Major axis light profile fit with a 
S\'ersic spheroid (red curve, $R_{\rm e,maj}=9\arcsec.5$ and
$n_{\rm maj}=2.12$) and an anti-truncated disc (bent blue line,
$\mu_0=18.27$, $R_{\rm bend,maj}=92\arcsec$, $h_{\rm inner,maj}=31\arcsec = 1.67$~kpc, and
$h_{\rm outer,maj}=65\arcsec = 3.51$~kpc). 
This `hot' outer disc model gives a bulge (aka spheroid) magnitude
of 10.39 mag and a total galaxy magnitude of 8.74 mag, as mentioned in Section~\ref{Sec_N4697}.
}
\label{Fig_4697-new}
\end{figure}

\begin{figure}
  \begin{center}
\includegraphics[trim=0.0cm 0cm 0.0cm 0cm, width=0.85\columnwidth, angle=0]{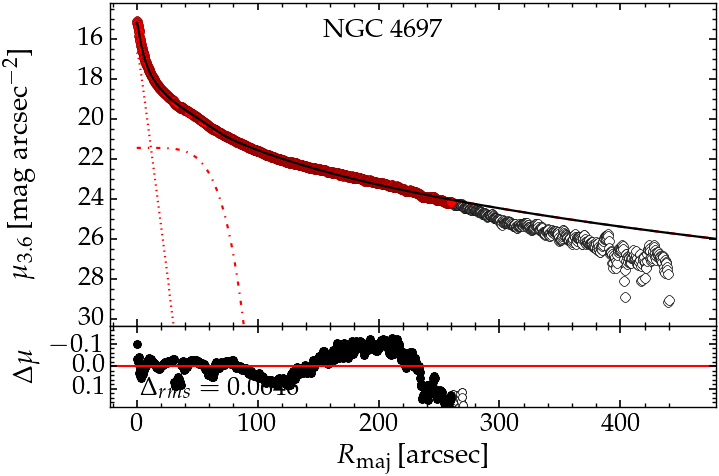} 
\end{center} 
  \caption{Sub-optimal (and rejected) fit to the major axis of NGC~4697. The
    dominant S\'ersic spheroid (red curve) has $R_{\rm e,maj}=80\arcsec$, 
    $n_{\rm maj}=3.6$, and contains 91.5 per cent of the total extrapolated
    flux.  Attempts to fit such a dominant S\'ersic spheroid with embedded
    discs \citep{2016ApJS..222...10S} results in an overestimation of the
    light at large radii.
  }
\label{Fig_N4697-old}
\end{figure}

\begin{figure*}
\begin{center}
\includegraphics[trim=0.0cm 0cm 0.0cm 0cm, height=0.24\textwidth, angle=0]{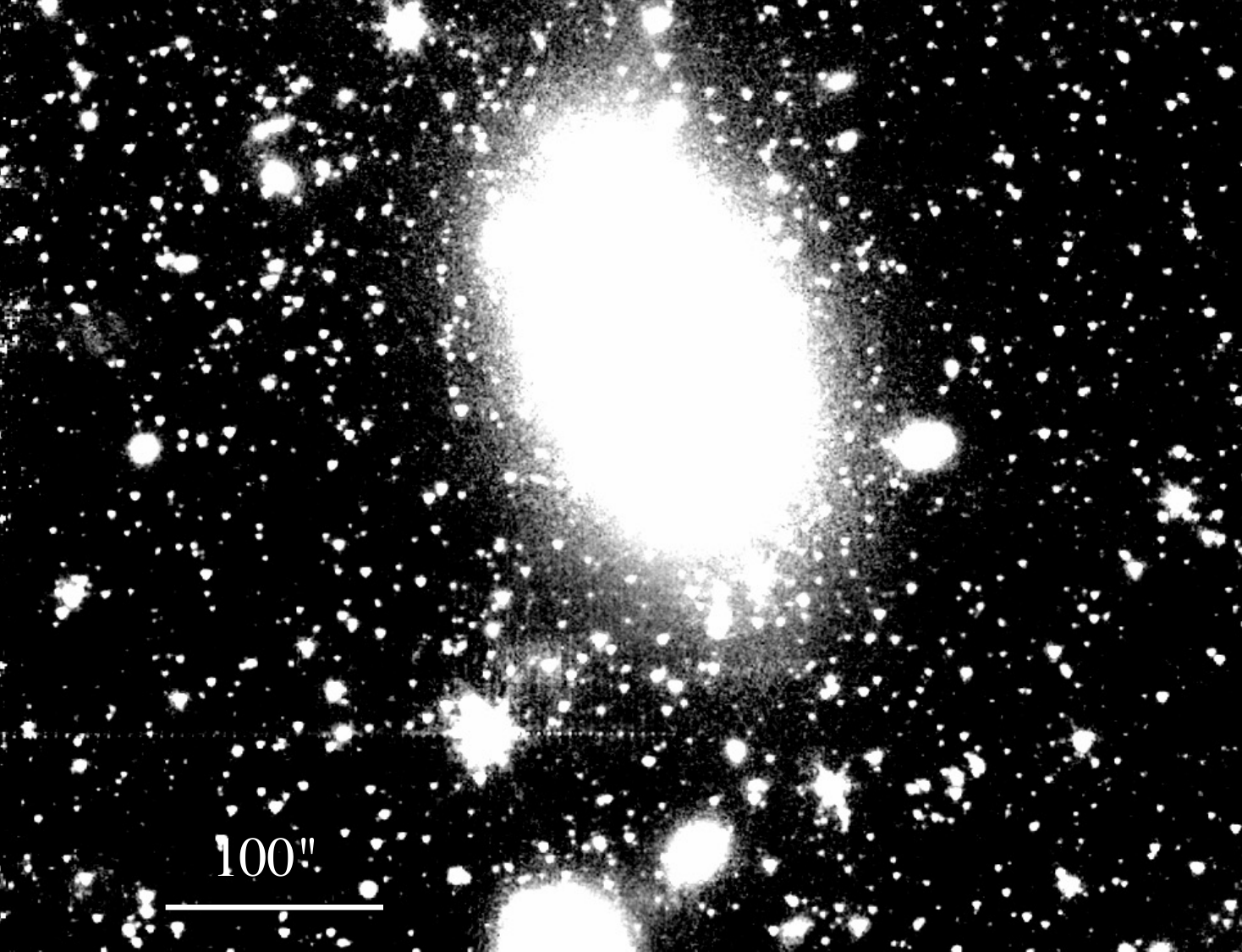} 
\includegraphics[trim=0.0cm 0cm 0.0cm 0cm, height=0.24\textwidth, angle=0]{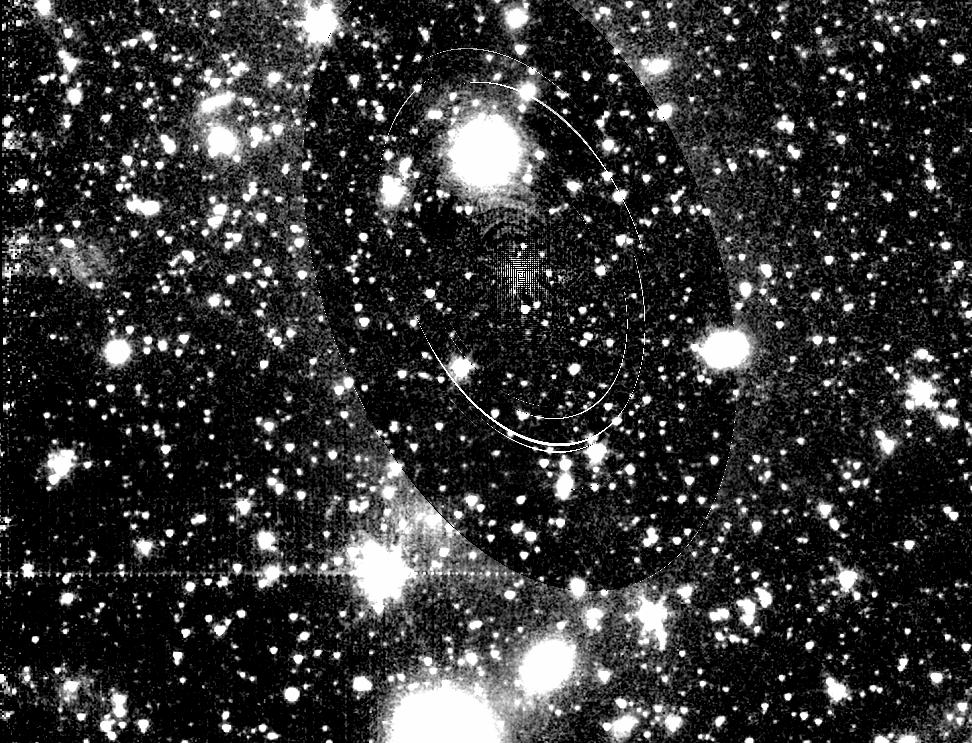} 
\includegraphics[trim=0.0cm 0cm 0.0cm 0cm, height=0.24\textwidth, angle=0]{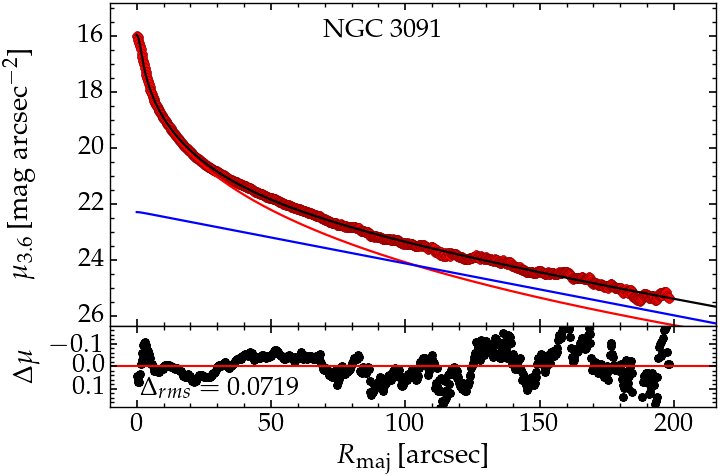} 
\end{center}
\caption{Similar to Fig.~\ref{Fig_N3379} but for NGC~3091.
  Image courtesy of {\it SHA}. 
The scale bar is 100$\arcsec$ $=$ 27.6~kpc long. 
  S\'ersic bulge (red curve: $\mathfrak{M}_{3.6} = 10.73$ mag, $n_{\rm
    maj}=3.47$, $R_{\rm e,maj}=24\arcsec.1$) plus exponential disc
  (straight blue line: $\mathfrak{M}_{3.6} = 11.88$ mag, $h_{\rm maj}=59\arcsec.3$).  }
\label{Fig_N3091} 
\end{figure*}

\begin{figure}
\begin{center}
$
\begin{array}{c}
\includegraphics[trim=0.0cm 0cm 0.0cm 0cm, width=0.85\columnwidth,
  angle=0]{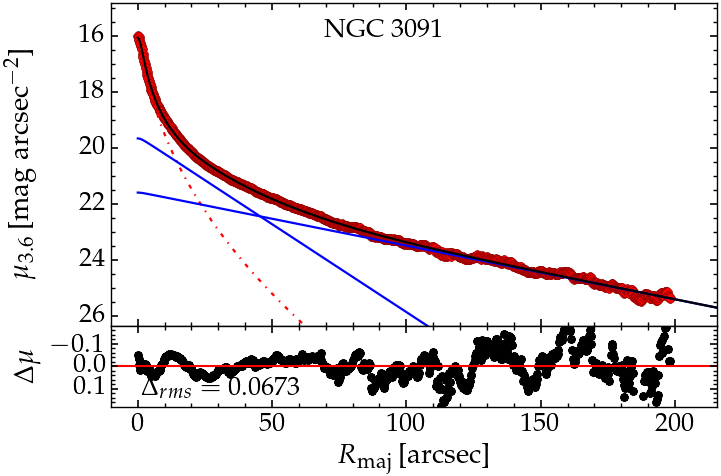} \\
  \includegraphics[trim=0.0cm 0cm 0.0cm 0cm, width=0.85\columnwidth,
  angle=0]{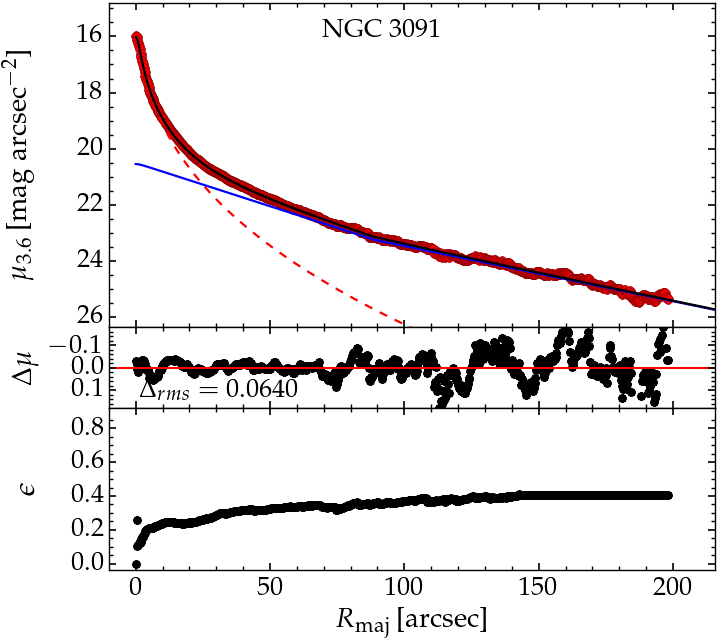} \\
\end{array}
$
\end{center}
\caption{
  Alternative (preferred) fits for NGC~3091. 
Upper panel: An intermediate-scale disc 
  ($\mu_0=19.57$ mag arcsec$^{-2}$, $h=17\arcsec.3 = 4.8$~kpc) and
  large-scale disc ($\mu_0=21.56$ mag arcsec$^{-2}$, $h=57\arcsec = 15.7$~kpc).  
  plus a S\'ersic bulge (red curve:  
  $\mu_{\rm e}=18.75$ mag arcsec$^{-2}$, $n_{\rm maj}=2.17$, and $R_{\rm
    e,maj}=7\arcsec.0$). 
  Lower panel: S\'ersic spheroid (red curve: $R_{\rm e,maj}=11\arcsec.4$,
  $\mu_{\rm e}=19.43$ mag arcsec$^{-2}$, and $n_{\rm maj}=2.67$) plus
  anti-truncated disc
  (bent blue line: $\mu_0=20.51$ mag arcsec$^{-2}$, $h_{\rm inner,maj}=35\arcsec = 9.7$~kpc, $R_{\rm
    bend,maj}=88\arcsec$, $h_{\rm outer,maj}=55\arcsec =15.2$~kpc). 
  The disc magnitude is 11.12 mag, and the total galaxy magnitude is 10.42 mag.
}
\label{Fig_3091-e}
\end{figure}

\subsubsection{NGC~4697}
\label{Sec_N4697}

A {\it Spitzer/IRAC1} 3.6~$\mu$m image of NGC~4697 was taken from S$^4$G. 
Again, a basic but inadequate S\'ersic-bulge $+$ exponential-disc model was
created (Fig.~\ref{Fig_N4697}). The snake-like residual light profile suggests there
is more going on.  
The first bump in the ellipticity profile peaks at 9$\arcsec$ 
\citep{2016ApJS..222...10S}, and 
\citet{2014ApJ...780...69L} reported two embedded discs in this galaxy. 
\citet[][their figure~C1]{2011MNRAS.414.2923K} report rotation of $\pm$131 km
s$^{-1}$ over the inner 30$\arcsec$, which had been shown to extend to
$\sim$90$\arcsec$ \citep{2008MNRAS.385.1729D} along the major axis.
From the final analysis (preferred fit) performed here (Fig.~\ref{Fig_4697-new}),
the total galaxy magnitude at 3.6~$\mu$m is 8.74 mag.
Breaking from the traditional interpretation that NGC~4697 is an E galaxy, it
is an ES or S0 galaxy.  \citet{Graham:Sahu:22b} regarded it as an ES galaxy,
but if the outer structure is rotating sufficiently fast, it may be considered
an S0 galaxy.  As with NGC~4278 and NGC~4494 (previously mentioned but not
in the current sample), \citet{2017MNRAS.467.4540B} report an 
upturn in the spin parameter for NGC~4697 beyond $\sim$100$\arcsec$, 
which is roughly the galaxy's major axis half-light radius \citep{2017MNRAS.464.4611F}. 
In addition to a thin, intermediate-scale disc, captured by the inner, steeper part
of the anti-truncated disc model in Fig.~\ref{Fig_4697-new}, 
NGC~4697 is modelled with a thick (``hot'') large-scale disc, captured by the
outer part of this anti-truncated disc model.
This double disc structure is supported by the kinematics maps of \citet[][see their
  figure~14]{2014ApJ...791...80A}, where the data extends to 146$\arcsec$
along the major axis, and the inner disc extends to $100\pm20\arcsec$. 
Fitting an anti-truncated disc model to capture the two discs
gives a spheroid magnitude of 10.39 mag and a $B/T$ ratio of 0.22.
At a luminosity distance of 11.3$\pm$0.7~Mpc
\citep{2001ApJ...546..681T, 2002MNRAS.330..443B, 2019Natur.567..200P}, 
and using $M/L=0.70$ \citep{Graham:Sahu:22a}, the
logarithm of the spheroid's stellar mass is 10.20~dex. 

If a disc becomes too `hot', it
will  resemble a spheroid, and this may be what we are witnessing in
NGC~4697, where the outer component visually resembles a spheroid but rotates
more like a disc.
\citet{1987ApJ...312..514C} characterised NGC~4697 as a rapidly
rotating E galaxy with an embedded weak disc.  \citet{2014ApJ...780...69L} labelled the
weak disc as embedded within an `envelope' instead of a spheroid.
Although NGC~4697 is only a BCG, it appears to be an S0 galaxy (Fig.~\ref{Fig_4697-new}) 
rather than an ES galaxy (Fig.~\ref{Fig_N4697-old}), as modelled in \citet{2016ApJS..222...10S}. 
From the kinematic maps of \citet[][see their figure~14]{2014ApJ...791...80A}, 
the $V/\sigma$ ratio beyond the inner disc may be around 1, reflective of a
heated disc (with $\sigma > 50$ km s$^{-1}$ that is 
transitioning to a spheroid. To err on the side of caution, this
galaxy is grouped with the S0 galaxies and thus excluded from the E/ES sample. 
NGC~821 and NGC~3377 appear as more evolved versions of NGC~4697, such that
their outer component (hot disc now spheroid) do not have quite the
same level of `spin' \citep[][their figure~11]{2017MNRAS.467.4540B}.

\begin{figure*}
\begin{center}
\includegraphics[trim=0.0cm 0cm 0.0cm 0cm, height=0.24\textwidth, angle=0]{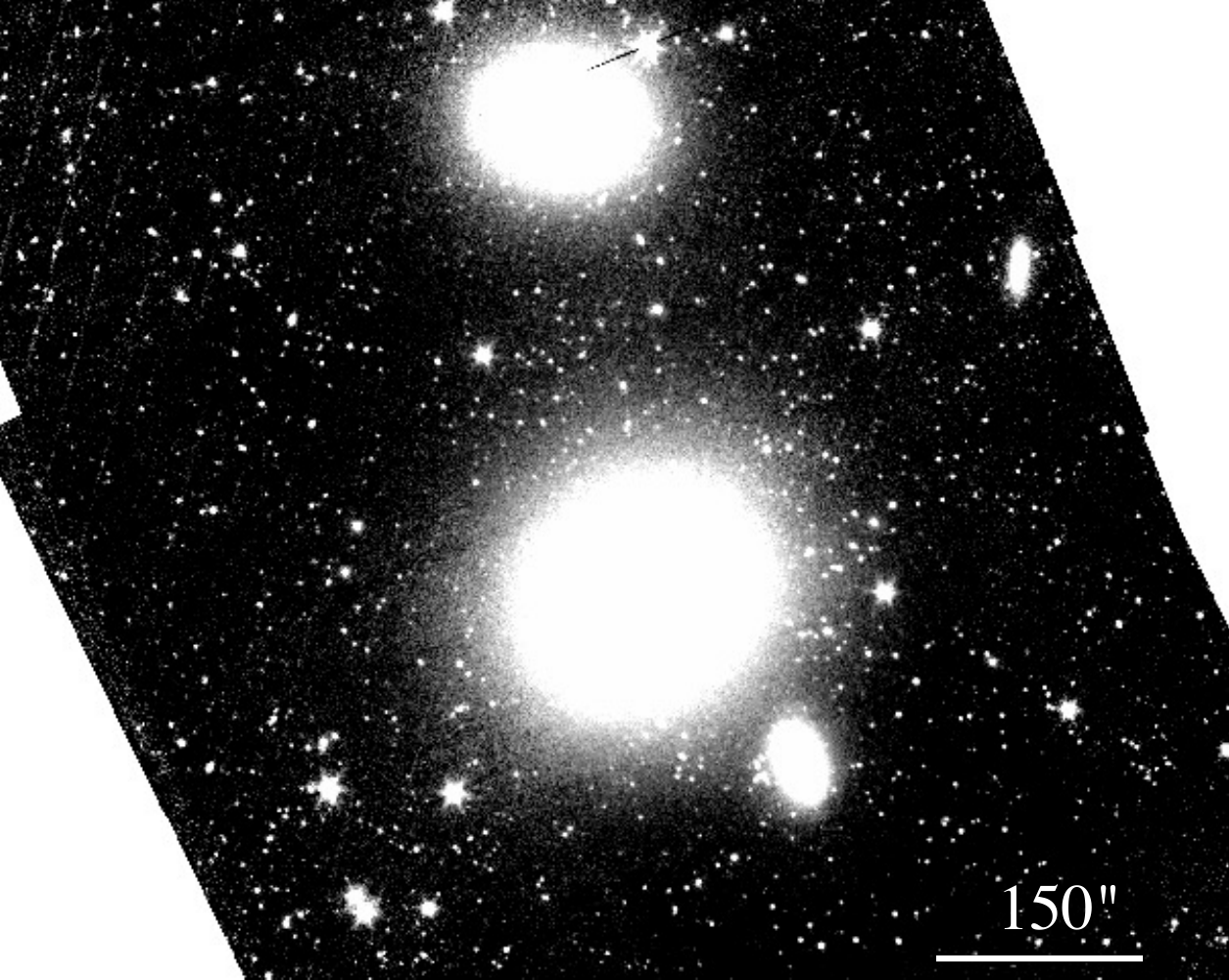}
\includegraphics[trim=0.0cm 0cm 0.0cm 0cm, height=0.24\textwidth, angle=0]{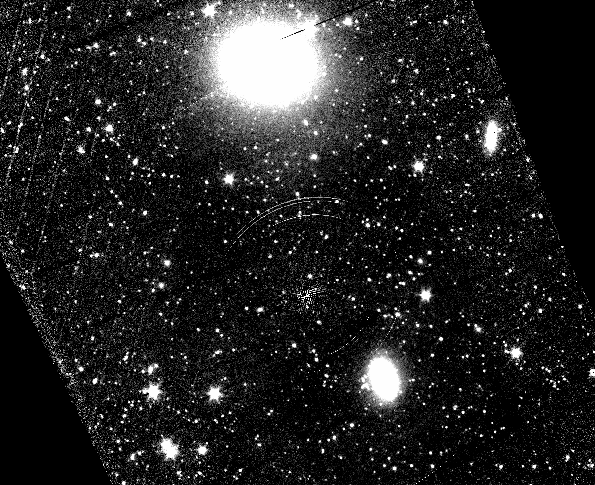}
\includegraphics[trim=0.0cm 0cm 0.0cm 0cm, height=0.24\textwidth, angle=0]{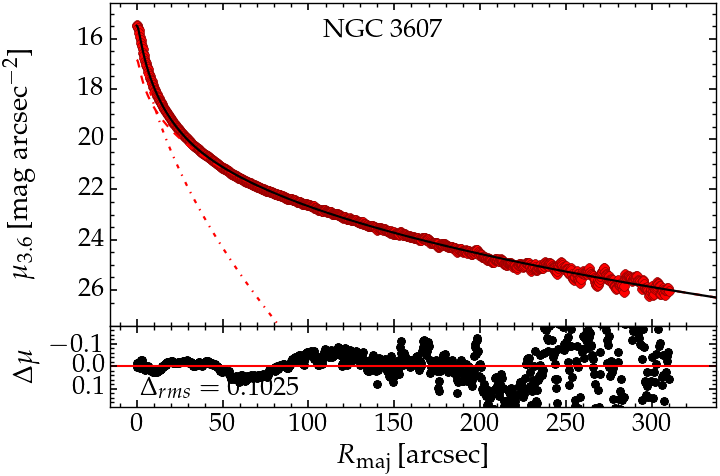}
\end{center}
  \caption{Similar to Fig.~\ref{Fig_N3379} but for NGC~3607. 
    Image courtesy of S$^4$G. 
The scale bar is 150$\arcsec$ $=$ 18.0~kpc long. 
A S\'ersic model has been used for the large-scale spheroidal component
($n_{\rm maj}=3.06$, $R_{\rm e,maj}=59\arcsec.8$, and $\mu_{\rm e}=21.52$ mag
arcsec$^{-2}$) and
also for the inner disc component(s):
$n_{\rm maj}=1.80$, $R_{\rm e,maj}=8\arcsec.4 = 1.0$~kpc, $\mu_{\rm e}=18.29$ mag
arcsec$^{-2}$.
NGC~3605 is in the lower portion of the image, while NGC~3608 is in the upper
portion. 
  }
\label{Fig_N3607}
\end{figure*}

\begin{figure}
\begin{center}
\includegraphics[trim=0.0cm 0cm 0.0cm 0cm, width=0.85\columnwidth, angle=0]{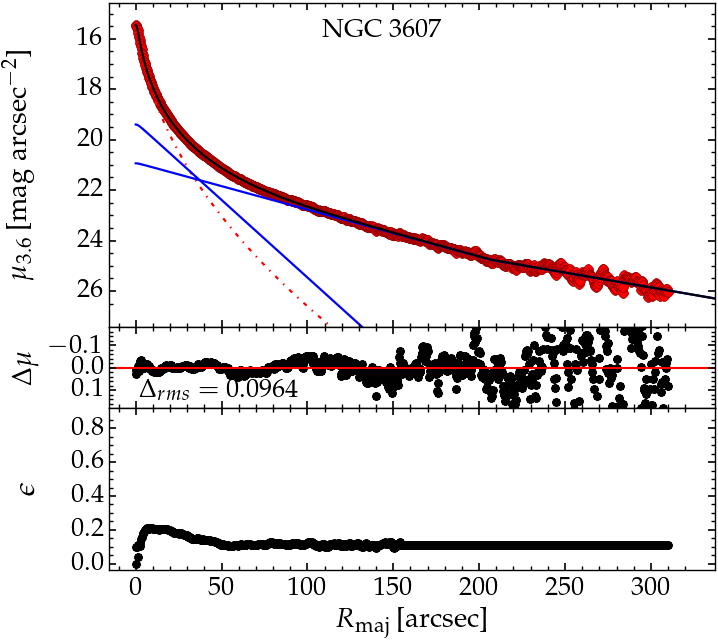}
\end{center}
\caption{Alternative (questionable and not preferred) fit for NGC~3607
  obtained by adding a large-scale anti-truncated disc (bent blue line:
  $\mu_0=20.91$ mag arcsec$^{-2}$, $h_{\rm inner,maj}=58\arcsec.4 = 7.0$~kpc,
  $R_{\rm bend,maj}=207\arcsec$, $h_{\rm inner,maj}=92\arcsec.4 = 11.1$~kpc),
  plus two intermediate-scale discs, the smaller of which is fit with a
  S\'ersic function (red, inner-most dot-dashed curve: $n_{\rm maj}=2.0$,
  $R_{\rm e,maj}=10\arcsec.3 = 1.24$~kpc, and $\mu_{\rm e}=18.28$ mag
  arcsec$^{-2}$) matching the strongly rotating component in the kinematic map
  \citep{2011MNRAS.414.2923K} and the second with an exponential disc
  (straight blue line: $\mu_0=19.32$ mag arcsec$^{-2}$, $h=17\arcsec.7 =
  2.1$~kpc) matching the ellipticity profile.  }
\label{Fig_N3607-disc}
\end{figure}

\subsubsection{NGC~3091}

NGC~3091 was shown to rotate fast by \citet{1989ApJ...344..613F}, reaching a
rotational speed of 96 km s$^{-1}$ by 14$\arcsec$, as confirmed by
\citet{1992AandA...262...52B}.  Furthermore, its ellipticity profile
asymptotes to 0.4 at $R_{\rm maj}=200\arcsec$, suggesting a large-scale disc
rather than (solely) an intermediate-scale disc is present.
Moreover, upon closer inspection, E4 galaxies are almost always found to be S0 galaxies.
The light profile from NGC~3091 is fit with a S\'ersic-bulge plus an
exponential-disc in Fig.~\ref{Fig_N3091}. The total 3.6~$\mu$m galaxy
magnitude is 10.41 mag.  However, once again, a better fit for NGC~3091
contains an anti-truncated disc (Fig.~\ref{Fig_3091-e}), giving a revised
total magnitude of 10.42 mag.  In passing, it is noted that the 15~kpc
scalelength of the outer half of the disc is large and may be indicative that
something else is afoot.  Possibly, there is some contribution of
halo/envelope, as NGC~3091 resides in a compact group of galaxies
\citep{1982ApJ...255..382H} where encounters may strip stars from galaxies and
start to build the intragroup light.
Curiously, \citet{2007MNRAS.378.1575S} discovered a tendency for the envelopes
of BCGs to be exponential, a result subsequently seen by \citet{2008A&A...483..727P}.
Furthermore, the cores in some of these galaxies, built from infalling dense
galaxy nuclei \citep{2010ApJ...725.1707G}  rather
than just massive black holes, may be great enough that the central spheroid takes
on a low S\'ersic index in a process dubbed `galforming' \citep{Graham:Sahu:22b}.
Also worth bearing in mind are the planes-of-satellites
around massive S galaxies \citep[e.g.][and references
  therein]{2020MNRAS.491.3042P, 2024MNRAS.528.2805K} that are, arguably, related
to pre-merger encounters \citep{2022MNRAS.513..129B}.

Using a Galactic-extinction-corrected $(B-V)_{\rm Vega}$ colour of 0.96 mag
gives $M_\star/L_{3.6} = 0.84$ \citep[][their equation~4]{Graham:Sahu:22a}.
At a luminosity distance of 58.6$\pm$10.9~Mpc \citep{2007A&A...465...71T}, the
galaxy's 3.6~$\mu$m absolute magnitude $\mathfrak{M}_{3.6}= -23.43$~mag, and,
as before, using $\mathfrak{M}_{\odot,3.6} = 6.02$ \citep[AB
  mag:][]{2018ApJS..236...47W}, the logarithm of the galaxy stellar mass is
11.70 dex, 0.16 dex less than reported in \citep{Graham:Sahu:22a} based on a
single S\'ersic fit.  Having a 'fast rotating' ETG with such a high stellar
mass is unusual.  The logarithm of the spheroid mass is roughly half this
value, at 11.38~dex when using the anti-truncated disc model, and it drops to
11.25 dex when additionally including a nuclear component.  The residual and
ellipticity profile suggests a nuclear ($\sim$2$\arcsec$) disc-like component
may exist.  Fitting for this changed the anti-truncated disc magnitude to
10.98 mag and dimmed the bulge magnitude to 11.55 mag ($n_{\rm
  maj}=2.36\pm0.14$, and $R_{\rm e,maj}=11\arcsec.4\pm0.72$).

\subsection{A couple of ellicular galaxies (ES)}
\label{Sec-App-ESe}

\subsubsection{NGC~3607}

The ES galaxy NGC~3607 is a merger product \citep{2009A&A...505...73R}
with a dust-rich inner disc. 
The kinematics presented by \citet{2011MNRAS.414.2923K} reveal that the 
fast-rotating inner disc, which extends to $R_{\rm maj}\approx15\arcsec$, is embedded
within a larger, slower-rotating component.
The behaviour of this inner embedded disc in the kinematic map is similar to that
seen in 
NGC~3377, NGC~4473, and other ES galaxies \citep{2014ApJ...791...80A}. 

Fig.~\ref{Fig_N3607} provides a basic decomposition into two components, with a
large-scale spheroid plus an inner component for this ES galaxy's embedded disc(s). 
The galaxy's total apparent magnitude is 9.27~mag. 
Conceivably, the upturn in the outer light profile ($\gtrsim$200$\arcsec$) may
be due to contamination from two neighbouring galaxies.  Nonetheless,
performing a fit within $R_{\rm maj}=220\arcsec$ and again extrapolating the
optimal fitted model to infinity dims the galaxy magnitude by just 0.06 mag.
Using the initial brighter magnitude, along with a luminosity distance of 25~Mpc and
an $M/L$ ratio of 0.75, yields $\log(M_{\rm \star,gal})= 11.37$~dex, which is
0.09~dex smaller than the value given in \citet{Graham:Sahu:22a}. 

The elevation in the ellipticity
profile extends to 45$\arcsec$ \citep{2016ApJS..222...10S}, suggesting an 
additional larger (than 15$\arcsec$) component is present. 
Some 17 per cent of the flux is 
assigned to the inner component (Fig.~\ref{Fig_N3607}) associated with the spikes in the rotation
profile (within $R_{\rm maj}\sim15\arcsec$) and the ellipticity profile (within
$R_{\rm maj}\sim45\arcsec$). 
These two features are drawn out more clearly in Fig.~\ref{Fig_N3607-disc}, where the galaxy
has been fit with a questionable large-scale anti-truncated disc.
This fit yields the same total galaxy magnitude as obtained with the fit shown
in Fig.~\ref{Fig_N3607} but is not preferred because a large-scale disc does not match the
kinematic or ellipticity profiles at large ($\gtrsim$45$\arcsec$) radii.
The kinematic and ellipticity data sets NGC~3607 apart from NGC~3091.
The 
fit shown in Fig.~\ref{Fig_N3607-disc} has been included to help visualise how a heated thick outer `disc' can 
produce a dynamically-hot spheroid-dominated (Fig.~\ref{Fig_N3607}). 
The decomposition in Fig.~\ref{Fig_N3607-disc} only applies 
{\em if} NGC~3607 has an overlooked face-on large-scale disc. 
\citet{2011MNRAS.415.2158R} report a depleted core with a break radius 
$R_b=0\arcsec.22$ in NGC~3607.

\subsubsection{NGC~4291}

\begin{figure*}
\begin{center}
\includegraphics[trim=0.0cm 0cm 0.0cm 0cm, height=0.27\textwidth, angle=0]{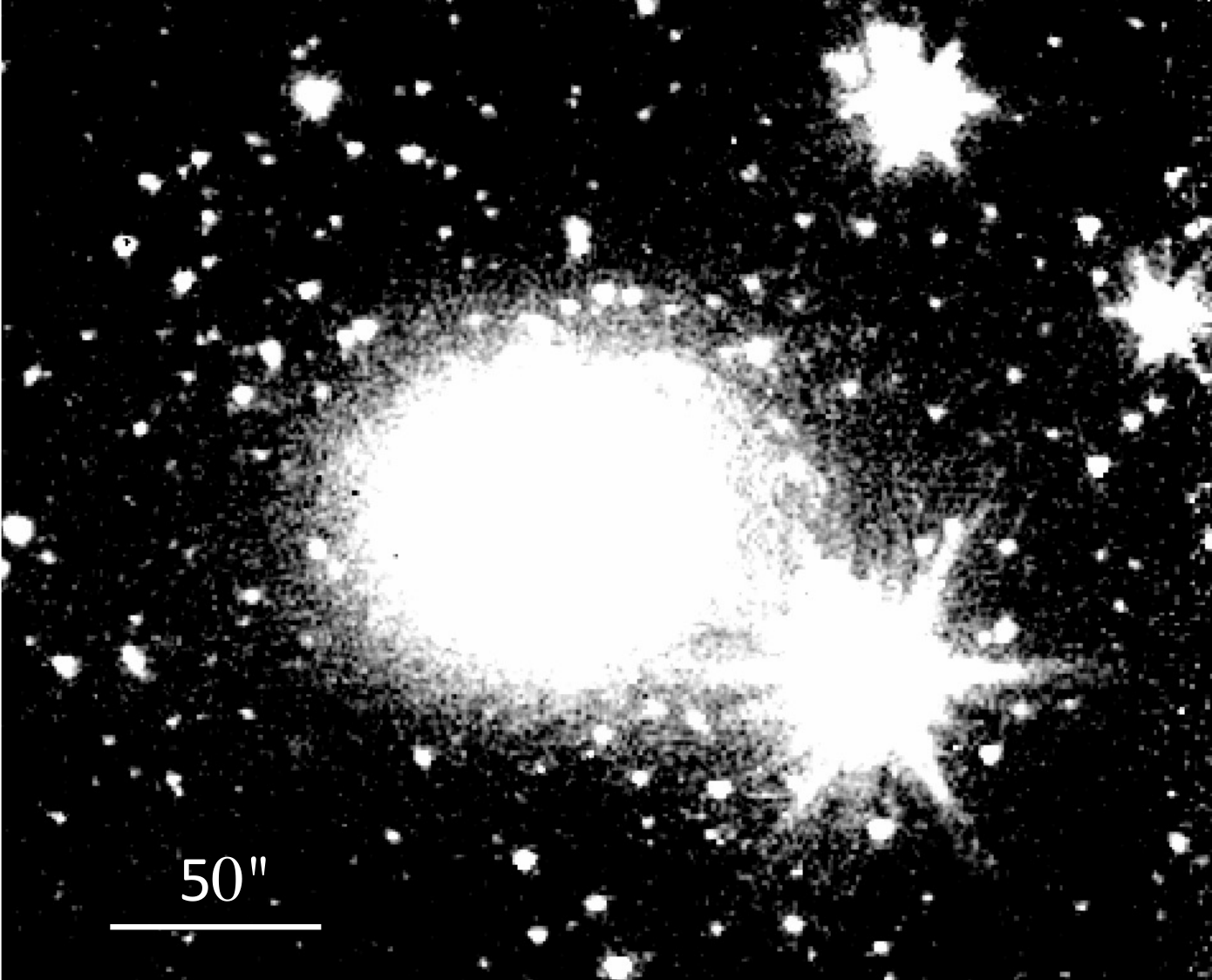} 
\includegraphics[trim=0.0cm 0cm 0.0cm 0cm, height=0.27\textwidth, angle=0]{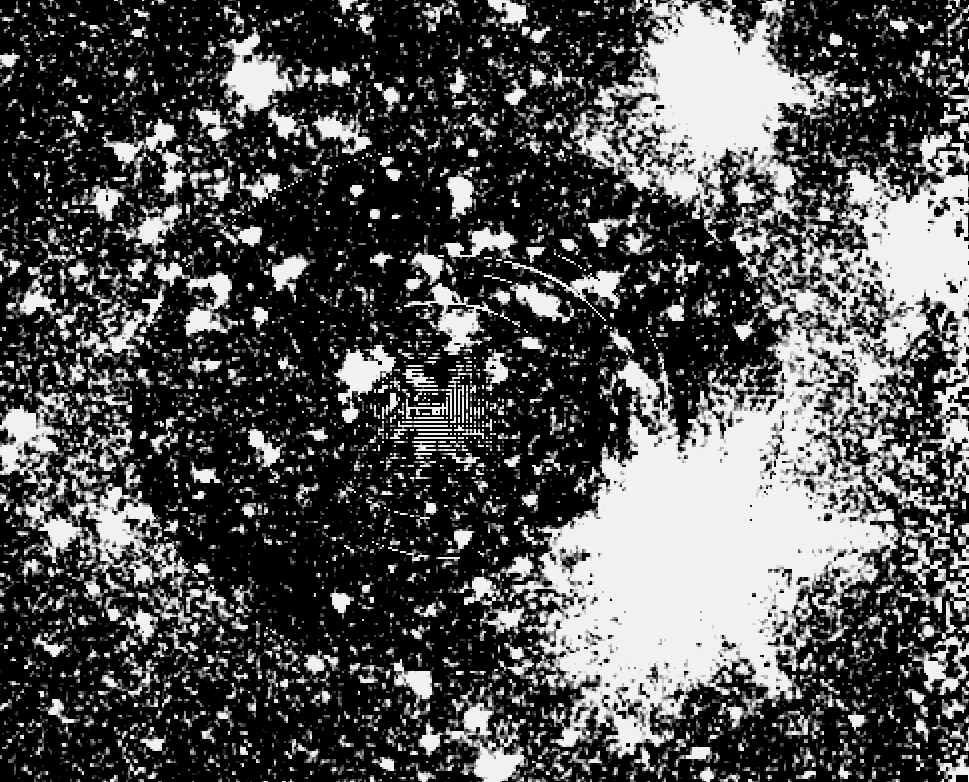} 
\includegraphics[trim=0.0cm 0cm 0.0cm 0cm, height=0.27\textwidth, angle=0]{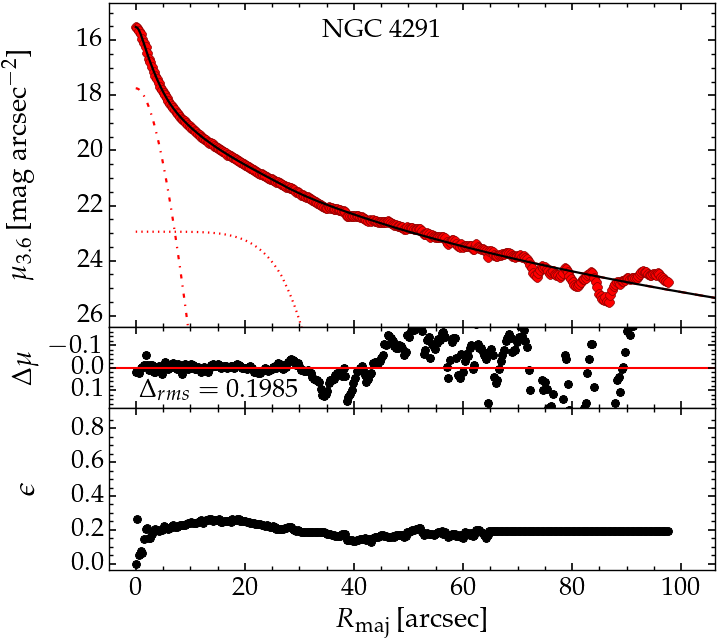}
\end{center}
  \caption{NGC~4291 retires its E galaxy designation to become an ES galaxy
    --- image courtesy of {\it SHA}. 
The scale bar is 50$\arcsec$ $=$ 6.05~kpc long. 
    The spheroid-to-total ratio is 0.91. Two inner components fit with
    S\'ersic functions are detected, 
    with the ellipticity profile suggesting a somewhat face-on disc in accord
    with the rotation reported in the literature. The dominant spheroid has
    $n_{\rm maj}=3.65$, $R_{\rm e,maj}=15\arcsec.2 = 1.84$~kpc, and $\mu_{\rm e}=20.03$
    mag arcsec$^{-2}$. }
\label{Fig_N4291} 
\end{figure*}

\citet{1989AJ.....98..147J} report $V/\sigma=76/273 = 0.28$ for NGC~4291,
which makes it a `fast (inner)  rotator', although the rotation falls from 100 km
s$^{-1}$ at 3$\arcsec$ to 40 km s$^{-1}$ at 12$\arcsec$, before peaking again
at $\sim$15$\arcsec$ and then falling back down to $\sim$50 km s$^{-1}$ by
20$\arcsec$ \citep[see also][]{1994MNRAS.269..785B}. Therefore, there appear 
to be two components causing this rotation.
This pattern is reflected in the ellipticity profile, which rises rapidly to 
peak at $\sim$15$\arcsec$ before slowly declining. 
Indeed, \citet{2003ApJ...596..903P} speculated that NGC~4291 may be an ES galaxy.

The surface brightness profile is well represented by a dominant spheroid plus
two inner components (Fig.~\ref{Fig_N4291}).  The total 3.6~$\mu$m magnitude is 
10.96 mag, and the spheroid magnitude is 11.06 mag. 
At a luminosity distance of 25.2$\pm$3.7~Mpc, and with $B-V=0.927$ and
$M_\star/L_{3.6}=0.78$, one has $\log(M_{\rm \star,sph})= 10.68$ dex, which is
0.12 dex smaller than the total galaxy stellar mass reported by
\citep{Graham:Sahu:22a} that was  based on a single-S\'ersic fit. 
This reduction is due 
in part (0.04 dex) to some reassigned mass into the two extra components and 
in part (0.08 dex) because the revised spheroid S\'ersic index declined
(from 4.2 to 3.6 on the major axis),   
and thus the large-radius tail of the S\'ersic profile falls away more quickly, 
i.e., there is a little less extrapolated light at large radii.  

The light profile for NGC~4291 appears to be a particularly good example of
how an anti-truncated disc can morph into a single spheroidal structure 
well described by the S\'ersic function.\footnote{With {\it Hubble Space
  Telescope}  resolution, NGC~4291 has a depleted core
with a break radius equal to 0$\arcsec$.3 \citep{2014MNRAS.444.2700D}.}  There
remains a hint of a double disc structure in the light profile.
The next galaxy that is remodelled is
NGC~5077.  Although it has a similar light profile to NGC~4291, little to no trace of
a double disc structure is apparent in its light profile.  Nonetheless, the
light profiles are still better modelled by including an inner and
faint intermediate-scale component.

\subsection{A couple of elliptical galaxies (E)}
\label{Sec-App-E}

\subsubsection{NGC~5077}

NGC~5077 was thought to possibly display somewhat fast (inner) and slow (outer) rotation
as far back as \citet{1991ApJ...373..369B}.
While there is no clear stellar rotation within 
the inner $\sim$10$\arcsec$, from $\sim$10-28$\arcsec$
there is systematic (ring?) rotation along the major axis,
albeit at just 30-40 km s$^{-1}$ \citep{2021AandA...650A..34R}. 
As such, this is not on par with the previous ES galaxies. 
Looking at the gas, rather than the stellar kinematics, 
NGC~5077 has a high-rotation ($\sim$300 km s$^{-1}$) polar gas disc that has been
accreted \citep{1991ApJ...373..369B, 2021AandA...650A..34R}.
The accreted system may contribute to the additional components detected
during the analysis of the light profile.
The accretion event may also have damaged/eroded the inner portion of a
previous intermediate-scale stellar disc aligned with the major axis.
Such accretion events can invoke star formation in (what becomes disturbed) E
galaxies \citep{2024arXiv240504166B}.

\begin{figure*}
\begin{center}
\includegraphics[trim=0.0cm 0cm 0.0cm 0cm, height=0.27\textwidth,
  angle=0]{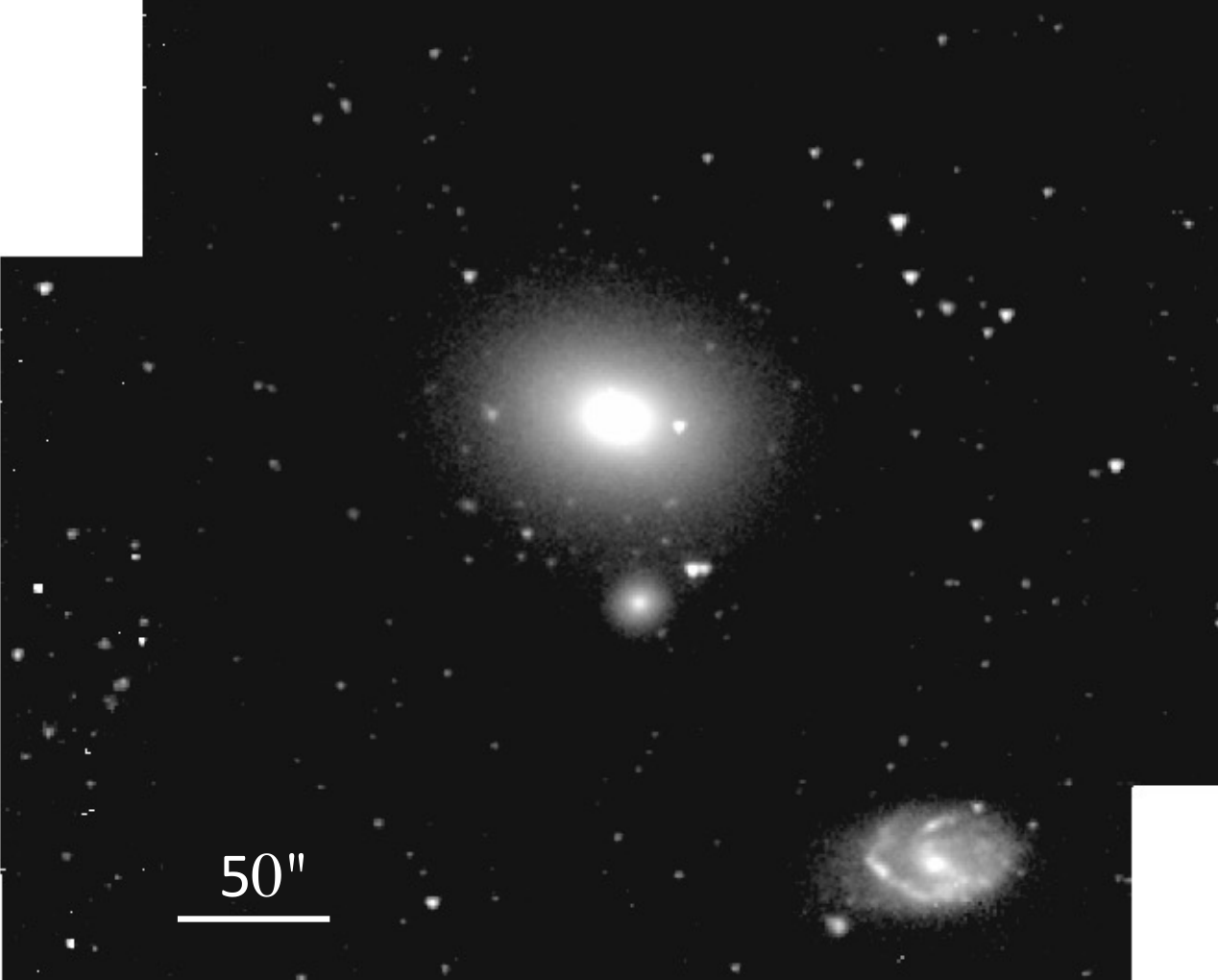} 
\includegraphics[trim=0.0cm 0cm 0.0cm 0cm, height=0.27\textwidth,
  angle=0]{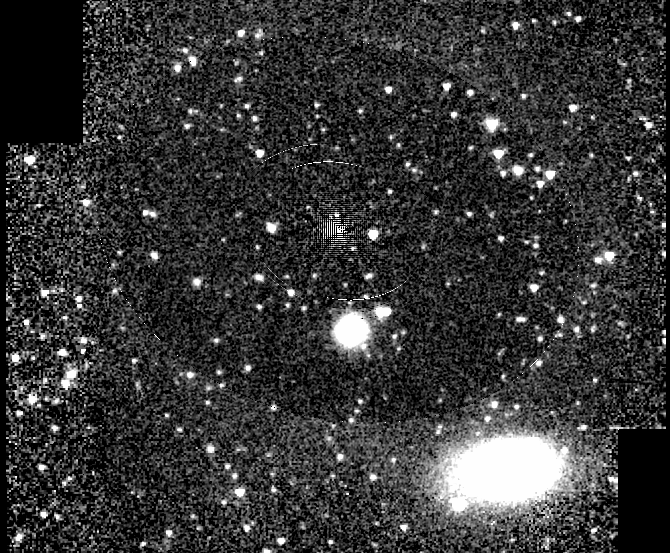}
\includegraphics[trim=0.0cm 0cm 0.0cm 0cm, height=0.27\textwidth,
  angle=0]{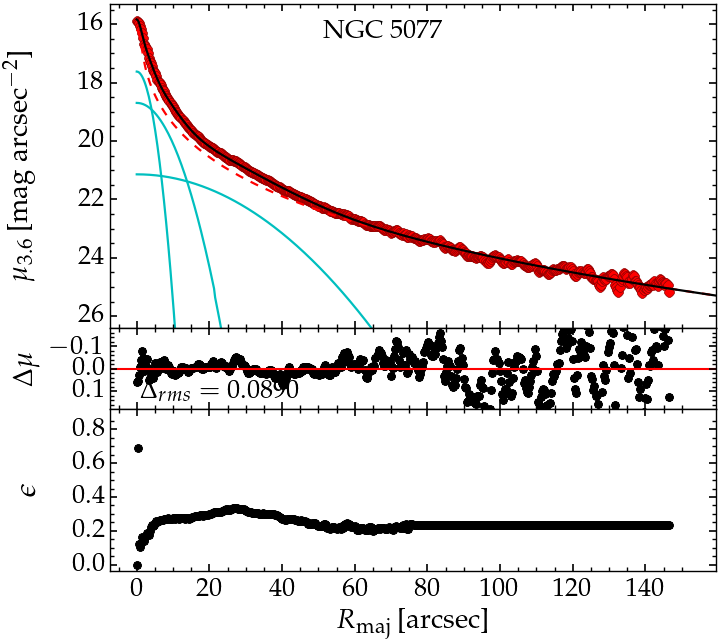} 
\end{center}
\caption{Similar to Fig.~\ref{Fig_N3379}, but for NGC~5077, which retains
  its E galaxy designation, although it arguably may be considered an ES
  galaxy, like NGC~4291.
   Image courtesy of {\it SHA}. 
The scale bar is 50$\arcsec$ $=$ 9.45~kpc long. 
   The light profile is decomposed into a S\'ersic spheroid (dashed red curve:
   $n_{\rm maj}=4.40$, $R_{\rm e,maj}=39\arcsec.7$, and $\mu_{\rm e}=21.87$ mag
   arcsec$^{-2}$), a broad Gaussian ring aligned with the major axis and having 
   a peak ring-to-spheroid flux occurring at $\sim$28$\arcsec$ 
  \citep{2021AandA...650A..34R}, and two (near) nuclear components associated with
  unrelaxed accreted gas/dust aligned with the minor axis and is the reason
  why this is not considered an ES galaxy.  
  The spheroid-to-total flux ratio is $0.80^{+0.05}_{-0.10}$.
}
\label{Fig_N5077} 
\end{figure*}

NGC~5077 has a total 3.6~$\mu$m magnitude equal to 10.49 mag, a luminosity
distance of 39.8$\pm$7.4~Mpc \citep{2007A&A...465...71T}, a
Galactic-extinction-corrected $(B-V)_{\rm Vega}$ colour of 0.99 mag and an
$M_{\star}/L_{3.6}=0.90$, giving $\log(M_{\rm \star,gal}/M_\odot)=11.37$ dex,
just 0.04 dex less than reported in \citet{Graham:Sahu:22a}.  From the
decomposition shown in Fig.~\ref{Fig_N5077}, $B/T=0.8$ and the spheroid
stellar mass is 11.27 dex. Arguably, some of the accreted material should now
be considered a part of the spheroid and a logarithmic mass between 11.27 and
11.37 dex used for the spheroid. A possible depleted core, with a major axis break radius
of 0$\arcsec$.2, was identified by \citet{2004AJ....127.1917T}.

\begin{figure*}
\begin{center}
\includegraphics[trim=0.0cm 0cm 0.0cm 0cm, height=0.23\textwidth, angle=0]{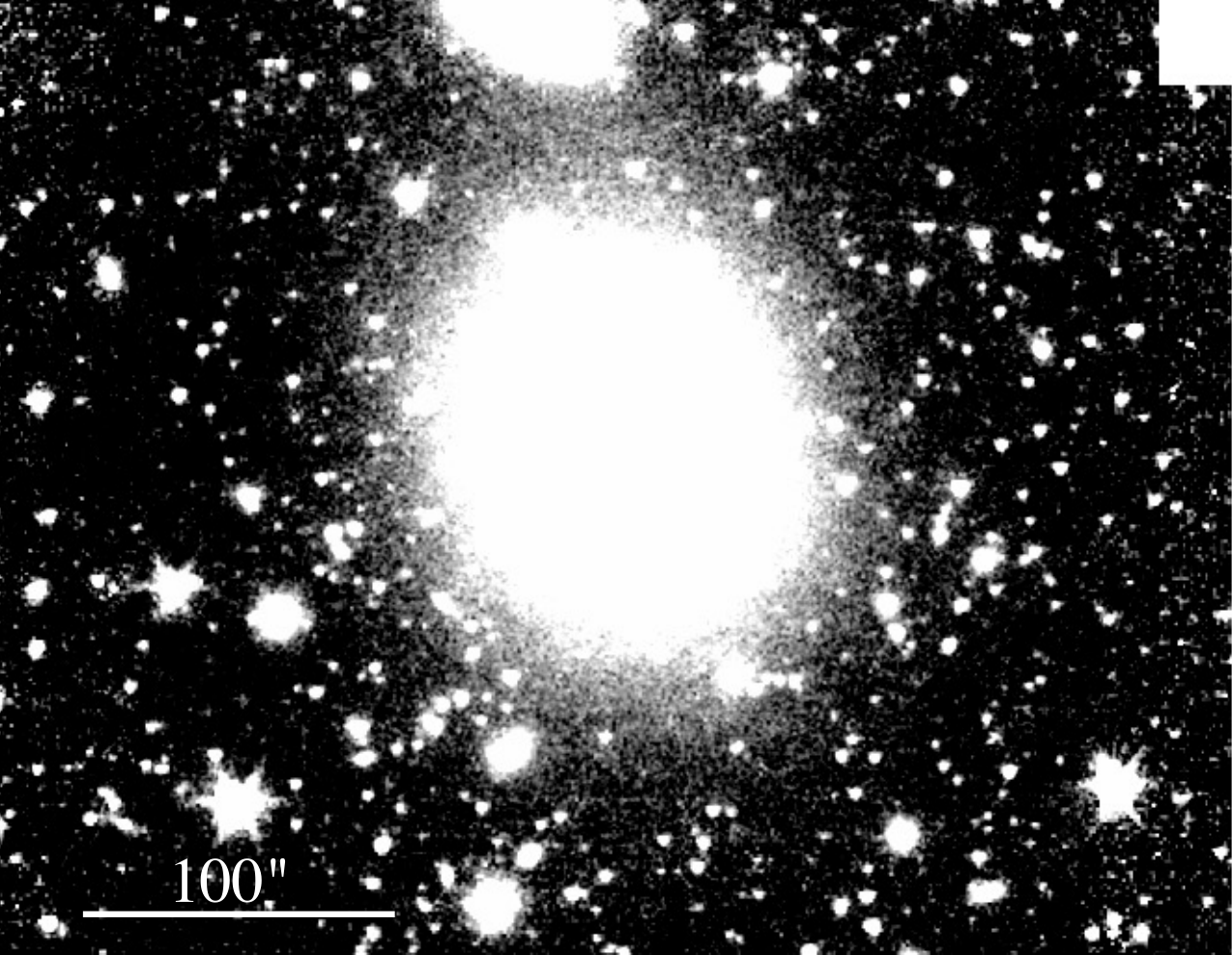} 
\includegraphics[trim=0.0cm 0cm 0.0cm 0cm, height=0.23\textwidth,
  angle=0]{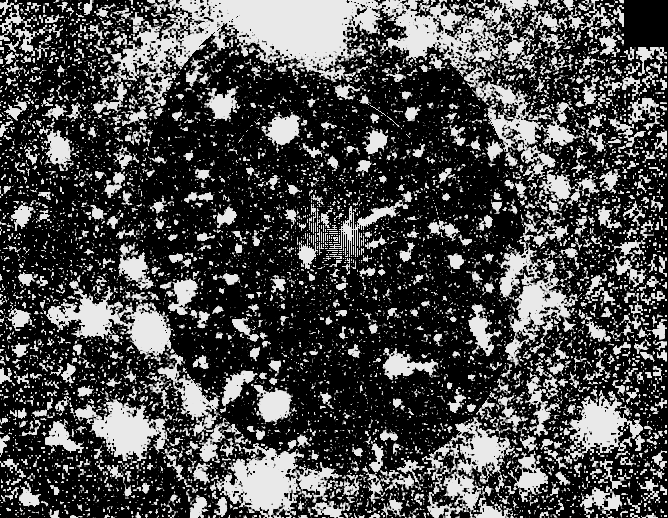}
\includegraphics[trim=0.0cm 0cm 0.0cm 0cm, height=0.23\textwidth,
  angle=0]{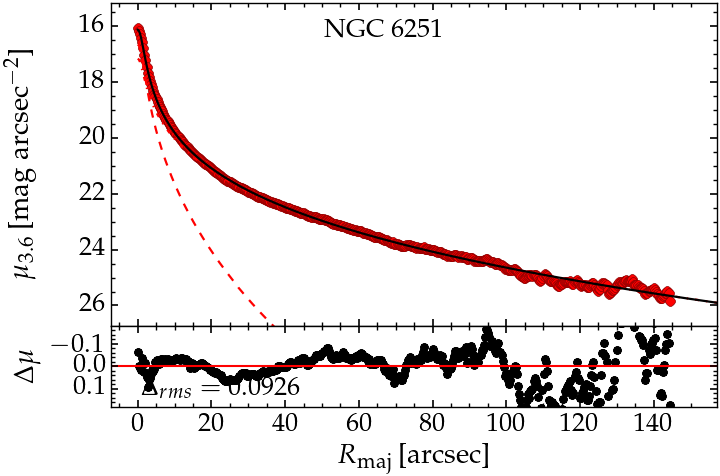} 
\end{center}
\caption{Similar to Fig.~\ref{Fig_N3379} but for the apparently `slow rotator'
  NGC~6251, which retains
  its E galaxy designation.   Image courtesy of {\it SHA}. 
The scale bar is 100$\arcsec$ $=$ 48.3~kpc long. 
A poorly constrained inner S\'ersic function
(dashed red curve: $n_{\rm maj}=1.5$, $R_{\rm e,maj}=3\arcsec.5$, and $\mu_{\rm
  e,maj}=19.57$ mag arcsec$^{-2}$) 
is included to accommodate the feature associated with the dust lane/disc(?); which contains
4 per cent (when $n=0.5$) to 10 per cent (when $n=2.5$) of the total flux,
although this latter value is undoubtedly an upper limit due to the extrapolation of the
S\'ersic model. Arguably, NGC~6251 could be considered an ES galaxy. 
The dominant spheroid has $n_{\rm maj}=4.64$, $R_{\rm e,maj}=36\arcsec.5$, and
$\mu_{\rm
  e,maj}=22.29$ mag arcsec$^{-2}$. 
The AGN jet is visible in the residual image. 
}
\label{Fig_N6251}
\end{figure*}

\begin{figure}
\begin{center}
\includegraphics[trim=0.0cm 0cm 0.0cm 0cm, width=0.85\columnwidth, angle=0]{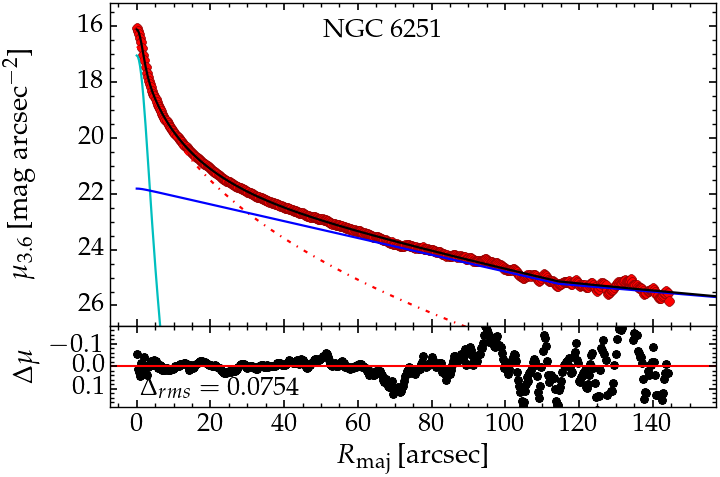}
\end{center}
\caption{Alternative (questionable and not preferred) fit for the `slow
  rotator' NGC~6251 obtained by adding an anti-truncated disc
  model (bent blue line: $\mu_0=21.78$ mag arcsec$^{-2}$, $h_{\rm inner,maj}=36\arcsec$,
  $R_{\rm bend,maj}=114\arcsec$, $h_{\rm outer,maj}=98\arcsec$)
  and using a S\'ersic function with $n=0.5$, i.e.\ a
  Gaussian (cyan curve: $\mu_0=15.53$ mag arcsec$^{-2}$,
  full-width-half-maximum $=1\arcsec.26$), for the inner component, yields a S\'ersic bulge (red curve) with
  $\mathfrak{M}_{3.6,sph} = 12.02$ mag, $n_{\rm maj}=2.67$, and $R_{\rm e,maj}=10\arcsec.37$.
  The `disc', or rather `halo', magnitude in this fit is 11.76 mag, and the total magnitude is 11.09 mag.
  }
\label{Fig_N6251-disc}
\end{figure}

\subsubsection{NGC~6251}
\label{Sec-6251}

The supergiant E radio galaxy NGC~6251 \citep{1977MNRAS.181..465W} 
is remodelled in Fig.~\ref{Fig_N6251}.
This Seyfert~2 galaxy has a one-sided jet --- reported at radio wavelengths by \citet{1977MNRAS.181..465W}
and \citet{1984ApJS...54..291P} --- that 
can be seen in the residual image (Fig.~\ref{Fig_N6251}, middle panel). 
The dominant spheroid's S\'ersic index equals 4.0, and its apparent magnitude
equals 11.29~mag.  The galaxy's total magnitude is
$\mathfrak{M}_{3.6,gal} = 11.18$~mag.  At a distance of 105~Mpc, and with
$M/L=0.73$, one has $\log(M_{\rm \star,gal}/M_\odot)=11.84$~dex, in good
agreement with the value of 11.87$\pm$0.15~dex reported by \citet{Graham:Sahu:22a}.  The
spheroid mass is slightly smaller, at 11.80~dex.  Consequently, NGC~6251
resides rightward of the main E galaxy distribution in the $M_{\rm
  bh}$--$M_{\rm \star,gal}$ diagram (Fig.~\ref{Fig-M-M-M}).  This situation
could arise if NGC~6251 experienced multiple 
mergers, more like a BCG than a regular E galaxy, with dry mergers driving it
rightward of the steep quasi-quadratic $M_{\rm bh}$--$M_{\rm \star,gal}$
relation for E galaxies built by, on average, one major dry merger.

In an exploration for a large-scale disc, or perhaps a captured/assimilated
outer rotating component or a halo of stars, 
Fig.~\ref{Fig_N6251-disc} presents a potential but questionable solution.
This is akin to what was done with NGC~3607 in Fig.~\ref{Fig_N3607-disc}. 
The exponential disc scalelength is 36$\arcsec$ (17~kpc) before anti-truncation at
$R_{\rm maj}=114\arcsec$ increases the scalelength to 98$\arcsec$ (47~kpc). 
The reduced spheroid in Fig.~\ref{Fig_N6251-disc} has 
 $\log(M_{\rm \star,gal}/M_\odot)=11.50$~dex. 
If applicable, NGC~6251 would join NGC~3091, reclassified as an S0 with an
unusually large galaxy stellar mass around $5\times10^{11}$~M$_\odot$. 
Should extended 
kinematic information become available and reveal  rotation in NGC~6251 at 
$R_{\rm maj} \gtrsim 20$-30$\arcsec$, it might help solidify the
disc-to-spheroid transition process as a route to the `S\'ersicification' of E galaxies. 
However, 
the exponential structure in Fig.~\ref{Fig_N6251-disc} is very extended, reminiscent of the
halos/envelopes around cD and BCGs that were discovered to have a tendency for
exponential light profiles \citep{2007MNRAS.378.1575S}. Although, despite residing
in the $M_{\rm bh}$--$M_{\rm \star,sph}$ diagram 
next to the second brightest Fornax Cluster galaxy, NGC~1399,
NGC~6251 is
in a poor cluster \citep{2000MNRAS.317..105W, 2011MNRAS.412.2433C}
and is, therefore, not expected to be surrounded by a halo of stars
stripped from other galaxy cluster members.

\subsection{A brightest cluster galaxy (BCG)}
\label{Sec-App-BCG}

\subsubsection{NGC~4486 (M87)}

\begin{figure*}
\begin{center}
\includegraphics[trim=0.0cm 0cm 0.0cm 0cm, height=0.27\textwidth, angle=0]{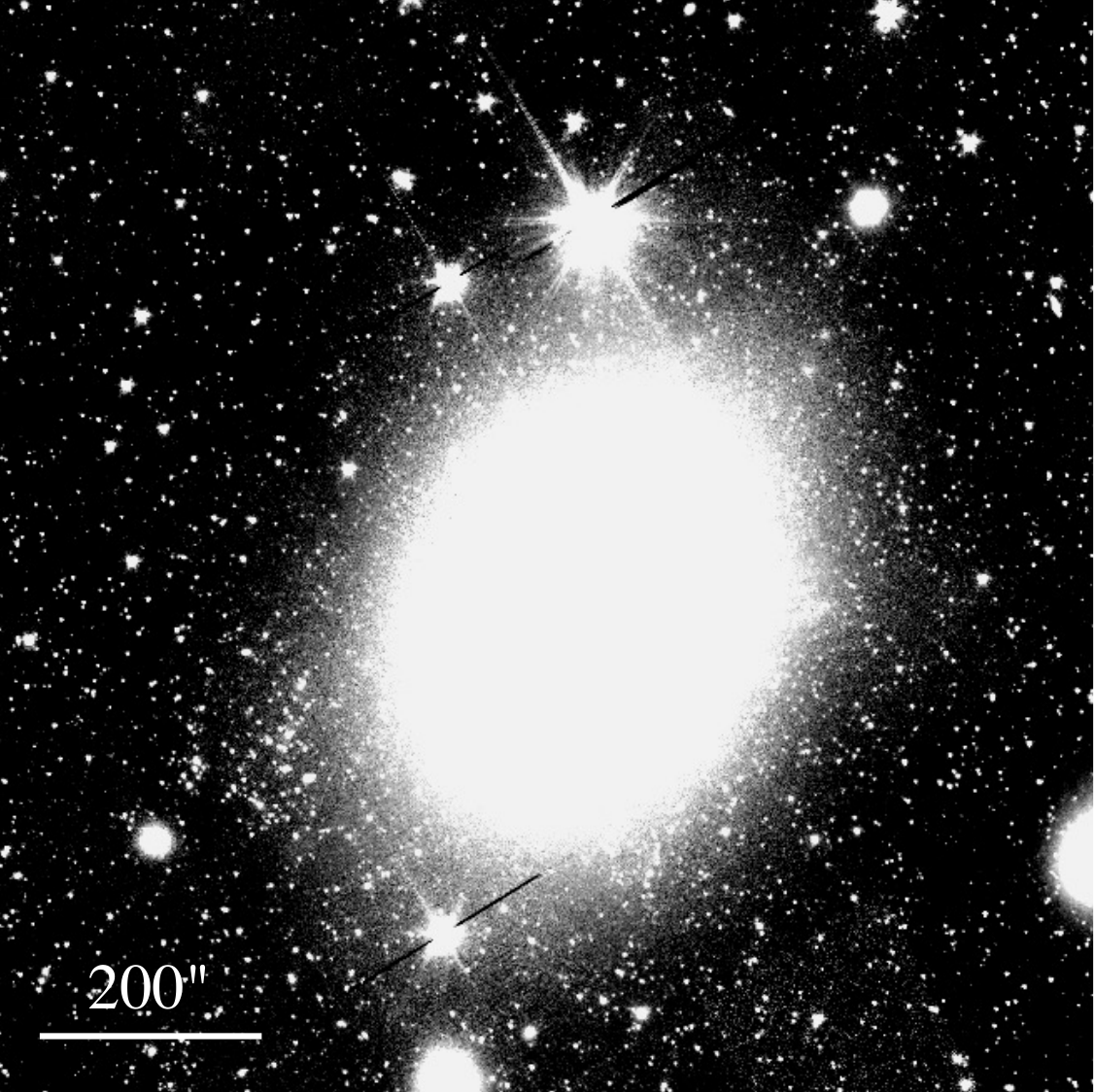}
\includegraphics[trim=0.0cm 0cm 0.0cm 0cm, height=0.27\textwidth, angle=0]{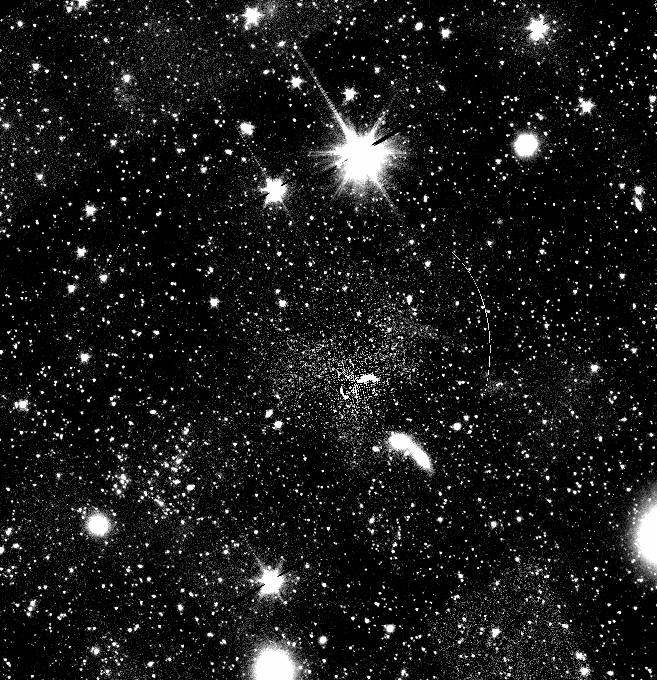}
\includegraphics[trim=0.0cm 0cm 0.0cm 0cm, height=0.27\textwidth, angle=0]{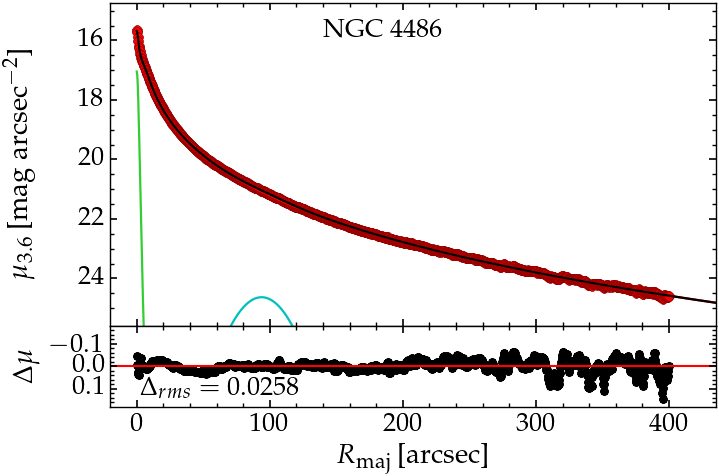}
\end{center}
\caption{Similar to Fig.~\ref{Fig_N3379} but for NGC~4486, which retains
  its E galaxy designation.   Image courtesy of S$^4$G.
The scale bar is 200$\arcsec$ $=$ 16.2~kpc long.
The light profile is decomposed into a central point-source
($\mathfrak{M}_{3.6,psf} = 15.04$ mag) plus a core-S\'ersic spheroid ($n_{\rm
  maj}=4.83$, $R_{\rm e,maj}=100\arcsec.4$, $R_{\rm b,maj}=6\arcsec.7 =
0.54$~kpc, $\gamma_{\rm maj}=0.37$, and $\mathfrak{M}_{3.6,sph} = 7.99$ mag)
and a Gaussian $\mathfrak{M}_{3.6,Gauss} = 12.22$ mag) to accommodate the
extra light centred around 94$\arcsec$.  The galaxy magnitude
$\mathfrak{M}_{3.6,gal} = 7.96$ mag. }
\label{Fig_N4486}
\end{figure*}

NGC~4486 is modelled in Fig.~\ref{Fig_N4486} with a core-S\'ersic function \citep{2003AJ....125.2951G},
a small central point source. and an extended excess (perhaps an
undigested meal) centred around
94$\arcsec$ along the major axis.  The galaxy's magnitude is 7.96~mag.
At a luminosity distance of 16.8$\pm$0.8~Mpc \citep{2019ApJ...875L...1E}, and using
$M/L=0.80$ (based on $B-V=0.94$), one can derive $\log(M_{\rm
  \star,gal}/M_\odot) = 11.58$~dex, 0.09~dex smaller than reported in \citet{Graham:Sahu:22a}.

\bsp    
\label{lastpage}
\end{document}